\documentclass[a4paper,11pt]{article}
\usepackage{jheppub}
\usepackage{amsmath,amssymb,amscd,braket,amsfonts,mathrsfs}
\usepackage{color}
\usepackage[table]{xcolor}
\usepackage{graphicx}
\usepackage{slashed}
\usepackage{amsthm}
\usepackage{verbatim}
\usepackage{mathtools}
\usepackage{bbold}
\usepackage[linesnumbered,ruled]{algorithm2e}
\usepackage{caption}
\usepackage{subcaption}

\newcommand{\bea}{\begin{eqnarray}}
\newcommand{\eea}{\end{eqnarray}}
\newcommand{\be}{\begin{equation}}
\newcommand{\ee}{\end{equation}}
\newcommand{\ba}{\begin{align}}
\newcommand{\ea}{\end{align}}

\newcommand{\floor}[1]{\lfloor#1\rfloor}

\title{
The pseudospectra of black holes in AdS
}

\author[a]{Bradley Cownden,}
\emailAdd{brad.cownden@ucd.ie}

\author[a]{Christiana Pantelidou}
\affiliation[a]{School of Mathematics and Statistics, University College Dublin, Belfield, Dublin 4, Ireland}
\emailAdd{christiana.pantelidou@ucd.ie}

\author[b]{and Miguel Zilhão}
\affiliation[b]{Centre for Research and Development in Mathematics and Applications (CIDMA),\\ 
Department of Mathematics, University of Aveiro, 3810-193 Aveiro, Portugal}
\emailAdd{mzilhao@ua.pt}

\abstract{We study the stability of quasinormal modes (QNMs) in electrically charged black brane spacetimes that asymptote to AdS by means of the pseudospectrum. Methodologically, we adopt ingoing Eddington–Finkelstein coordinates to cast QNMs in terms of a generalised eigenvalue problem involving a non-selfadjoint operator; this simplifies the computation  significantly  in comparison with previous results in the literature. 
Our analysis reveals spectral instability for (neutral) scalar as well as gravitoelectric perturbations. This indicates that the equilibration process of perturbed black branes is sensitive to external perturbations. Particular attention is given on the hydrodynamic modes, which are found to be the least unstable. In contrast with computations in hyperboloidal coordinates, we find that the pseudospectral contour lines cross to the upper half plane. This indicates the existence of pseudo-resonances as well as the possibility of transient instabilities. We also investigate the asymptotic structure of pseudospectral contour levels and we find remarkable universality across all sectors, persistent in the extremal limit.}

\begin{document}
\maketitle

\section{Introduction}

For self-adjoint operators present in non-dissipative systems,  the spectral theorem guarantees the stability of the spectrum under perturbations, meaning that small-scale perturbations of the operator will lead to a small movement of the operator's eigenvalues on the complex plane. In contrast, in the case of dissipative systems such as black holes, the spectral theorem is not applicable leading to lack of completeness in the set of eigenfunction and their orthogonality, potentially giving rise to instabilities. Such instabilities will manifest themselves as strong sensitivity of the eigenvalues (i.e.\ large eigenvalue migrations) to small-scale perturbations of the operator.

Mathematical tools for detecting such instabilities without the need to compute eigenvalues have been extensively used in the literature, including fields like quantum mechanics and hydrodynamics. One such tool is the pseudospectrum \cite{trefethen2005spectra}. A  particularly interesting application of the pseudospectrum comes from the study of the stability of fluid flows~\cite{treften90hydro}. It is well known that fluid flows that are laminar (stable) at low speeds become unstable and then turbulent at higher speeds.  This phenomenon was traditionally investigated by linearising the Navier-Stokes equation and testing for unstable eigenvalues of the linearised problem, but the results agreed poorly in many cases with experiments. Considerations of the pseudospectrum of the linearised problem resolved this tension~\cite{treften90hydro}, by suggesting that small perturbations of the smooth flow may be amplified substantially by linear mechanisms. Motivated by the seminal work of Nollert and Price \cite{Nollert:1998ys,Nollert:1996rf} (see also~\cite{Daghigh:2020jyk}), the pseudospectrum was recently introduced in the context of black hole physics in  asymptotically flat~\cite{Jaramillo:2020tuu,Jaramillo:2021tmt,Destounis:2021lum,alsheikh:tel-04116011,Boyanov:2023qqf,Gasperin:2021kfv}, in asymptotically anti-de~Sitter (AdS)~\cite{Arean:2023ejh} and in asymptotically de~Sitter (dS) geometries~\cite{Sarkar:2023rhp,Destounis:2023nmb}, aiming to provide a better understanding of the structural stability of quasinormal modes (QNMs).

In the case of asymptotically flat spacetimes, other than a conceptual problem in black hole physics, the study of the pseudospectrum also has implications for the black hole spectroscopy program: given that, in the near future, high-precision gravitational wave observations from binary black hole mergers will allow for a detailed inspection of the spectra of astrophysical black holes during the ringdown phase~\cite{Nollert:1999ji,lisaconsortiumwaveformworkinggroup2023waveform}, it is necessary to understand if the QNM spectrum itself is affected by the astrophysical environment, quantum corrections, or other generic modifications.  The pseudospectrum analysis  for the Schwarzschild and Reissner–Nordstr\"om~(RN) black holes was carried out in \cite{Jaramillo:2020tuu} and \cite{Destounis:2021lum} respectively, and found spectral instability of scalar and polar gravitoelectric quasinormal modes in subextremal and extremal black holes. The response of the Schwarzschild black hole spectrum to specific potential perturbations was studied in \cite{Cheung:2021bol,Jaramillo:2020tuu,Konoplya:2022pbc}. It was shown that overtones generically deviate at an increasing rate from their non-deformed limits, while the fundamental mode is stable under deformations localised close to the horizon~\cite{Konoplya:2022pbc,Jaramillo:2020tuu} but unstable under perturbations that are localised further away~\cite{Cheung:2021bol,Jaramillo:2020tuu}. This raised concerns about the robustness of the spectroscopy program, which let to a detailed study of the physical behaviour of time-domain signals~\cite{Berti:2022xfj}.  The pseudospectrum was also studied for exotic horizonless objects~\cite{Boyanov:2022ark}.

In the case of asymptotically AdS black branes in 5 dimensions, the pseudospectrum for neutral scalar and transverse electromagnetic perturbations of the planar Schwarzschild black hole was studied in~\cite{Arean:2023ejh}.  The aim of~\cite{Arean:2023ejh} was to shed a new light on the behaviour of strongly coupled quantum many-body systems modelled by the gauge/gravity duality. Compared to the asymptotically flat case, dissipation in AdS is reduced given that the AdS boundary works as a box not allowing energy to escape to infinity. However, due to the presence of the event horizon, the system is still dissipative and the spectral operator is still non-self adjoint. As such, similarly to the asymptotically flat case, the pseudospectrum revealed spectral instability.

 While the computation we perform in this paper follows the computations already done in the literature, an important difference is the coordinate basis used. In asymptotically flat spacetimes the pseudospectrum computation is carried out in hyperboloidal coordinates~\cite{Jaramillo:2020tuu, Destounis:2021lum,Boyanov:2022ark}. Other than compactifying the domain, this set of coordinates has the added benefit of making manifest ingoing boundary conditions at the horizon and outgoing boundary conditions at future null infinity. In asymptotically AdS, \cite{Arean:2023ejh} used a set of ``regular compact coordinates'' that asymptote to ingoing Eddington–Finkelstein coordinates close to the horizon. This made manifest the ingoing boundary conditions at the horizon, while at the boundary of AdS they imposed that the source for the scalar field vanishes.  In both of these cases, the equations of motion for the perturbations were second order in time and a time-reduction procedure was employed to cast the spectral problem as a regular eigenvalue problem. In contrast, our calculation employs Eiddington-Finkelstein coordinates throughout the spacetime leading to a problem that is first order in time and produces a generalised eigenvalue problem when solving for the quasinormal frequencies. Note that in \cite{Arean:2023ejh} the authors also considered formulating the problem in purely ingoing Eddington–Finkelstein coordinates, but ultimately decided to proceed using the above coordinate system.

In this work, we expand the previous work done for asymptotically AdS black holes by (i) considering perturbations of a massless scalar as well as electromagnetic and gravitational perturbations around the Reissner–Nordstr\"om (RN) black hole and (ii) formulating the pseudospectrum computation as a generalised eigenvalue problem in ingoing Eddington–Finkelstein coordinates.  The QNM spectrum  for black holes/branes in AdS has been extensively studied in the literature~\cite{Policastro:2002se,Kovtun:2005ev}, including more mathematical work~\cite{Warnick:2013hba}. Thanks to the coordinate system used, ingoing boundary conditions at the horizon become manifest.\footnote{In addition to ingoing boundary conditions at the horizon, we also need to ensure the absence of sources at infinity.}  Moreover, the equations of motion for the  perturbation are now first order in time and in fact the time derivative appears as a mixed $(v,z)$ derivative, where $v$ in the ingoing Eddington–Finkelstein coordinate and $z$ is a compactified radial coordinate. Solving for the spectrum can then be recast as a generalised eigenvalue problem instead of a standard one~\cite{trefethen2005spectra}. Computations of the pseudospectrum in the case of generalised eigenvalue problems are less developed in the literature~\cite{alsheikh:tel-04116011,trefethen2005spectra}. In particular, there are different definitions of the pseudospectrum depending on which operators are perturbed and one needs to resort to physical arguments in order to select the definition most suitable to the problem at hand. In our case, we perturb only the physically meaningful part of the operator. After this work appeared on the arXiv, recent work on the convergence of the pseudospectrum in AdS~\cite{Boyanov:2023qqf} showed that the pseudospectrum convergence properties present a better behaviour in null coordinates than in hyperboloidal slicing and they are consistent with Warnick’s theorems~\cite{Warnick:2013hba} for AdS black hole QNMs (when an appropriate norm is considered). It is currently not understood what is the underlying reason behind the lack of convergence in the case of hyperboloidal coordinates.

Our results for the scalar field are presented in section~\ref{ssec:scalar} as function of the momentum $k$ and charge $Q$ (or equivalently, the temperature $T$). In particular, we see that all modes in the spectrum are unstable for sufficiently large perturbations, with the fundamental mode being the least unstable. A novel observation for these spectra, seen also in the gravitoelectric case, is that the pseudospectral contour lines cross in the upper half plane when subject to perturbations of size~$< \mathcal{O}(1)$. This translates into an unstable perturbed spectrum and suggest the existence of transient instabilities and the possibility of pseudo-resonances~\cite{Boyanov:2022ark,Jaramillo:2022kuv}, both of which are mechanisms that give rise to non-linear instabilities. 

The pseudospectrum for metric and gauge field perturbations is computed in section~\ref{ssec:metricandgauge}. Note that in the case of the AdS-RN black hole the gravitational and electromagnetic perturbations do not decouple from each other. However, it is well understood that they can be split into three sectors: the scalar, vector and tensor\footnote{The tensor channel is present only in bulk dimensions $D\geq 5$.} sectors,  depending on how the perturbations transform under rotations on the plane transverse to the momentum. Each of the scalar and vector sectors contain two channels: in the scalar sector these are the sound and charge diffusion channels; in the vector sector these are the shear and transverse gauge channels. The tensor sector, on the other hand, only includes one channel and in fact behaves as a neutral massless scalar. Thus, its pseudospectrum is captured by the computation of section~\ref{ssec:scalar}. Our pseudospectral analysis demonstrates generic instability in all channels, with higher overtones more sensitive to perturbations. Particular attention is given to the channels that contain  hydrodynamic modes, namely modes that reside at the origin of the complex frequency plane at zero momentum; these are the shear, sound and charge diffusion. We find that these modes require very strong perturbations in order to be pushed to instability. Just like for scalar perturbations, we also observe that the pseudospectral contour lines cross to the upper half plane, suggesting possible transient black hole instabilities as well as pseudo-resonances~\cite{Boyanov:2022ark,Jaramillo:2022kuv}.  

The following two comments are now in order. With regards to the potential transient instabilities mentioned above, further analysis is required to check if these actually correspond to exponential growth of the evolution operator along the lines of~\cite{embree2017pseudospectra,Boyanov:2022ark,Jaramillo:2022kuv}. It should be noted that previous pseudospectrum computations in hyperboloidal coordinates do not present any such indications~\cite{Arean:2023ejh}. This highlights the need to understand the dependence of the pseudospectrum on the spacetime slicing. With regards to the nature of the instabilities observed, in order to understand this better we have also studied the response of the spectrum to specific potential perturbations. This indicated similar behaviour to asymptotically flat spacetime~\cite{Konoplya:2022pbc,Jaramillo:2020tuu}, namely that it is a ``high-wavenumber'' phenomenon.

 We have also investigated the asymptotic behaviour of pseudospectral contour levels at large real values of the spectral frequency, finding universality to be generic across different sectors. Similar universality was seen in 4-dimensional asymptotically flat spacetimes with either Schwarzschild~\cite{Jaramillo:2020tuu, Jaramillo:2021tmt} or RN black holes~\cite{Destounis:2021lum}. Crucially, our preliminary analysis indicates that the pseudospectral contour levels exhibit polynomial behaviour in the real part of the frequency, in contrast to the logarithmic behaviour noted in asymptotically flat spacetimes. This difference may arise from the imposition of reflective boundary conditions on the AdS boundary, as opposed to the Dirichlet conditions employed in asymptotically flat backgrounds. However, as we will discuss,  asymptotic analysis of the pseudospectrum in AdS is complicated by the necessity of additional terms in the energy inner product to ensure regularity in the complex plane as the distance from the real axis is increased~\cite{Boyanov:2023qqf}. In order to draw a firm conclusion, further examination of these effects is required.

The rest of the paper is organised as follows. In section~\ref{sec:spectra} we give the formal definition for the spectrum and pseudospectrum for a standard eigenvalue problem as well as a generalised eigenvalue problem
and in section~\ref{sec:spectra} we motivate our choice of the \emph{energy norm} as the fundamental scale of the system. In section~\ref{sec:implementation} we describe details of our numerical implementation, including the discretisation method used and the matrix representation of the energy norm in terms of the Gram matrix. Our results  for the pseudospectrum of the 5-dimensional planar AdS-RN black brane in ingoing Eddington–Finkelstein coordinates are then presented in section~\ref{sec:ResultsRN}. In particular, in section~\ref{ssec:scalar} we discuss the pseudospectrum of a neutral, massless scalar field, while in~\ref{ssec:metricandgauge} we consider gravitoelectric perturbation in all three channels. The universality of the asymptotic behaviour of the pseudospectral contour lines is discussed in~\ref{ssec:asymptBehaviour}. Finally, we conclude with an outlook and future directions in section~\ref{sec:Conclusions}. There are also two appendices. In appendix~\ref{sec:SAdS} we study the scalar spectrum of global AdS$_4$, where we find spectral stability associated with the lack of energy dissipation, while in appendix~\ref{sec:convergence} we discuss the numerical convergence of our QNM computation.

%%%%%%%%%%%%%%%%%%%%%%%%%%%%%%%%%%%%%%%%%%%%%%%%%%%%%%%%%%%%%%%%%%%%%%%%%%%%%%%%%%%%%%

\section{Spectra and pseudospectra}
\label{sec:spectra}

The notion of \emph{spectrum} of an operator is a familiar one in many areas of Physics, Mathematics and Engineering. It can be understood as a generalisation of the set of eigenvalues of a matrix. Recall that for a matrix $A$, its eigenvalues are the values of $\lambda$ that satisfy the equation
\[
\det(A-\lambda \mathbb{1})=0
\]
where $\mathbb{1}$ is the identity matrix.
If $\lambda$ is an eigenvalue of $A$, the operator $A-\lambda \mathbb{1}$ is not one-to-one and therefore its inverse $(A-\lambda \mathbb{1})^{-1}$ is not defined. 
We then define the \emph{spectrum of A} as
\begin{equation}
\label{eq:spectrum-def}
    \sigma(A) = \{ 
    \lambda \in \mathbb{C} :
    \lambda \mathbb{1} - A \text{ is not invertible}
    \}.
\end{equation}
It is also useful to define the \emph{resolvent} $R_A(\lambda) = (\lambda \mathbb{1} - A)^{-1}$ of the operator $A$ and note that if the spectrum $\sigma(A)$ is empty $R_A(\lambda)$ is defined everywhere in the complex plane.

The spectrum is an intrinsic property of an operator, and it is an incredibly useful concept.
Nevertheless, there are classes of problems where the associated methods may fail to give a useful answer. In particular, it is only for \emph{normal} operators $N$ satisfying $N N^\dagger = N^\dagger N$, where ${}^\dagger$ denotes the adjoint, that one can show that the spectrum of the operator is \emph{stable}.  
For non-normal operators, in general, small perturbations to the operator can drastically alter its spectrum.

To study such problems, Trefethen and collaborators~\cite{reddy1990lax, trefethen2005spectra} introduced the concept of \emph{pseudospectrum}. To motivate its definition let us start by noting that, from definition~\eqref{eq:spectrum-def}, whenever $\lambda$ is an eigenvalue of $A$, the operator $\lambda \mathbb{1} - A$ is singular. Equivalently, the resolvent $R_A(\lambda) = (\lambda \mathbb{1} - A)^{-1}$ is not defined.

Let us then define the \emph{$\epsilon$-pseudospectrum} as the set of points in the complex plane such that~\cite{trefethen2005spectra}
\begin{equation}
\label{eq:eps-pseudospectrum}
    \sigma^\epsilon(A) =
    \{\lambda\in\mathbb{C}: 
    \|R_A(\lambda)\| = 
    \|(\lambda \mathbb{1} - A)^{-1}\| > 1/\epsilon\} 
\end{equation}
for some choice of norm $\| \cdot \|$. Note that $\|R_A(\lambda)\| = \infty$ whenever $\lambda$ is an eigenvalue of $A$ and thus, with this convention, $\sigma^0(A) = \sigma(A)$.

It can also be shown~\cite{reddy1990lax} that the definition~\eqref{eq:eps-pseudospectrum} is equivalent to 
\begin{equation}
\label{eq:pseudospectrum-def2}
    \sigma^\epsilon(A) =
    \{ 
    \lambda \in \mathbb{C}, \, 
    \exists \, \delta A \text{ with } 
    \| \delta A \| < \epsilon :
    \lambda \in \sigma(A+\delta A) \},
\end{equation}
which means that the points in $\sigma^\epsilon(A)$ are the eigenvalues of an operator $A+\delta A$, that is, of some $\epsilon$-perturbation of the operator $A$. Note however that, while the spectrum is an intrinsic property of the operator, the pseudospectrum is not, as it depends also on the choice of the operator norm. Indeed, the choice of norm is crucial since it quantifies how large are the perturbations $\delta A$, and thus a small perturbation in a given norm could actually correspond to a large perturbation under another norm -- see also the relevant discussion in~\cite{Jaramillo:2020tuu,Gasperin:2021kfv}.

A related measure of the stability of a spectrum is the \emph{condition number} $\kappa$ of each eigenvalue. Let us consider a matrix $A \in \mathbb{C}^{N \times N}$ with a set of distinct eigenvalues $\{\lambda_1, \ldots, \lambda_N\}$, which implies the existence of a set of right eigenvectors ($v_j$) and left eigenvectors ($u_j$) that satisfy
\begin{align}
    \label{eigensystem}
    A v_j = \lambda_j v_j \qquad \text{and} \qquad u^*_j A = \lambda_j u^*_j \, .
\end{align}
Under the perturbation $A \to A(t) = A + t \,\delta A$ for\footnote{The definition of the matrix and vector norms will become central to our discussion in later sections. For now we leave the definition of $||\cdot||$ general.} $\|\delta A\| = 1$ and $t \in \mathbb C, |t| < 1$, the perturbed eigenpair $(\lambda_1(t), v_1(t))$ satisfy
\begin{align}
    \label{pert eigensystem}
    A(t) v_1(t) = \lambda_1(t) v_1(t) \, .
\end{align}
Expanding $\lambda_1(t)$ and $v_1(t)$ as Taylor series in $t$~\cite{trefethen2005spectra},
\begin{align}
    \lambda_1(t) &= \lambda_1 + \alpha_1 t + \alpha_2 t^2 + \ldots \\
    v_1(t) &= v_1 + \sum^N_{j=1} \left(t \beta^{(1)}_j + t^2 \beta^{(2)}_j + \ldots \right) v_j
\end{align}
we can take the inner product of~\eqref{pert eigensystem} with $u^*_1$ and see that
\begin{align*}
    \lambda_1(t) \langle u^*_1, v_1(t)\rangle = \lambda_1(t) \langle u^*_1, v_1\rangle  &= \langle u^*_1, A(t) v_1(t) \rangle \\
    &= \langle u^*_1, A v_1(t) \rangle + t \langle u^*_1, \delta A v_1(t) \rangle \\
    &= \lambda_1 \langle u^*_1, v_1\rangle  + t \langle u^*_1, \delta A v_1(t) \rangle \, .
\end{align*}
Thus, the perturbation of $A$ has moved the eigenvalue $\lambda_1$ by
\begin{align}
    |\lambda_1 - \lambda_1(t)| &= |t| \frac{|\langle u^*_1, \delta A v_1(t) \rangle|}{|\langle u^*_1, v_1\rangle|} \leq |t| \frac{\|u_1\|\,\|v_1(t)\|}{|\langle u^*_1, v_1 \rangle |} \notag \\
    &= |t| \frac{\|u_1^*\| \, \| v_1\|}{|\langle u^*_1, v_1\rangle |} + \mathcal{O}(t^2) \, ,
\end{align}
where we have used the Cauchy-Schwarz inequality and the fact that $\|\delta A\| = 1$. Defining the condition number of the $j$-th eigenvalue by
\begin{align}
    \label{condition number}
    \kappa_j = \frac{\|u_j\| \, \| v_j\|}{|\langle u^*_j, v_j \rangle |}
\end{align}
we see that the magnitude of $\kappa_j$ gives us an indication of the stability of $\lambda_j$ as it determines whether the shift in the eigenvalue remains of the order of the size of the perturbation $|t|$. Indeed, for normal operators with $A^\dagger = A$, $u_j$ is colinear with $v_j$ and so $\kappa_j = 1$. However, non-normal operators, i.e.\ unstable operators, may have non-colinear ${(u_j, v_j)}$, resulting in arbitrarily small values for $|\langle u^*_j, v_j \rangle|$ and arbitrarily large values of $\kappa_j$.

\subsection{Pseudospectra for matrix pencils}
\label{sec:PSmatrix-pencils}

The definitions \eqref{eq:eps-pseudospectrum} and \eqref{eq:pseudospectrum-def2} stem from an eigenvalue problem of the form $A v = \lambda v$.
As we will see, the 
computation of the spectrum (and, correspondingly, also of the pseudospectrum) for black holes in AdS becomes more natural in ingoing Eddington–Finkelstein coordinates.

This coordinate system naturally leads to a \emph{generalised} eigenvalue problem, in the form
\begin{equation}
\label{eq:gen-eigenvalue}
    A v = \lambda B v.
\end{equation}
The matrix $A - \lambda B $ is also known as a \emph{matrix pencil $(A,B)$}.

When considering the pseudospectrum in the form of 
a generalised eigenvalue problem, as we will do in this work, the available techniques are not as well-developed as for the standard eigenvalue case.
Herein we will follow the approach of reference~\cite{van1997pseudospectra}, which we now outline.

Take the generalised eigenvalue problem, equation~\eqref{eq:gen-eigenvalue}, 
with $B$ potentially singular. The (generalised) eigenvalues of the matrix pencil $(A,B)$ are the set of complex numbers $\lambda$ for which $\det(A-\lambda B) = 0$. This set of eigenvalues, as usual, is called the \emph{spectrum} and is denoted by $\sigma(A,B)$.  Analogously, the \emph{$\epsilon$-pseudospectrum of the matrix pencil $(A,B)$} is defined as~\cite{riedel1994generalized,van1997pseudospectra}
\begin{equation}
\label{eq:pencil_PS1}
    \sigma^\epsilon(A,B) =
    \{\lambda\in\mathbb{C}: 
    \|(\lambda B - A)^{-1}\| > 1/\epsilon\} 
\end{equation}
for some choice of norm, which naturally reduces to~\eqref{eq:eps-pseudospectrum} when $B=\mathbb{1}$. Furthermore, it is shown in~\cite{van1997pseudospectra} that definition~\eqref{eq:pencil_PS1} is equivalent to

\begin{equation}
\label{eq:pencil_PS2}
    \sigma^\epsilon(A,B) =
    \{ 
    \lambda \in \mathbb{C}, \, 
    \exists \, \delta A \text{ with } 
    \| \delta A \| < \epsilon :
    \lambda \in \sigma(A+\delta A,B) \}.
\end{equation}
In other words, $\sigma^\epsilon(A,B)$ is the actual spectrum for some $\epsilon$-perturbation of $A$, and we recover the (usual) definition~\eqref{eq:pseudospectrum-def2} for the pseudospectra of an operator $A$ when $B=\mathbb{1}$. This definition does not consider perturbations to the operator $B$ and is therefore not as general as others considered in the literature~\cite{trefethen2005spectra}. However, as we will argue later on, for our present purposes this is the definition of interest since all relevant physical perturbations will be fully contained within the operator $A$.

The notation of the condition number also extends to matrix pencils. While defining a condition number in the case of a singular $B$ matrix is more subtle (see~\cite{Stewart1975GershgorinTF, James1993} for discussions on this topic), when $B$ is non-singular the definition of a condition number follows from applying the mechanism outlined in section~\ref{sec:spectra} to the matrix product $B^{-1}A$.

\subsection{Energy norm}
\label{ssec:QNMs and energy norm}

Following the discussion of the previous section, we need to define an appropriate norm~$\| \cdot \|$ to quantify the size of perturbations. A natural definition is to use the \emph{energy norm} associated with the energy of each mode present in the spectrum~\cite{Gasperin:2021kfv,Jaramillo:2020tuu}. 

An intuitive way to obtain such a norm is as follows. 

We start by considering the equation governing the spectrum of interest, which in a wide range of cases (and in particular for all cases we will consider) reduces to a complex scalar field that 
obeys the wave equation in 1+1 Minkowski spacetime with a potential $V$.\footnote{In the context of black hole physics, it can be shown that a wide class of perturbations around a black hole background take this form. The coordinate $y$ is identified with the tortoise coordinate $y=r_*$ and the potential  depends, among other things, on the concrete black hole solution being perturbed and the specific type of perturbation~\cite{Berti:2009kk}. 
} In Cartesian coordinates $(t,y)$ this reads
\begin{equation}
\label{eq:wave}
    \left( \square_{_2} - V \right) \phi(t,y) = 0 ,
\end{equation}
where $\square_{_2}=-\partial^2_t + \partial^2_{y}$.
 Multiplying the latter by $\partial_t \, \bar \phi$, where ${}~\bar~{}~$ denotes complex conjugation, and summing with its complex conjugate, we  obtain~\cite{bartnik1996wave}
\[
  \partial_t \left(
        \partial_t \bar \phi \, \partial_t \phi
        + \partial_{y} \bar \phi \, \partial_{y} \phi
        + V \, \bar \phi \phi
\right) = 
  \partial_{y} \left(
              \partial_t \bar \phi \, \partial_{y} \phi 
            + \partial_t \phi \, \partial_{y} \bar \phi
        \right).
\]
Integrating both sides in $y$,
we find that solutions of equation~\eqref{eq:wave} obey
\begin{equation}
\label{eq:dEdt}
    \frac{d}{dt} E(t) =               
             \frac{1}{2} \partial_t \bar \phi \, \partial_{y} \phi \Big|_{{y = a}}^{{y = b}}
            + \frac{1}{2} \partial_t \phi \, \partial_{y} \bar \phi \Big|_{{y = a}}^{{y = b}}
\end{equation}
where
\begin{equation}
\label{eq:energynorm0}
    E(t) = \frac{1}{2} \int_{a}^{b}  \left(  \partial_t \bar \phi \, \partial_t \phi
        + \partial_{y} \bar \phi \, \partial_{y} \phi
        + V \, \bar \phi \phi
        \right) dy
\end{equation}
and $a,b$ are the boundaries of our domain. $E(t)$ can be interpreted as the ``total energy'' contained in a spatial slice $t=\text{constant}$, and we see from equation~\eqref{eq:dEdt} that the energy is conserved if there are no modes entering or leaving the physical domain at $y=a,b$.

Another way of obtaining the energy norm that does not rely on a coordinate system was introduced in~\cite{Jaramillo:2020tuu}. There one starts by defining the usual stress-energy tensor of the scalar field $\phi$ 
\begin{equation}
    T_{ab} = \frac{1}{2} \nabla_a \bar \phi \nabla_b\phi
    + \frac{1}{2} \nabla_a \phi \nabla_b \bar \phi
    - \frac{\eta_{ab}}{2} \left(
      \nabla_c \phi \nabla^c \bar \phi + V \phi \bar \phi
    \right) ,
\end{equation}
where $\eta_{ab}$ is the 2-dimensional Minkowski metric (in arbitrary coordinates).
We can then define the energy on a hypersurface $\Sigma$ 
through
\begin{equation}
\label{eq:Edef}
    E = \int_{\Sigma} T_{ab} \, \xi^a \kappa^b \, d \Sigma \,.
\end{equation}
$\xi$ is the timelike Killing vector associated with stationarity  
and $\kappa$ is the normal vector to the hypersurface $\Sigma$. 
In $(t,y)$ coordinates and when the hypersurface $\Sigma$ is chosen to correspond to a $t=\text{constant}$ slice, $\xi = \partial_t $ and it is easy to show that~\eqref{eq:Edef} reduces to expression~\eqref{eq:energynorm0}. 

It is illuminating to also consider~\eqref{eq:Edef} on a null hypersurface.\footnote{ In the context of black holes, a natural choice for a null surface is a $v=\text{constant}$ slice, where $v$ corresponds to the ingoing Eddington–Finkelstein coordinate defined as $v=t+y$, where $y=r_*$ is the tortoise coordinate.}
In particular, let us consider the coordinate system $(v,Y)$ where
\begin{align}
   v=t+y\,,\qquad Y=y
\end{align}
and the metric element reduces to
\[
ds^2 = -dv^2 + 2 dv \, dY\,.
\]
Let us also choose the hypersurface $\Sigma_v$ to correspond to  $v=\text{constant}$. In this case,  $\xi = \partial_t = \partial_v$ and $\kappa$ is the normal vector to $\Sigma_v$. Since $\kappa_a = -\partial_a v$, we have $\kappa = -\partial_Y$. We then have $T_{ab} \, \xi^a \kappa^b = - T_{v\small{Y}} = \frac{1}{2} \partial_Y \phi \partial_Y \bar \phi + \frac{1}{2} V \phi \bar \phi $. Finally, since $d\Sigma_v = -\kappa_v dY = dY$, we obtain~\eqref{eq:energynorm1}.
\begin{equation}
\label{eq:energynorm1} 
    E[\phi] = \frac{1}{2} \int_{\Sigma_v}  \left( 
        \partial_{Y} \bar \phi \, \partial_{Y} \phi
        + V \, \bar \phi \phi
        \right) dY.
\end{equation}
This expression motivates the definition of the scalar product
\begin{equation}
    \label{eq:scalar-prod-AdS}
    \langle \phi_1, \phi_2 \rangle_{_E}
    =  \int_a^b \left(  \partial_Y \bar \phi_1 \, \partial_Y \phi_2 + V \bar \phi_1 \, \phi_2 \right) dY ,
\end{equation}
where $a,b$ are now the boundary points in the $Y$ coordinate.

\section{Matrix formulation}
\label{sec:implementation}

In this section we give details about the discretisation approach we follow in order to construct the spectrum and the pseudospectrum of our systems. Sections~\ref{sec:grid} and~\ref{ss:chebyshev int} follow the approach of~\cite{Jaramillo:2020tuu}, which we reproduce herein for completeness. 

\subsection{Choosing the grid}
\label{sec:grid}

Based on the geometry of the problem, different choices of abscissa are available~\cite{Grandclement:2007sb}. The most common choices are the Gauss-Chebyshev points, given by ${x^{N}_i = \cos(\pi (2i - 1) / 2N)}$, ${i = 1, \ldots, N}$, and the Chebyshev-Lobatto points, ${x^N_i = \cos( \pi i /N), \, i = 0, \ldots, N}$. Note that the Chebyshev-Lobatto abscissa includes the endpoints $x = \pm 1$ while the Gauss-Chebyshev do not; this is relevant whenever the functions being evaluated are singular at the ends of the grid. 
In this work we choose to use the Chebyshev-Lobatto points (denoted $\bar x_i$) in part because we have specific boundary conditions that must be enforced and also because of a convenient overlap between a grid of $N$ points, $\bar x^N_i$, and a grid of $\bar x^{2N}_i$ that allows for easy interpolation. This will be important during the numerical evaluation of the energy inner product, as we will explain below.

First, for completeness, we note that there are known expressions for the discretised first and second derivative operators in terms of simple functions of the gridpoints $\bar x_i$ when using Chebyshev collocation methods. In particular, the first and second derivative matrices for the Chebyshev-Lobatto abscissa are
\begin{align}
    \mathbb{D}^{(1)} &= 
        \begin{dcases}
            -\frac{2N^2 + 1}{6} & \text{if } i=j=0 \\
            \frac{2N^2 + 1}{6} & \text{if } i=j=N \\
            - \frac{\bar x_j}{2(1-\bar x_j^2)} & \text{if } 0 < i = j < N \\
            \frac{\kappa_i}{\kappa_j} \frac{(-1)^{i-j}}{(\bar x_i - \bar x_j)} & \text{if } i \neq j
        \end{dcases}
\end{align}
and
\begin{align}
    \mathbb{D}^{(2)} &= 
        \begin{dcases}
            \frac{N^4 - 1}{15} & \text{if } i = j = 0, N \\
            -\frac{1}{(1-\bar x_i^2)^2} - \frac{N^2 - 1}{3(1-\bar x_i^2)} & \text{if } 0 < i = j < N \\
            \frac{2}{3} \frac{(-1)^j}{\kappa_j} \left[ \frac{(2N^2 + 1)(1 - \bar x_j) - 6}{(1-\bar x_j)^2 }\right] & \text{if } i = 0,~ 0 < j \leq N \\
            \frac{2}{3} \frac{(-1)^{N+ j}}{\kappa_j} \left[ \frac{(2N^2 + 1)(1 + \bar x_j) - 6}{(1+\bar x_j)^2 }\right] & \text{if } i = N,~ 0 \leq j < N \\
            \frac{(-1)^{i-j}}{\kappa_j} \left[ \frac{\bar x_i^2 + \bar x_i \bar x_j - 2}{(1-\bar x_i^2)(\bar x_j - \bar x_i)^2 }\right] & \text{if } i \neq j,~ 0 < i < N
        \end{dcases}
\end{align}     
where
\[
\kappa_i = 
        \begin{dcases}
            2, & i = 0,~N \\
            1, & 0 < i < N
        \end{dcases}
\]

\subsection{Chebyshev integration}
\label{ss:chebyshev int}

Numerical evaluation of the energy inner product defined in~\eqref{eq:scalar-prod-AdS} involves evaluating integrals of the form
\begin{align}
    \mathcal{I}(f) = \int^1_{-1} f(x) dx \, .
\end{align}
To do so, we first expand the function in a basis of Chebyshev polynomials\footnote{Exact equivalence is achieved in the limit $N \to \infty$, but at finite $N$ the relation remains approximate.},
\begin{align}
	f(x) \simeq f_N(x) = \frac{c_0}{2} + \sum^{N}_{i=1} c_i T_i(x) \, ,
\end{align}
and then, using properties of the $T_i(x)$, find that
\begin{align}
    \int^{1}_{-1} f(x) dx \approx c_0 - \sum_{k=1}^{\floor{N/2}}\frac{c_{2k}}{4k^2 - 1} \, .
\end{align}
More practical, however, is the writing of the coefficients $c_i$ in terms of the function values at the collocation points $\bar x$:
\begin{align}
    \label{eq:spectral coeff at coll}
    c_i = \frac{(2-\delta^i_N)}{2N} \left( f(\bar x_0) + (-1)^i f(\bar x_N) + 2\sum_{j=1}^{N-1} f(\bar x_j) T_i(\bar x_j)\right) \, .
\end{align}
Combining these expressions, the integral $\mathcal{I}(f)$ can be written as
\begin{align}
    \mathcal{I}(f) \approx \mathcal{I}_N(f) = \int^1_{-1} f(x) dx = f^T(\bar x) \cdot C(\bar x)
\end{align}
where $f^T(\bar x)$ is a row vector of the function evaluated on the $\bar x$ abscissa and the vector $C$ has entries
\begin{align}
    C_{i} = \frac{2}{\kappa_i N} \left[ 1 - \sum_{k=1}^{\floor{N/2}} \frac{T_{2k}(\bar x_i) (2 - \delta^N_{2k})}{4k^2 - 1} \right] \, .
\end{align}

When the integrand is a product of functions, each of which is described in terms of an interpolation over the basis of Chebyshev functions, the integral
\begin{align}
    \mathcal{I}(f(x)g(x)) = \int^1_{-1} f(x) g(x) d\mu 
\end{align}
with respect to the weight $d\mu = \mu(x) dx$ is a simple extension of the integral of a single function,
\begin{align}
	\mathcal{I}(f(x)g(x)\mu(x)) \simeq f^T(\bar x) \cdot C_\mu(\bar x) \cdot g(\bar x)
\end{align}  
where $C_\mu(\bar x)$ is now a diagonal ${(N+1)\times(N+1)}$ matrix with components
\begin{align}
\label{Gauss quad}
	(C_\mu)_{ii} = \frac{2 \mu(\bar x_i)}{\kappa_i N} \left[ 1 - \sum_{k=1}^{\floor{N/2}} \frac{T_{2k}(\bar x_i) (2 - \delta^N_{2k})}{4k^2 - 1}\right] \, .
\end{align}
Thus, an inner product of the form in~\eqref{eq:scalar-prod-AdS} can be approximated by the discrete expression 
\begin{equation}
    \label{eq:scalar-prod-G}
    \langle \phi_1, \phi_2 \rangle_{_G}
    = (\phi_1^*)^i \, G_{ij} \, (\phi_2)^j
    = \phi_1^* \cdot G \cdot \phi_2
    ,
\end{equation}
where ${\phi}^* = \bar\phi^t$. The matrix $G$ is known as the \emph{Gram matrix}, corresponding to the inner product of interest.

Finally, we address an important technical note regarding integral quadrature and grid interpolation: integral quadrature over two functions -- each of which are approximated by polynomials of degree $N+1$ -- requires a grid of $2N + 2$ points to be spectrally accurate~\cite{boyd2013chebyshev}. Therefore, when evaluating an inner product of the form~\eqref{eq:scalar-prod-G}, where $\phi_1$ and $\phi_2$ are each approximated by $N + 1$ Chebyshev polynomials, the Gram matrix $G$ containing the quadrature weights must be of size $(2N+2)\times(2N+2)$. Practically speaking, we achieve this by evaluating~\eqref{Gauss quad} at double the spectral resolution and then interpolating down to $(N+1)\times(N+1)$ by the application of an interpolation matrix $\mathfrak{I}$ via $\mathfrak{I}^t \cdot G \cdot \mathfrak{I}$. The form of $\mathfrak{I}$ can be determined by evaluating the expression for the spectral coefficients in~\eqref{eq:spectral coeff at coll} at the double-resolution collocation points, thereby giving the elements of the (Chebyshev-Lobatto) interpolation matrix
\begin{align}
    \mathfrak{I}_{\bar{i} i} = \frac{1}{\kappa_i N}\left( 1 + \sum_{k=1}^N (2 - \delta_{k,N})T_k(\bar{x}_{\bar{i}}) T_k (\bar x_i)\right) \, ,
\end{align}
where $\bar{i} \in [0,2N+1]$.

\subsection{Reduction of energy norm to the $\ell^2$ norm}
\label{ss: e-norm to L2-norm}

Once discretised and interpolated by the methods described above, we arrive at the Gram matrix $G$ for the energy inner product described by~\eqref{eq:scalar-prod-G}. Since $G$ will be Hermitian and positive definite, it admits a Cholesky decomposition of the form $G = F^* F$.  We can then follow~\cite{trefethen2005spectra,reddy1993pseudospectra} to express the (discretised) energy norm $\| \cdot \|^2_{_G} = \langle \cdot, \cdot \rangle_{_G} $ (which we will refer to as the $G$ norm) as an $\ell^2$ norm\footnote{For completeness, note that the $\ell^2$ matrix norm (induced by the vector $\ell^2$ norm) is given by the largest singular value of the matrix, i.e.\ the spectral norm.}:
\begin{align}
    \| M \|^2_{_G} & = \underset{x \in \mathbb{C}^n}{\max} 
    \frac{\| M x \|^2_{_G}}{\|x\|^2_{_G}} 
    = \underset{x \in \mathbb{C}^n}{\max}
        \frac{ \langle M x, Mx \rangle_{_G} }{\langle x, x \rangle_{_G}}
    = \underset{x \in \mathbb{C}^n}{\max}
        \frac{x^* M^* F^* F M x}{x^* F^* F x} \notag \\
    & = \underset{y \in \mathbb{C}^n}{\max}
        \frac{y^* {F^{-1}}^*  M^* F^* F M F^{-1} y}{y^* y}
    =  \underset{y \in \mathbb{C}^n}{\max}
        \frac{ \| F M F^{-1} y \|^2_{_2} }{ \|y\|^2_{_2}}
    = \| F M F^{-1} \|^2_{_2} \, .
\end{align}
When applied to the norm of the resolvent $R_{A,B}(\lambda) \equiv \left( \lambda B - A \right)^{-1}$ used in equation~\eqref{eq:pencil_PS1}, we can straightforwardly see that
\begin{equation}
    \left\| \left( \lambda B - A \right)^{-1} \right\|_{_G}
    = \left\| \left( \lambda \tilde B - \tilde A \right)^{-1} \right\|_{_2}
\end{equation}
with $\tilde B=F B F^{-1}$,  $\tilde A=F A F^{-1}$, and thus the $\epsilon$-pseudospectrum~\eqref{eq:pencil_PS1} in the $G$ norm $\sigma^\epsilon_{_G}$ can be expressed as the ``usual'' $\epsilon$-pseudospectrum in the $\ell_2$ norm $\sigma^\epsilon_{_2}$,
\begin{equation}
\label{eq:sigmaG_from_sigma2}
    \sigma^\epsilon_{_G}(A,B) = \sigma^\epsilon_{_2}(\tilde A, \tilde B) .
\end{equation}

The advantage of this is that it can be shown (see, e.g.,~\cite{trefethen2005spectra,van1997pseudospectra}) that the $\epsilon$-pseudospectrum can then be computed from
\begin{equation}
\label{eq:pencil_PS3}
    \sigma^\epsilon_{_2}(\tilde A, \tilde B) =
    \{\lambda\in\mathbb{C}: s^{\min}(\lambda \tilde B - \tilde A) < \epsilon\} 
\end{equation}
where $s^{\min}$ denotes the smallest singular value\footnote{Numerically, this is achieved by using the randomized subspace method outlined in~\cite{halko2011finding} to compute only the smallest value rather than the entire set.}. This is the definition that we will be using for practical computations of the pseudospectrum, since it can be readily obtained from the singular value decomposition of the matrix $M \equiv \lambda F B F^{-1} - F A F^{-1}$.

%%%%%%%%%%%%%%%%%%%%%%%%%%%%%%%%%%%%%%%%%%%%%%%%%%%%%%%%%%%%%%%%%%%%%%%%
\section{Scalar, electromagnetic and gravitational perturbations for planar AdS-RN}
\label{sec:ResultsRN}

Let us now consider our physical setup, consisting of Einstein-Maxwell theory in $D = 5$ spacetime dimensions with action given by 
\begin{align}\label{eq:LagraEM}
S= \frac{1}{16 \pi G_5} \int d^{5}x \sqrt{-g} \left( R-2 \Lambda -\frac{1}{4} F_{\mu\nu}\, F^{\mu\nu}\right)
\end{align}
where $F\equiv dA$ and $\Lambda=-6$ is the cosmological constant. The variation of the action \eqref{eq:LagraEM} gives rise to the following field equations
\begin{equation}
\label{eq:eom}
\begin{aligned}
& R_{\mu\nu}-\frac{R}{2} g_{\mu\nu}-6g_{\mu\nu}-\frac{1}{2}\left(F_{\mu\rho}F_{\nu}{}^{\rho} -\frac{1}{4}\,g_{\mu\nu}F_{\lambda\rho}F^{\lambda\rho}\right) = 0\,, \\
& \nabla_{\mu}F^{\mu\nu} = 0\,.
\end{aligned}
\end{equation}
The above equations admit what is known as the AdS Reissner–Nordstr\"om (RN) black-brane solution.
Concretely, the metric and gauge field, in Poincaré coordinates, take the form
\begin{equation}
\begin{aligned}
    ds^2 & = -r^2 f(r) dt^2 + \frac{dr^2}{r^2 f(r)} 
    + r^2 \left( dx_1^2 + dx_2^2 + dx_3^2
    \right) \\
    A & =\mu\left(1 - \frac{r_\mathrm{h}^2}{r^2} \right) dt\\
    f&= 1-\frac{r_\mathrm{h}^4}{r^4}(1+Q^2)+\frac{r_\mathrm{h}^6}{r^6}Q^2\,,
\end{aligned}
\end{equation}
where $r_\mathrm{h}$ is the black hole horizon, $Q$ is the electric charge carried by the black hole and $\mu=\sqrt{3} \,r_\mathrm{h}\,Q$ is the chemical potential.  Without loss of generality, we will henceforth fix units where the horizon radius is at $r_\mathrm{h} = 1$ -- the explicit dependence can be reinstated through
\begin{equation}
r\to \frac{r}{r_\mathrm{h}},\qquad
    t \to t\, r_\mathrm{h}, \qquad
    x_i \to x_i \,r_\mathrm{h}, \qquad
    \mu \to \frac{\mu}{r_\mathrm{h}} \,.
\end{equation}

We can now perform the following coordinate transformation 
\begin{align}
    t = v -r_* \,,\qquad Y=r_*,
\end{align}
where $dY=dr_*=\tfrac{dr}{r^2 f(r)}$. For convenience, we also compactify the radial coordinate $r$ by introducing
\[
z = \frac{1}{r}
\]
so that our physical domain spans $z=0$ (the AdS boundary, $r\to \infty$) to $z=1$ (the black hole horizon, $r=1$). To fix notation, we will refer to this coordinate system $(v,z)$ as the ingoing Eddington–Finkelstein coordinates.

Overall, we obtain
\begin{equation}
\label{eq:norm_ansatz}
\begin{aligned}
ds^{2}&=\frac{1}{z^2}\left(- f(z) dv^2
-2 dv\, dz 
+dx_1^2+dx_2^2+dx_3^2\right),\\
A&=\mu(1-z^2)\, \left(dv+\frac{dz}{f(z)}\right)\,,
\end{aligned}
\end{equation}
where
\begin{equation*}
    f(z) = 1 - z^4 \left(1+Q^2\right) + z^6 Q^2.
\end{equation*}
The (dimensionless) temperature of the black hole is given by 
\begin{align}\label{tempexp}
\frac{T}{\mu} = \frac{1}{2\pi\,\mu}\left[2-Q^2\right]\,.
\end{align}
When $ Q=0$ the system is neutral and the bulk solution \eqref{eq:norm_ansatz} reduces to the AdS-Schwarzschild black brane, while $Q=Q_\mathrm{max}=\sqrt{2}$ corresponds to the extremal limit $T/\mu\to 0$.

\subsection{Massless neutral scalar perturbations}
\label{ssec:scalar}

We now consider a neutral, massless scalar field $\Phi$, in the background of the 5-dimensional planar AdS-RN solution described above. Such a perturbation obeys the Klein-Gordon equation
\begin{equation}
    \label{eq:KG0}
    \square \Phi = 0 \,,
\end{equation}
which can then be tackled in different coordinate systems.

\paragraph{Poincaré coordinates:} It is not difficult to show that, in Poincaré coordinates $(t,r)$, the above equation takes the form~\eqref{eq:wave}. To see that, we decompose the scalar as~\cite{Horowitz:1999jd}
\[
\Phi = \frac{1}{r^{\frac{3}{2}}} \phi(t,r) \, e^{i \vec{k} \cdot \vec{x} }\,
\]
and we substitute in equation \eqref{eq:KG0} to find 
\begin{align}
\label{eq:wave-AdS5}
&\left( -\partial^2_t + \partial^2_{r_*} - V(r) \right) \phi(t,r) = 0 , \\ 
&V(r) \equiv f(r) \left( k^2 + \frac{3}{4} f(r) 
+ \frac{3}{2} r \frac{df(r)}{dr} \right) \notag
\end{align}
and $\partial_{r_*} = r^2 f(r) \partial_r$.
These coordinates are not horizon-penetrating, however, and are therefore unsuitable to perform the pseudospectrum analysis. This fact has been noted also in~\cite{Arean:2023ejh}.

\paragraph{Ingoing Eddington–Finkelstein coordinates:} 
We consider instead ingoing Eddington–Finkelstein coordinates, which are regular at the black hole horizon and also allow one to reach the AdS boundary. With these coordinates, one can naturally 
impose the required boundary conditions at the two ends of the physical domain following surfaces of constant~$v$. Note that this is not possible in asymptotically flat spacetimes, where -- due to the different asymptotics -- surfaces of constant~$v$ lead one to past null infinity~$\mathscr{I}^-$ instead. Indeed, that is the major motivation for using the hyperboloidal coordinates originally developed in~\cite{Jaramillo:2020tuu} and used thereafter in the literature.\footnote{We thank Rodrigo Panosso Macedo for clarification of these points.} See e.g.\ reference~\cite{PanossoMacedo:2023qzp} for an overview of hyperboloidal coordinates and their applications. In reference~\cite{Arean:2023ejh}, the authors worked with ``compactified regular coordinates'' -- which is more similar to the hyperboloidal coordinates used in asymptotically flat cases.
We find this to be unnecessary, and prefer to keep the elegance and simplicity of the ingoing Eddington–Finkelstein system,
at the expense of dealing with a generalised eigenvalue problem (as opposed to a standard eigenvalue problem). As discussed in section~\ref{sec:PSmatrix-pencils}, this poses no problems in practice.

With the ansatz
\[
\Phi = z^{\frac{3}{2}} \phi(v,z) \, e^{i \vec{k} \cdot \vec{x} }\,,
\]
the Klein-Gordon equation \eqref{eq:KG0} becomes 
\begin{align}
    \label{eq:AdS5-wave-vz}
   & \left[
    -2\, \partial_{v z}
    + \partial_z \left( f(z) \partial_z \right)
    - \hat V(z)
    \right] \phi(v, z) = 0,\\
&\hat V(z)  \equiv \frac{V(z)}{ f(z)} 
  = k^2 
  + \frac{15}{4} \frac{ f(z)}{z^2}
  - \frac{3}{2} \frac{ f'(z)}{z}
  = k^2 + \frac{15}{4 z^2} 
  + \frac{9}{4} \left(1+ Q^2\right) z^2
  - \frac{21}{4} Q^2 z^4 \notag
\end{align}
where the prime ${}'$ denotes derivative with respect to $z$.

Since solutions to equation~\eqref{eq:AdS5-wave-vz} behave as $\phi \sim z^{5/2}$ close to the AdS boundary $z=0$, we
proceed to redefine
\begin{equation}
\phi(v,z)=z^{\frac{5}{2}} \psi (v,z).
\end{equation}

With this redefinition, the required boundary conditions reduce to demanding regularity of the fields at the two boundaries. The equation of motion now takes the form
\begin{align}
    \label{eq:AdS5-wave-vz2}
    &\left[
   - 2\, \partial_{v z}-\frac{5}{z} \partial_v
    + \partial_z \left( f(z) \partial_z \right)+\frac{5 f(z)}{z}\partial_z 
    - \hat U(z)
    \right] \psi(v, z) = 0,\\
&\hat U(z)  \equiv \frac{U(z)}{ f(z)} 
 = k^2 - 4\frac{f'(z)}{z}
  = k^2 + 8 z^2 (26(1+Q^2)-3Q^2 z^2)\,.\nonumber
\end{align}
The latter can be rewritten in the form of a generalised eigenvalue problem
\begin{equation}
L_2 \, \partial_v \psi = i L_1 \psi
\end{equation}
where
\begin{align}
i L_1 & = \left[ 
      \partial_z \left( z f(z)
       \partial_z
      \right)
      - \left( k^2 z - 4 f'(z) \right) \mathbb{1}
    \right]
    + 4 f(z) \partial_z ,  \\
L_2 &= 2 z \partial_z  + 5  \mathbb{1} .
\end{align}
Note that the $i L_1$ operator contains the full physics of the problem (thus justifying our definition~\eqref{eq:pencil_PS2} for the pseudospectrum) and that 
the term in square brackets has the structure of a Sturm–Liouville operator.

Equation~\eqref{eq:energynorm1} now takes the form
\begin{align}
    E[\psi] = \frac{1}{2} \int_0^1 dz
    \bigg[ & 
    z^5 f(z) \partial_z \bar{\psi} \, \partial_z \psi
    + \frac{5}{2} z^4 f(z)
    \left(
      \bar{\psi} \, \partial_z \psi
      +  \partial_z \bar{\psi} \, \psi
    \right) \notag \\
    &{}
    + z^3 \left(
       10 f(z) - \frac{3}{2}z f'(z) + k^2 z^2
    \right) \bar{\psi} \,\psi
    \bigg].
\end{align}
We note that every term in $L_1$, $L_2$ and $E[\psi]$ is explicitly regular for $z \in [0,1]$. This contrasts with the approach of~\cite{Arean:2023ejh}, where divergent terms at $z=0$ need special care.
We can now define the scalar product
\begin{align}
   \langle \psi_1, \psi_2 \rangle_{_E} = \frac{1}{2}  \int_0^1 dz
    \bigg[ & 
    z^5 f(z) \, \partial_z \bar{\psi}_1 \, \partial_z \psi_2
    + \frac{5}{2} z^4 f(z)
    \left(
      \bar{\psi}_1 \, \partial_z \psi_2
      +  \partial_z \bar{\psi}_1 \, \psi_2
    \right) \notag \\
    &{}
    + z^3
    \left(
        10 f(z) - \frac{3}{2}z f'(z) + k^2 z^2
    \right)
    \bar{\psi}_1 \,\psi_2
    \bigg]
\end{align}
as well as the energy norm
$\| \cdot \|^2_{_E} = \langle \cdot, \cdot \rangle_{_E}$. Following section~\ref{ss:chebyshev int}, we can also build its discretised counterpart 
\begin{equation}
    \langle \psi_1, \psi_2 \rangle_{_G}
    = \psi_1^* \cdot G^E \cdot \psi_2    ,
\end{equation}
with the Gram matrix
\begin{equation}
G^E = \frac{1}{2} \left( C_{V} +C_1 \cdot \mathbb{D}+\mathbb{D}^t \cdot C_1
  + \mathbb{D}^t \cdot C_2 \cdot \mathbb{D}
  \right)
\end{equation}
and 
$\left(C_a\right)_{ij} = \mu_a(z_i) \, W_i \, \delta_{ij}$ (no summing on $i$). $W_i$ are the quadrature weights introduced in section~\ref{ss:chebyshev int},
$\delta_{ij}$ is the Kronecker delta, and 
\begin{align*}
\mu_V(z) = z^3\left(
        10 f(z) - \frac{3}{2}z f'(z) + k^2 z^2
    \right), \qquad
\mu_1(z) = \frac{5}{2} z^4 f(z), \qquad
\mu_2(z) = z^5 f(z).
\end{align*}

\subsubsection{Results}

With the operators $L_1$, $L_2$ and the Gram matrix $G^E$ just defined, we are now ready to compute the $\epsilon$-pseudospectrum $\sigma^\epsilon$ of this system following the approach outlined in section~\ref{sec:implementation}, in particular equation~\eqref{eq:pencil_PS3}.

\begin{figure}[h]
    \centering
    \begin{subfigure}[t]{0.48\textwidth}
        \centering
        \includegraphics[width=\textwidth]{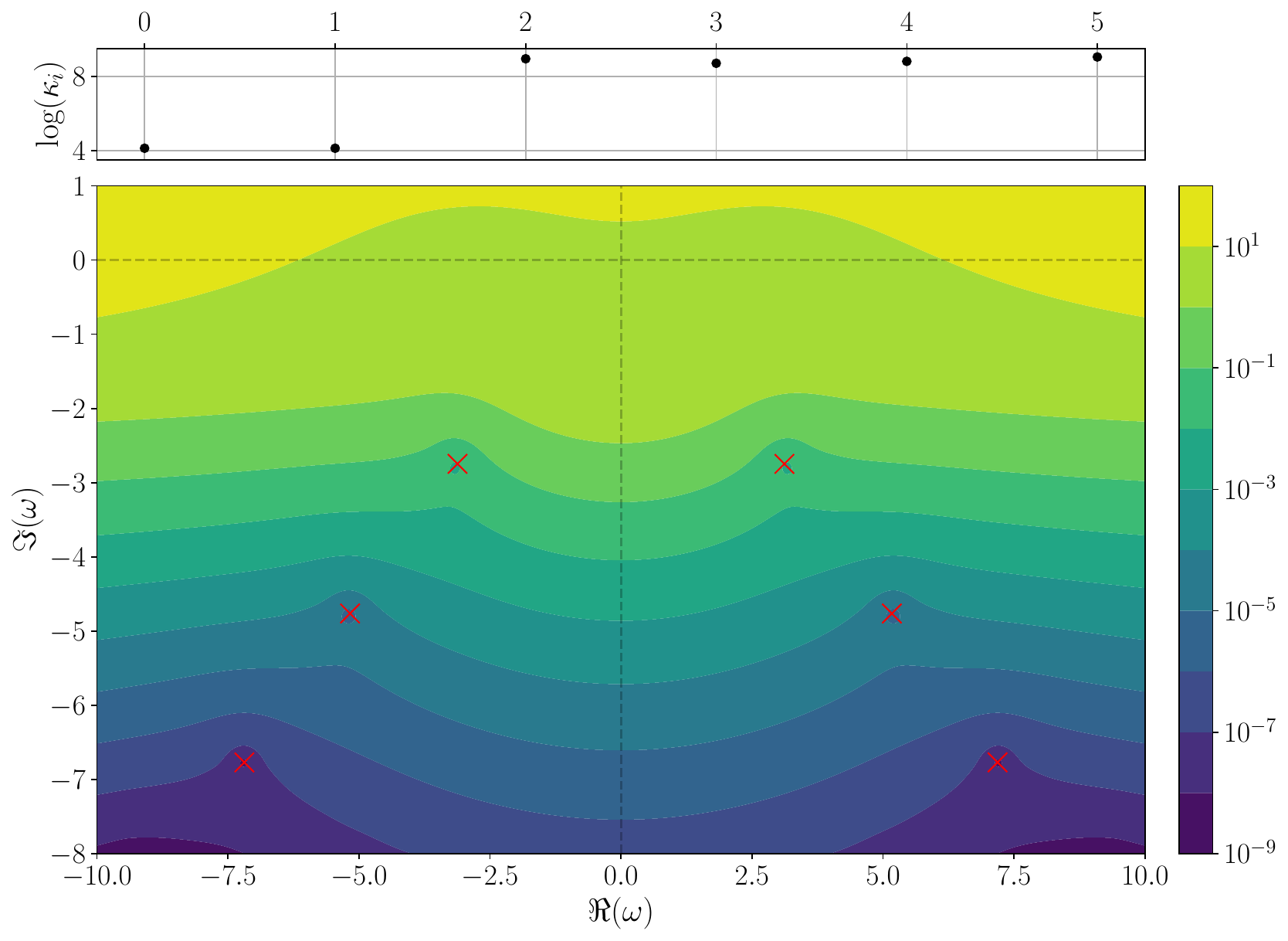}
        \caption{$k=0$, $Q=0$.}
        \label{fig:AdSplanark0q0}
    \end{subfigure}
    \;
    \begin{subfigure}[t]{0.48\textwidth}
        \centering
        \includegraphics[width=\textwidth]{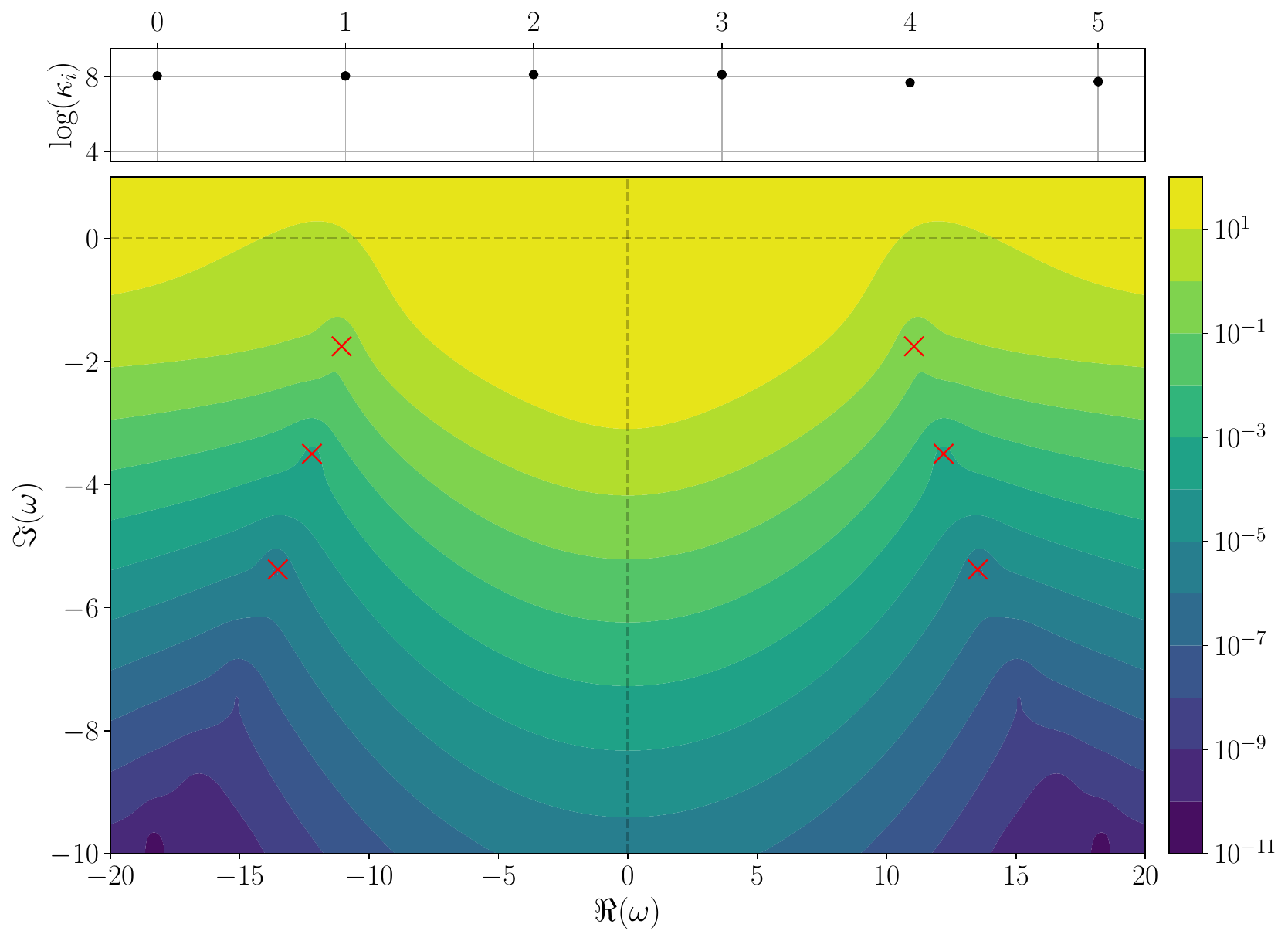}
        \caption{$k=10$, $Q=0$.}
        \label{fig:AdSplanark10q0}
    \end{subfigure}
    \\
    \begin{subfigure}[t]{0.48\textwidth}
        \centering
        \includegraphics[width=\textwidth]{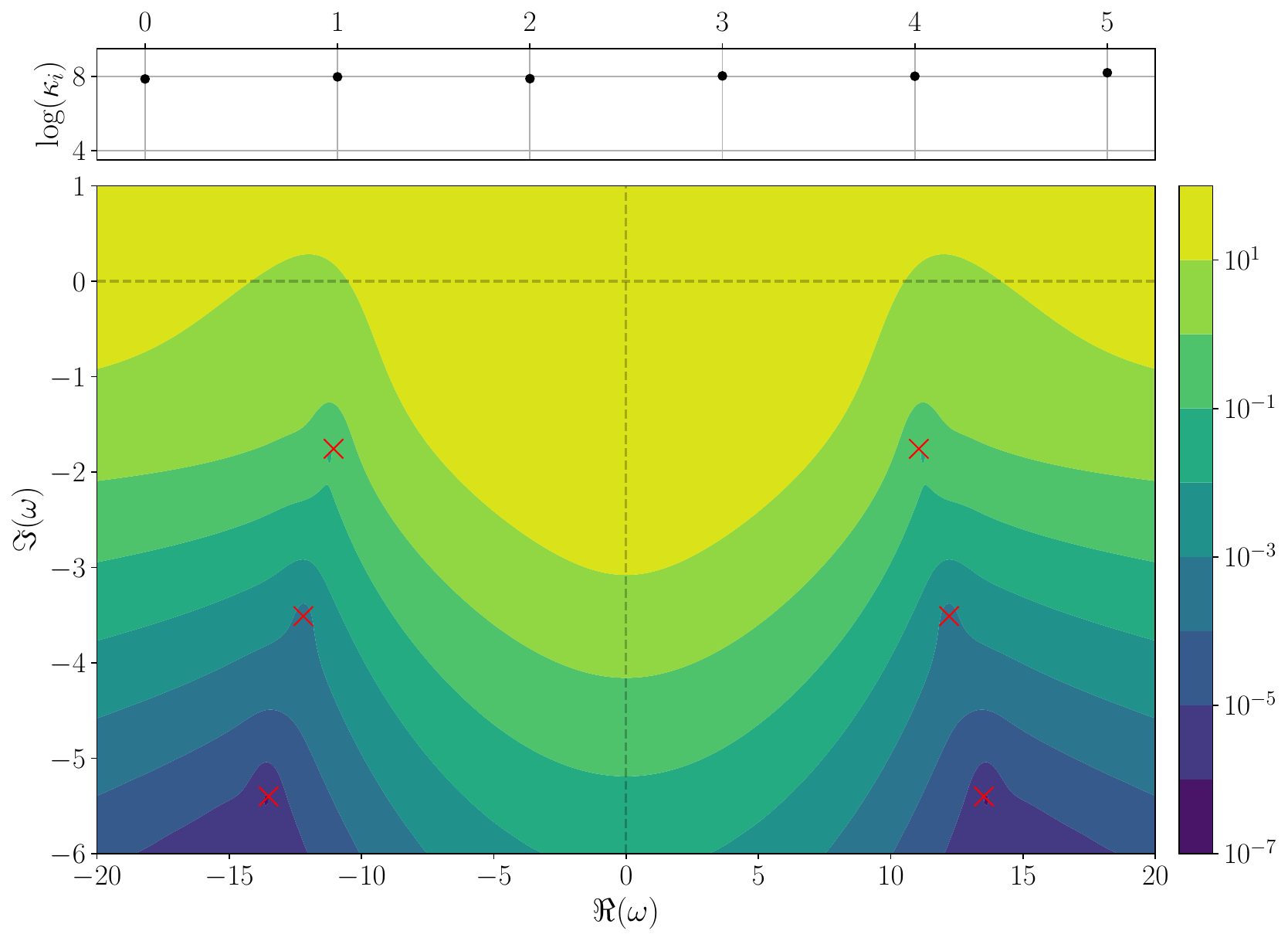}
        \caption{$k=10$, $Q=0.1$.}
        \label{fig:AdSplanark10q0.1}
    \end{subfigure}
    \;
    \begin{subfigure}[t]{0.48\textwidth}
        \centering
        \includegraphics[width=\textwidth]{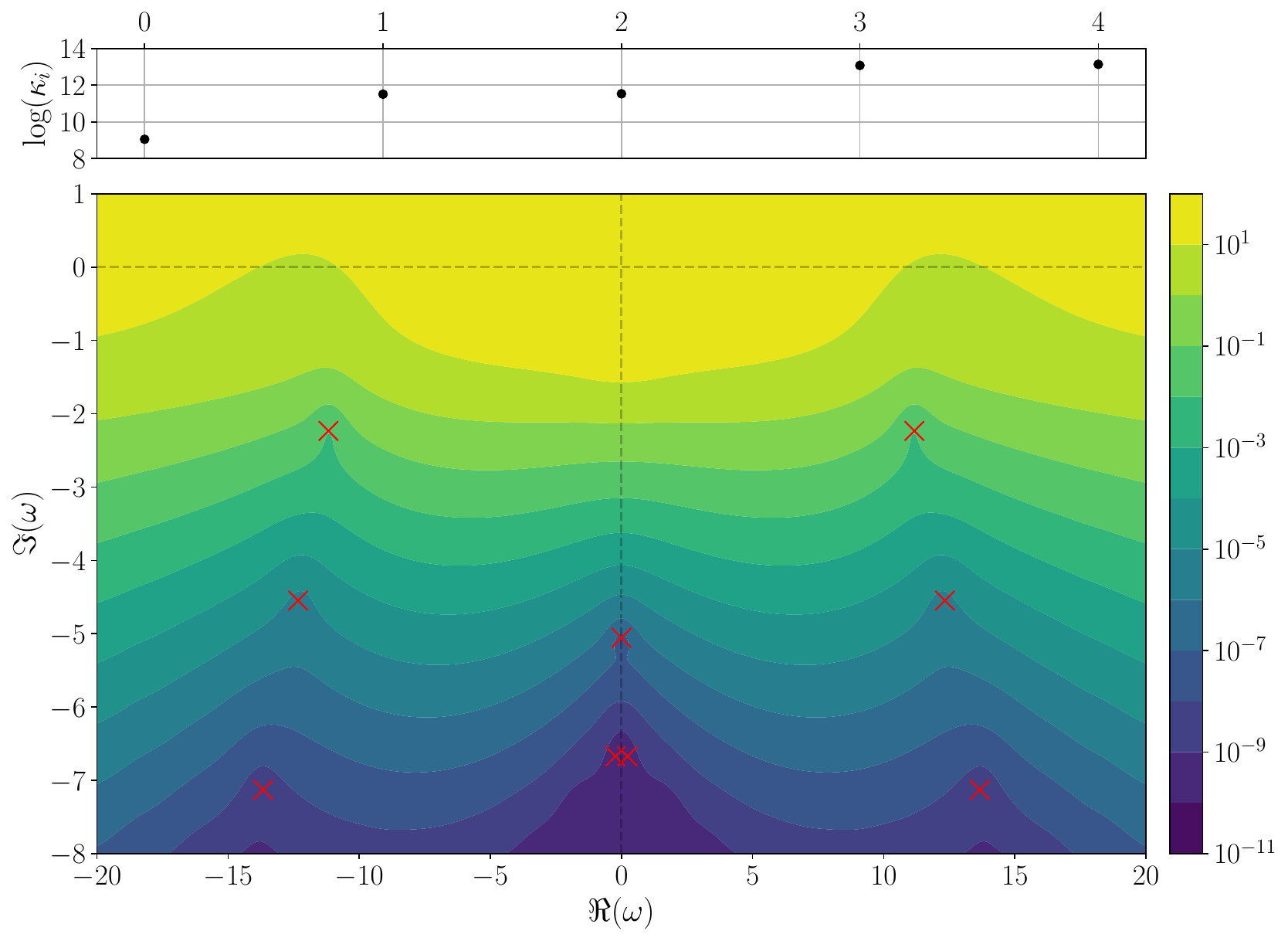}
        \caption{$k=10$, $Q=1 = 0.7 Q_\mathrm{max}$.}
        \label{fig:AdSplanark10q1}
    \end{subfigure}
    \caption{The $\epsilon$-pseudospectrum $\sigma^\epsilon$ for various planar black hole solutions coupled to a massless scalar field. Recall that the extremal solution is obtained for $Q_\mathrm{max}=\sqrt{2}$. Above each plot, the condition numbers $\kappa$ of the eigenvalues are plotted with increasing overtone number. The red crosses mark the spectrum points.}
    \label{fig:AdSplanarkq_scalar}
\end{figure}

\begin{figure}[thbp]
    \centering
    \includegraphics[width=0.48\textwidth]{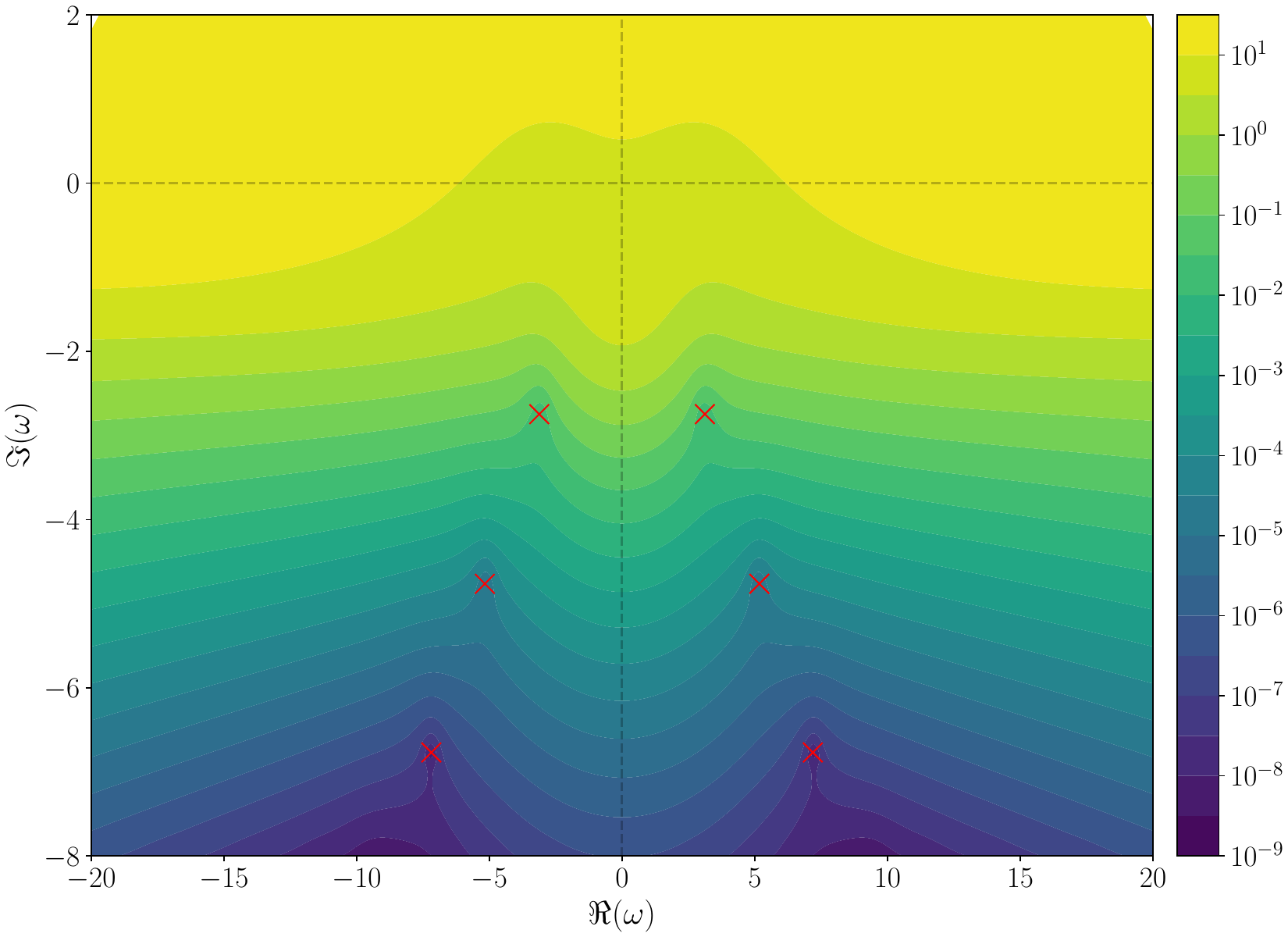}
    \hfill
    \includegraphics[width=0.48\textwidth]{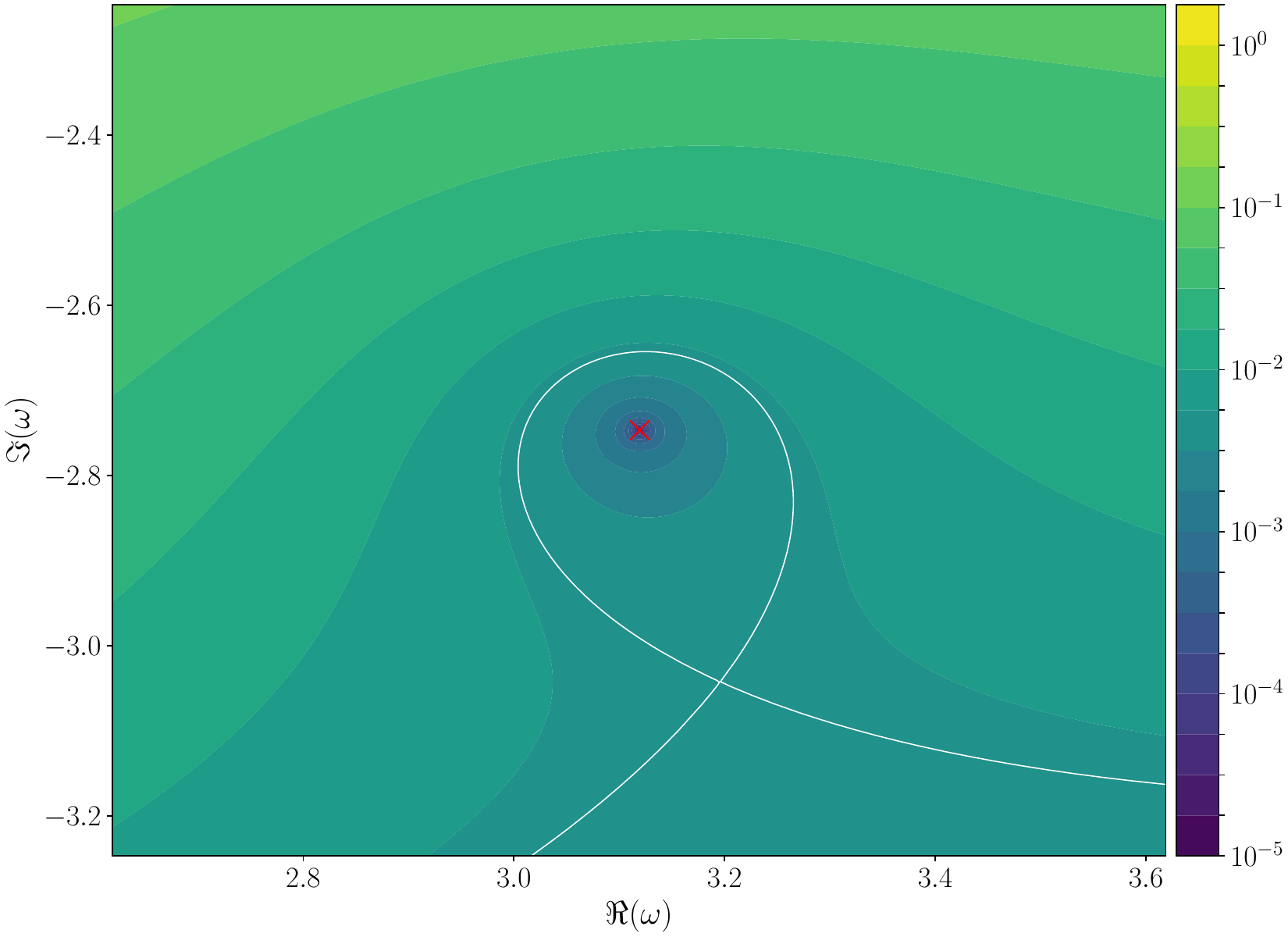}
    \caption{Zoomed-in and zoomed-out view of the $\epsilon$-pseudospectrum $\sigma^\epsilon$ for the $k=0, Q=0$ case from equation~\eqref{eq:norm_ansatz}. \emph{Left:} a wide view of the pseudospectrum demonstrates the ``open set" behaviour of contour lines away from the quasinormal modes. \emph{Right:} a zoomed view of the pseudospectrum near the fundamental mode $\omega_0 = 3.119 - 2.747 i$ shows that small perturbations around this mode remain close to the $\omega_0$ value. The highlighted contour line indicates the transition from circular to ``open set" contour lines.
    \label{fig:AdSplanark0q0_single}
    }
\end{figure}

Our results are displayed in figure~\ref{fig:AdSplanarkq_scalar} for a choice of $Q,k$, while a zoomed-in 
and a zoomed-out plot of the pseudospectrum for the uncharged $k=0$ case are shown in figure~\ref{fig:AdSplanark0q0_single}. The crosses correspond to the locations of the QNMs, where $\epsilon=0$. In figure~\ref{fig:AdSplanarkq_scalar} we also display the condition numbers 
%\footnote{\bc{Change needed} Except for figure~\ref{fig:AdSplanark10q0.99} where $\kappa_0, \kappa_1$ are the condition numbers for the fundamental mode, $\kappa_2$ is for the eigenvalue with $\Re(\omega) = 0$, and $\kappa_i,~i \geq 3$ are for the overtones.} 
$\kappa$ for each case, as defined in section~\ref{sec:spectra}. Recall that the condition number gives an indication of the relative size of the change in an eigenvalue with respect to the magnitude of a perturbation: quantitatively, 
for a completely stable spectrum $\kappa = 1$ for all modes; for non-conservative systems typically the condition number increases with increasing overtone, while for severely unstable spectra $\kappa \gg 1$ for all modes. Thus, the condition number can be used along with the pseudospectrum to understand the degree of instability of a spectrum. 

As can be observed from these figures, although the pseudospectrum still forms circular sets if we zoom arbitrarily close to the spectrum, its large scale global structure present open sets. These extend into large regions of the complex plane, indicating spectral instability. 
Interpreted through the gauge/gravity correspondence, this indicates spectral instability of the dual strongly coupled system \textemdash see section \ref{ssec:dual} for a discussion on the holographic interpretation.  More specifically, we see that for modes that reside further from the real axis, the value of $\epsilon$ needed in order to move the mode in the open set region reduces. This signals that the strength of instability increases with the overtone. %; for the dual theory, this implies that short-lived excitation are more unstable. 
 Conversely, this also means that the fundamental mode is the most stable of the QNMs. For the fundamental mode shown in Fig.~\ref{fig:AdSplanark0q0_single}, one needs to perturb the system by $\epsilon\sim 10^{-2.31}$ to see instability\footnote{In order to extract this number in a meaningful way, the pseudospectrum must be convergent in that region of the complex frequency plane. Following~\cite{Boyanov:2023qqf}, this requires the use of an $H^2$ norm.}.

Given the correlation between the distance from the real axis and the strength of the instability, we find that as the momentum $k$ increases and the modes move closer to the real axis, their stability increases. In a similar spirit, as the charge $Q$ increases, the modes move marginally further away from the real axis and thus their stability  marginally decreases.

We also observe that the pseudospectrum contour lines cross in the upper half plane. This corresponds to an unstable perturbed spectrum signaling the possibility of transients. In order to determine whether these transient instabilities indeed manifest themselves, one would need to compute the pseudospectral and numerical abscissa~\cite{embree2017pseudospectra,Boyanov:2022ark,Jaramillo:2022kuv}, which bound the norm of the evolution operator of the system from below and from above, respectively. If no exponential growth is observed, then the transients are not physical. We leave this for future work. However, we can already argue for the presence of pseudo-resonances~\cite{Boyanov:2022ark,Jaramillo:2022kuv} at the points where pseudospectrum contour lines cross the real axis into the upper half plane.

In view of~\cite{Boyanov:2023qqf}, comparison of our results with those of \cite{Arean:2023ejh} for $Q=0$ is difficult.  This is due to the different convergence behaviour of the pseudospectrum in null and hyperboloidal slicing. But even modulo the convergence issue, we do not expect to see a precise quantitative agreement since we are using a different definition for the pseudospectrum. Having said the above, a notable difference between the two computations is that the pseudospectrum contour lines in~\cite{Arean:2023ejh} do not cross in the upper half plane, implying absence of transients. It would be interesting to understand better the origin of this difference, and more generically the dependence of the pseudospectrum on the slicing.

%\cp{Flag} Finally, in the near-extremal limit we observed the existence of a series of equally-spaced quasinormal modes with zero real part, i.e.\ non-oscillatory modes. In figure~\ref{fig:AdSplanark10q0.99} we can see the first of these modes. Unlike the spectrum of the vector and scalar channel of gravitoelectric perturbations (see for example figure~\ref{fig:sheark1q1}), where non-oscillatory modes collect from $-i\infty$ to form a branch cut, these modes do not accumulate as we approach extremality. Curiously, they remain stable against both infrared and ultraviolet perturbations, as we will show in the next section.

%%%%%%%%%%%%%%%%%%%%%%%%%%%%%%%%%%%%%%%%%%%%%%%%%%%%%%%%%%%%%%%%%%%%%%%%%%%%%%%%%%%%%%%%%%%%%%%%%%%%%%%%%

\subsubsection{Testing the stability of scalar QNMs}

While the pseudospectrum allows us to visualize the stability of the QNM spectrum to general perturbations of order $\epsilon$, we may also be interested in how \emph{specific} perturbations to the background affect the spectrum~\cite{Cheung:2021bol}. These perturbations occur in the potential and can be of any form in general, although perturbations informed by physical processes are particularly important. For example, changes to the potential due to deviations from pure Schwarzschild geometry in the near-horizon region are expected to manifest as changes in the first few overtones~\cite{Konoplya:2022pbc}. With this in mind, we consider changes in the potential in~\eqref{eq:AdS5-wave-vz2} of the form $\hat U \to  \hat U + \delta \hat U$ with $\delta \hat U(z) = \epsilon \cos(\pi f (z-1))$. In terms of the generalised eigenvalue problem
\begin{align}
    \left( L_1 - \lambda L_2 \right) \psi = 0, \nonumber
\end{align}
this corresponds to a perturbation of the form $L_1 \to L_1 + \delta \hat U$.

It is important to note that adding a perturbation of size $\| \delta \hat U \|_{_E} \sim \epsilon$ relies on the definition of a norm just as the pseudospectrum does. Thus, to ensure that the size of $\delta \hat U$ remains of order $\epsilon$ relative to $L_1$ we must convert the energy norm to the 2-norm via the procedure outlined in subsection~\ref{ss: e-norm to L2-norm}, normalise $\|\delta \hat U\|_{_2}$, and finally add the overall perturbation to $L_1$.

\begin{figure}[!h]
     \centering
     \begin{subfigure}[t]{0.49\textwidth}
         \centering
         \includegraphics[width=\textwidth]{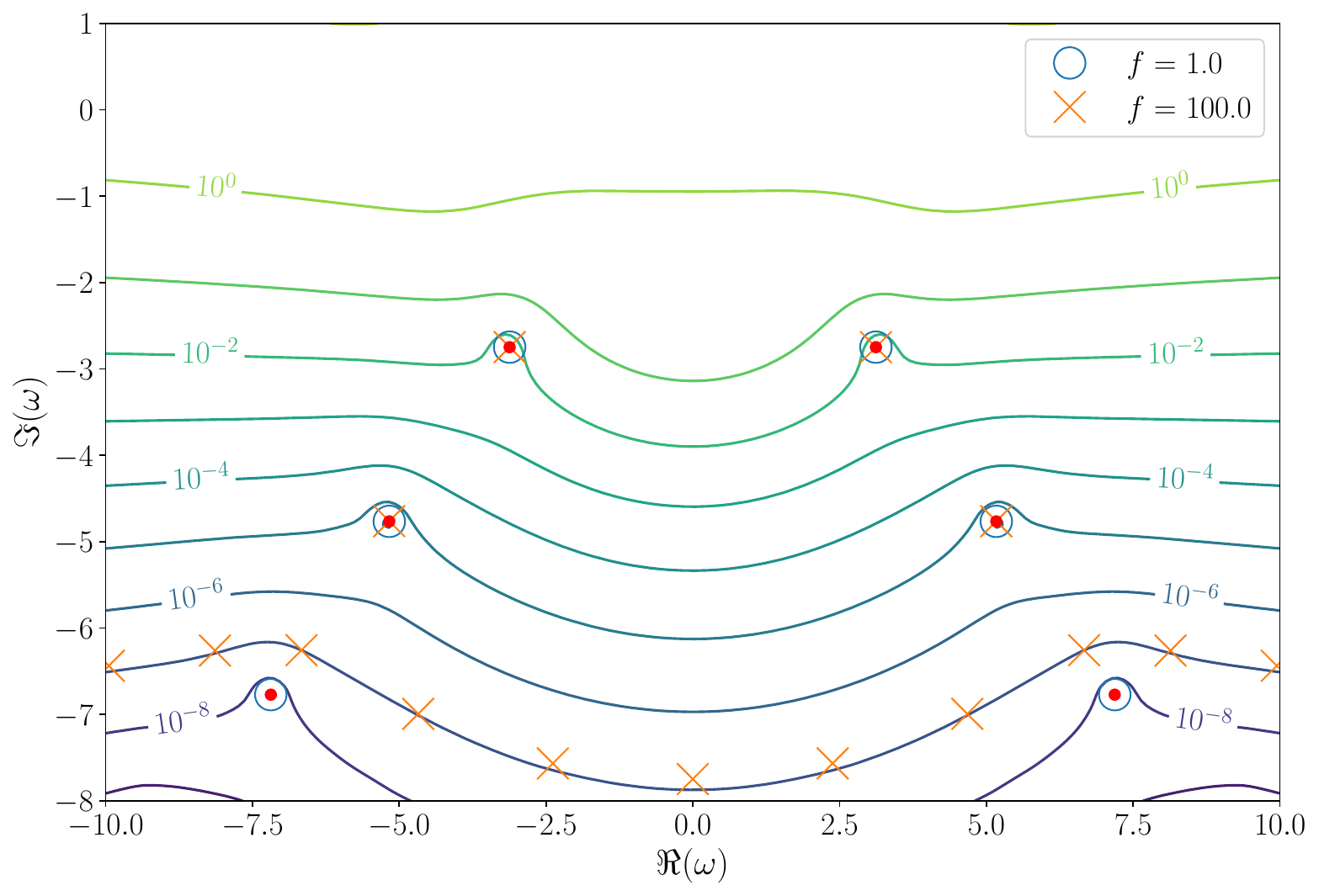}
    \end{subfigure}
    \hspace{-8pt}
    \begin{subfigure}[t]{0.49\textwidth}
         \centering
         \includegraphics[width=\textwidth]{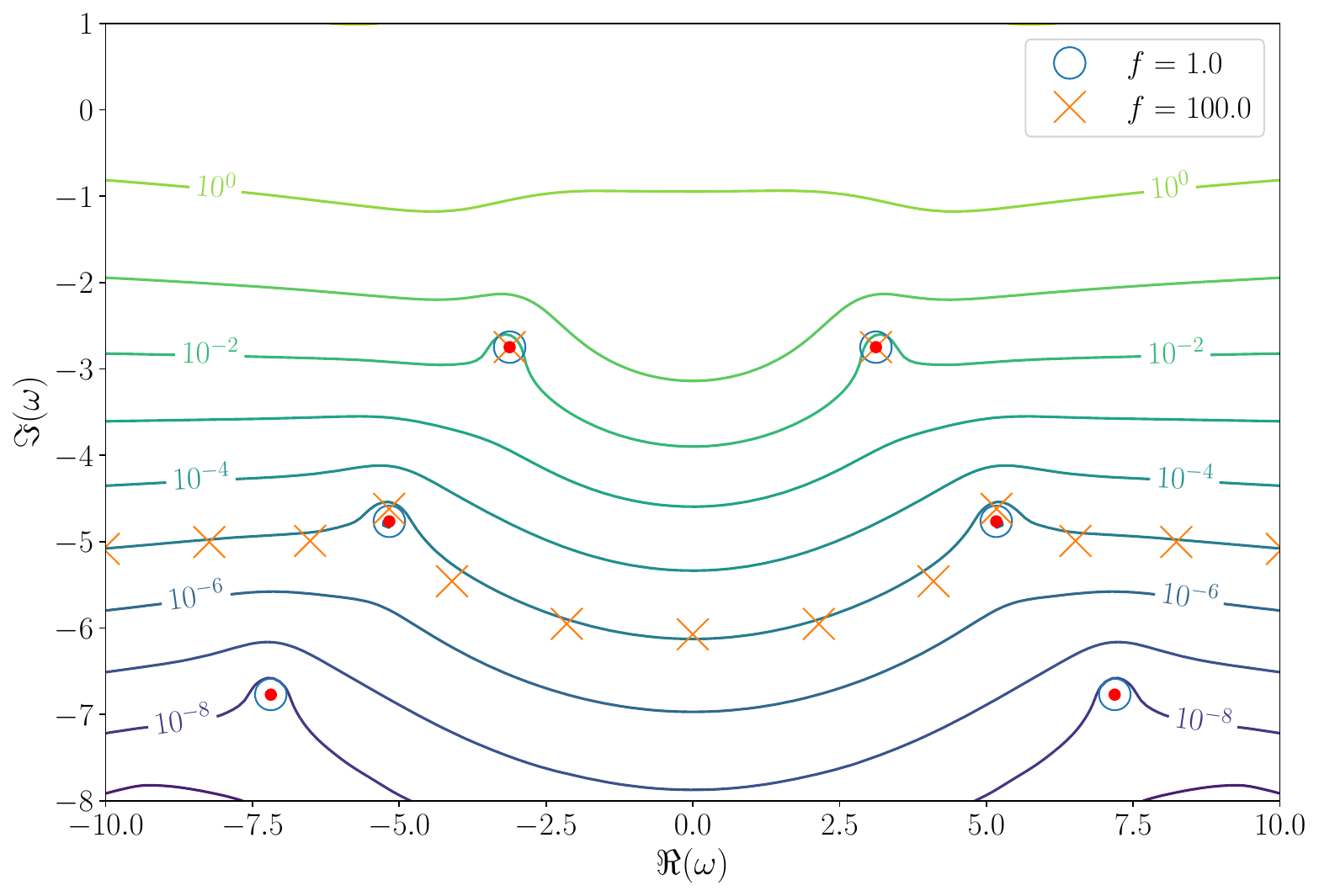}
    \end{subfigure}
    \\ %\vspace{-6pt}
    \begin{subfigure}[t]{0.49\textwidth}
         \centering
         \includegraphics[width=\textwidth]{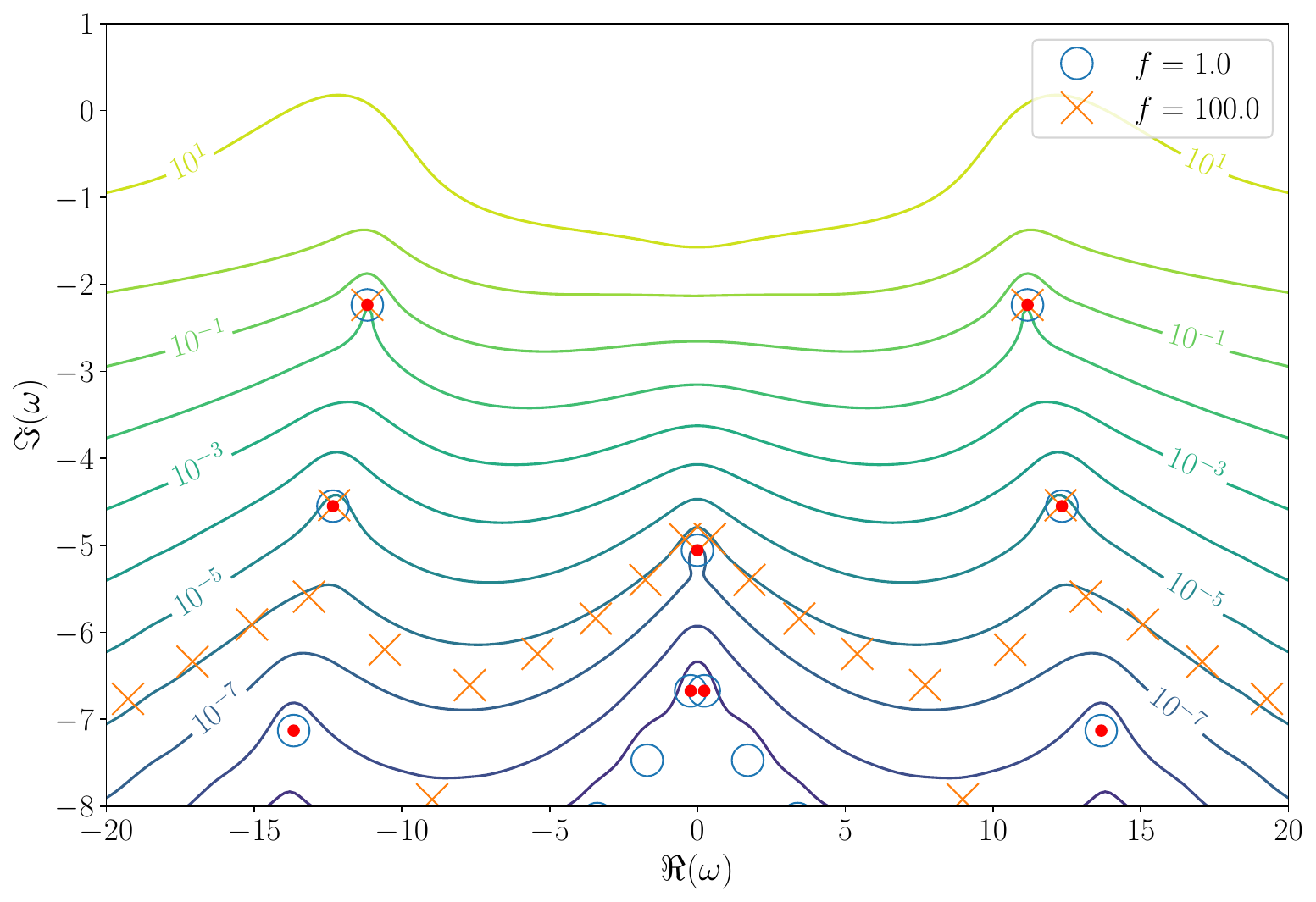}
    \end{subfigure}
    \hspace{-8pt}
    \begin{subfigure}[t]{0.49\textwidth}
         \centering
         \includegraphics[width=\textwidth]{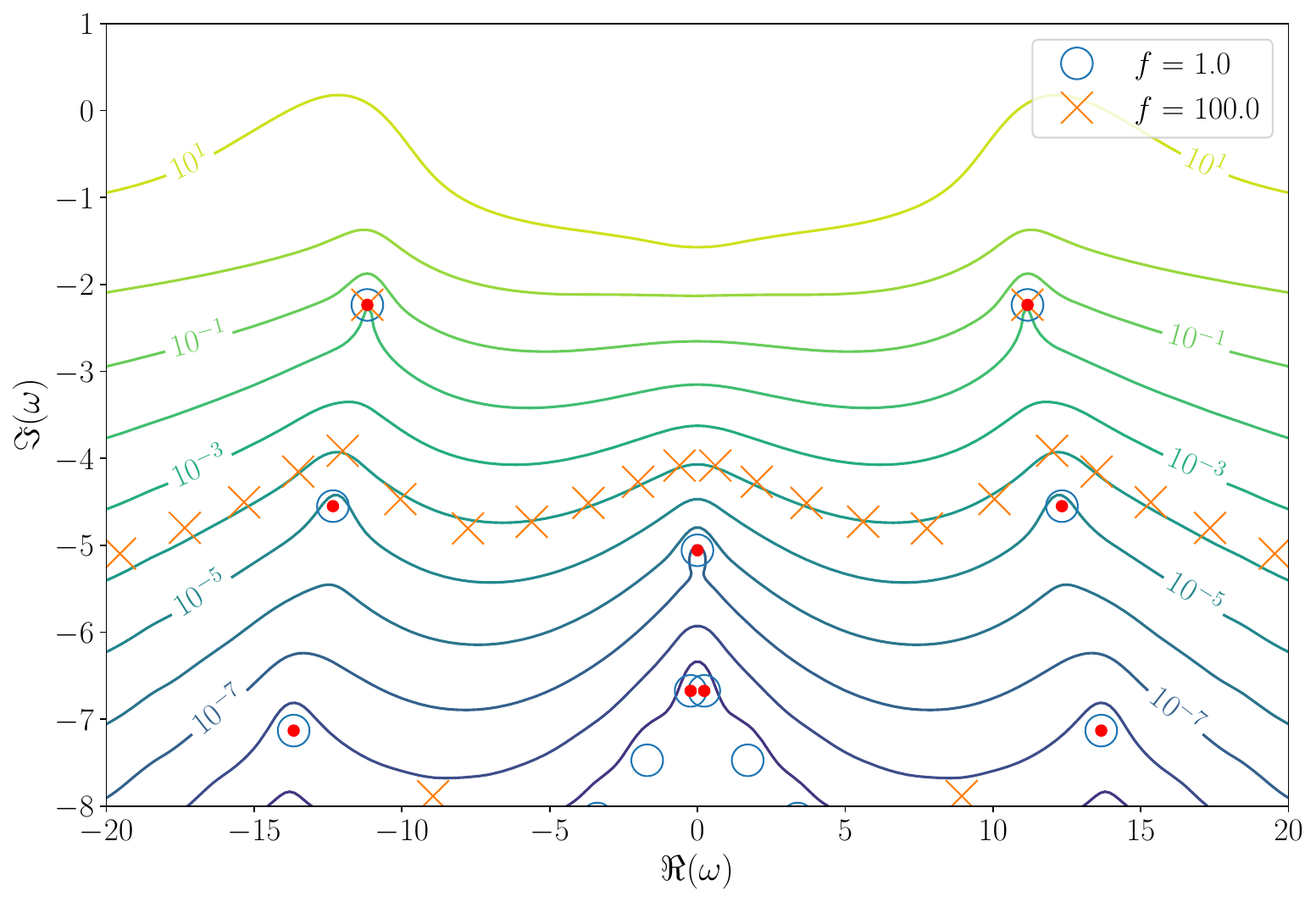}
    \end{subfigure}
    \caption{The perturbed scalar spectrum when ${\delta \hat U = \epsilon \, \cos(\pi f (z - 1))}$ for various AdS-RN configurations. \emph{Columns, left to right:} ${\epsilon = 10^{-8};~\epsilon = 10^{-6}}$. \emph{Rows, from top to bottom:} ${k=0,~Q=0}$; %${k=10,~Q=0}$; ${k=10,~Q=0.1}$;
    ${k=10,~Q=0.7Q_{\text{max}}}$.}
    \label{fig:massless_scalar_perts}
\end{figure}

In figure~\ref{fig:massless_scalar_perts}, we show the effects of adding a sinusoidal perturbation of magnitude $\epsilon = 10^{-8}$ (left column) and $\epsilon=10^{-6}$ (right column) to the potential. In each plot, the unperturbed spectrum is denoted by red dots; the spectra resulting from ${L_1 \to L_1 + \delta \hat U}$ are overlaid in circles ($f=1$) and crosses ($f=100$). Contour lines of the $\epsilon$-pseudospectra are included for reference, with the magnitudes indicated in the plot. We observe that for $f=1$ both the fundamental modes and the first few overtones that we can resolve numerically are stable. For $f=100$ we see that higher overtones are more prone to instability than the fundamental modes or lower overtones. Note that the shifted spectrum follows the contours of the $\epsilon$-pseudospectrum. We thus conclude that the instabilities we find are ``ultraviolet". This is in agreement with previous work in the literature in flat spacetime~\cite{Jaramillo:2020tuu,Konoplya:2022pbc} and in AdS~\cite{Konoplya:2023kem}.

%\cp{FLAG:} In the case of the near-extremal black hole, we can see that the non-oscillating mode is more stable against both short- and long-wavelength perturbations than the overtones, despite being located further from the imaginary axis. As shown in figure~\ref{fig:AdSplanark10q0.99}, this mode has a condition number several orders of magnitude less than the other QNMs in the spectrum. Hence, the increased stability of this mode is to be expected.

%%%%%%%%%%%%%%%%%%%%%%%%%%%%%%%%%%%%%%%%%%%%%%%%%%%%%%%%%%%%%%%%%%%%%%%%%%%%%%%%%%

\subsection{Metric and gauge field perturbations}
\label{ssec:metricandgauge}

We now move on to study gravitational  and electromagnetic fluctuations around the background configuration \eqref{eq:norm_ansatz}. 
Gravitational QNMs for the AdS-Schwarzschild and AdS-RN black branes were discussed previously in the literature~\cite{Policastro:2002se,Kovtun:2005ev}. In particular, for non-zero charge the electromagnetic and gravitational perturbations couple to each other. However, it is well known that these split into 3 sectors, depending on how they transform under the rotation group transverse to the momentum of the fluctuations: the tensor (for $D>4$), the vector and the scalar sectors. 

In what follows we will use the formalism of master equations, which was first derived for AdS-RN in~\cite{Kodama:2003kk} and later generalised in~\cite{Jansen:2019wag}. The idea is to construct gauge invariant combinations of the perturbations, and then combine them into master scalars that satisfy Klein-Gordon equations with certain potentials.
The tensor sector contains a single master scalar $\Phi_2$. The vector sector contains two, whose equations can be decoupled by taking certain linear combinations that we denote $\Phi_{1 \pm}$: these correspond to the shear channel (positive sign) and the transverse gauge field (negative sign).  Finally, the scalar sector contains two more master scalars that can be decoupled, which we denote $\Phi_{0 \pm}$: these correspond to the sound channel (positive sign) and the diffusion channel (negative sign). Out of these 5 types of fluctuations only the shear, sound and diffusion contain hydrodynamic modes, meaning modes that obey dispersion relations such that the frequency approaches zero as the momentum is decreased. 

Each of these master scalars satisfy a master equation of the form
\begin{align}\label{eq:mastereqs}
\square \Phi_i - Z_i(z) \Phi_i &= 0 \, ,
\end{align}
where $\square$ is the Laplacian on the spacetime \eqref{eq:norm_ansatz}. Without loss of generality, we decompose the master scalars into plane waves with the momentum pointing in the $x_1$ direction 
\begin{equation}
\Phi_i(v, z, x_1) = e^{i\,k x_1}  \Phi_i(v,z) \, ,
\end{equation}
where  $k$ is the momentum in the $x_1$ direction.

The potentials for the tensor, vector and scalar sectors can be written respectively as 
\begin{equation}
\begin{aligned}
Z_2(z) &=  0
\\
Z_{1\pm}(z) &= \frac{1}{2} (Z^{(1)}_{11} + Z^{(1)}_{22})\pm \frac{1}{2} \sqrt{ (Z^{(1)}_{11} - Z^{(1)}_{22})^2 +  4(Z^{(1)}_{12})^2}\\
Z_{0\pm}(z) &= \frac{1}{2} (Z^{(0)}_{11} + Z^{(0)}_{22})\pm \frac{1}{2} \sqrt{ (Z^{(0)}_{11} - Z^{(0)}_{22})^2 +  4(Z^{(0)}_{12})^2} 
\end{aligned}
\end{equation}
where 
\begin{equation}
\begin{aligned}
&Z^{(1)}_{11}=%
zf'- 3 f +z^4 a'^2\\
&Z^{(1)}_{22}=%k
3z f'-3f\\
&Z^{(1)}_{12}= |k| z^3 a' 
\end{aligned}
\end{equation}
and
\begin{equation}
\begin{aligned}
&Z^{(0)}_{11}=\frac{a'^2 z^2 }{\mathcal{D}^2}\left(18 f a'^2 z^2-9\mathcal{F} \left(
   2f+zf'\right)+36 k^2 f+4
   k^4 z^2\right)-2 (2f-zf')%
   \\
&Z^{(0)}_{22}=\frac{2 }{\mathcal{D}^2}\left(12 k^2 z^2 f a'^2+12 k^2\mathcal{F}
   \left(n f-zf'\right)-8 k^4\left(
  f- zf' \right)\right)%
  \\
&Z^{(0)}_{12}=\frac{ |k| z a' }{\mathcal{D}^2}\frac{2}{\sqrt{3}}\left(18 z^2 f
   a'^2+9\mathcal{F}(z)(2f -  z f')+12 k^2 f+4
   k^4 z^2\right)\\
&\mathcal{F}=f'/z\,,\quad \mathcal{D}=2k^2-3\mathcal{F}\,.
\end{aligned}
\end{equation}

Note that the tensor sector corresponds to a massless scalar with no potential, which is precisely what we studied in the previous subsection. Hence, in what follows we will focus only on the vector and scalar sectors. It is also important to emphasise that the two channels within the vector and the scalar sectors decouple from each other precisely because the following combination is constant 
\begin{align}
  & \frac{Z^{(i)}_{11}-Z^{(i)}_{2}}{Z^{(i)}_{12}}=-c_i \frac{1+Q^2}{|k|Q},\qquad i=0,1,
\end{align}
with $c_0=2, c_1=\tfrac{4}{\sqrt{3}}$. 

Following the same methodology as for the scalar field perturbations, we consider the following rescalings. First, we rescale the master scalars as 
\begin{align}
\Phi_i(v,z) = z^{\frac{3}{2}} \phi_i(v,z)
\end{align}
to bring the equations into the form~\eqref{eq:AdS5-wave-vz}
\begin{align} 
&-2f\partial_{vz}\phi_i+f\partial_z(f\partial_z\phi_i)-  V_i(z) \phi_i= 0 \\
&V_i(z)\equiv f \hat V_i(z) = \frac{f}{z^2}\,(Z_i(z) + k^2 z^2)+\frac{3}{2}\,\frac{f}{z^2}\left( -z f'+\frac{5}{2} f \right). \notag
\end{align}
However, as before, for the purposes of the numerical implementation we also consider the redefinition 
 \begin{align}
&\Phi_i(v,z) = z^{\Delta_i} \psi_i(v,z)\quad\text{ where }\quad\Delta_2=4, \quad \Delta_{1\pm}=3, \quad \Delta_{0\pm}=2
\end{align}
so that, given the absence of a source, the leading UV behaviour is $\psi= \mathrm{VEV}+\mathcal{O}(z)$. In this case the equations for the master scalars become
 \begin{align}
&-2f\partial_{vz}\psi_i-(2\Delta_i-3)\frac{f}{z}\partial_v\psi+f\partial_z(f\partial_z\psi_i) +(2\Delta_i-3)\frac{f^2}{z}\partial_z\psi-  U_i(z) \psi_i= 0\\
&U_i(z)\equiv f\hat{U_i}(z) =  \frac{f}{z^2}(Z_i(z) +k^2 z^2)+\Delta_i\,\frac{f}{z^2} \left(-z f'+(4-\Delta_i) f \right). \notag
\end{align}

The next step is to bring the above equation in the form
\begin{equation}
L_2 \, \partial_v \psi_i = i L_1 \psi_i
\end{equation}
where
\begin{align}
iL_1&=zf\partial^2_z+ z f'\partial_z +(2\Delta_i-3)f\partial_z- z \hat{U_i}(z)\\
L_2&=2 z\partial_{z}+(2\Delta_i-3) \mathbb{1} \,.
\end{align}
We again note that the tensor channel corresponds to a massless scalar and one can easily check that the above expressions agree with those in the previous subsection.

For the energy norm, our starting point is equation \eqref{eq:energynorm1}.
After performing the redefinition $\phi_i=z^{\Delta_i-3/2}\,\psi_i $, we obtain the following expression in terms of $\psi_i$,
\begin{equation}
 E[\psi_i] = \int dz \frac{z^{2 \Delta_i-3}}{2}
 \left(
 \left(
 \hat{V_i}+ (2 \Delta_i-3)^2 \frac{f}{4z^2}
 \right) \psi_i^2 + 
 \frac{f}{z}(2 \Delta_i-3) \psi_i \partial_z\psi_i +  f \partial_z\psi_i^2\right)
\end{equation}
We can now, in a matter entirely analogous to that of the previous subsection, extract our Gram matrix $G^E$ and compute the pseudospectrum.

\subsubsection{Results}

\begin{figure}[thb]
    \centering
    \begin{subfigure}[t]{0.48\textwidth}
        \centering
        \includegraphics[width=\textwidth]{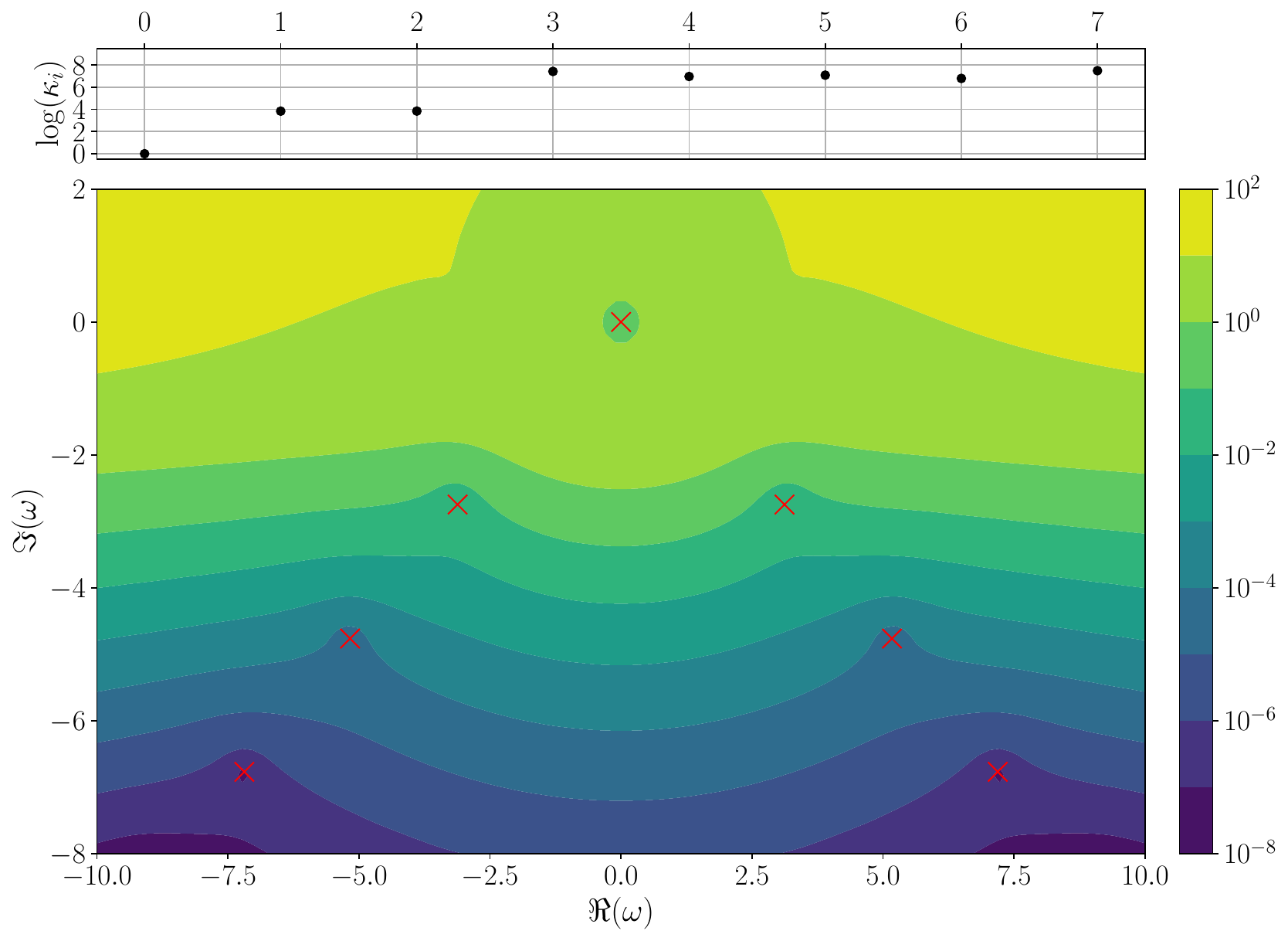}
        \caption{$k=0$, $Q=0$.}
        \label{fig:sheark0q0}
    \end{subfigure}
    \;
    \begin{subfigure}[t]{0.48\textwidth}
        \centering
        \includegraphics[width=\textwidth]{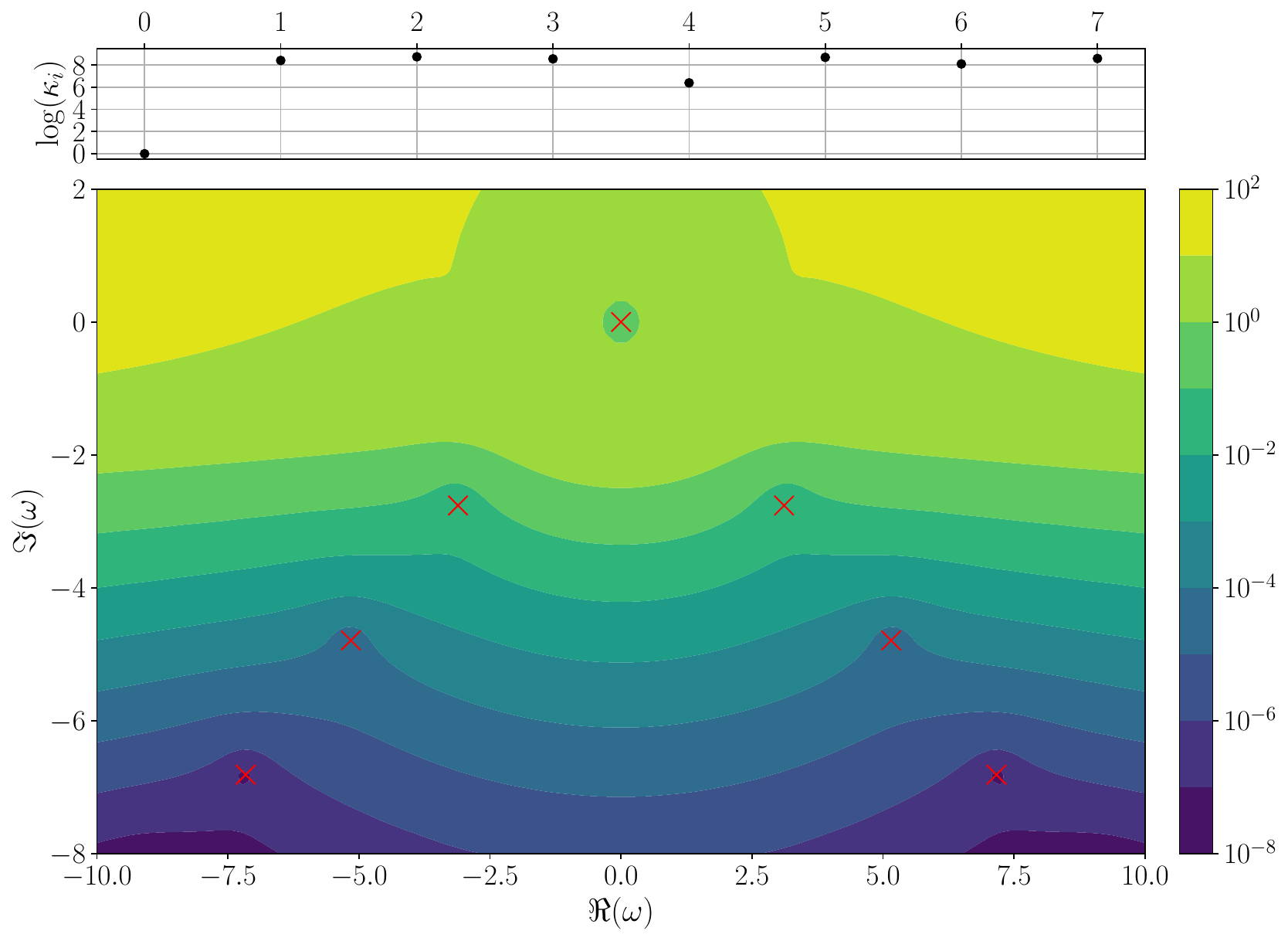}
        \caption{$k=0$, $Q=0.1$.}
        \label{fig:sheark0q0.1}
    \end{subfigure}
    \\
    \begin{subfigure}[t]{0.48\textwidth}
        \centering
        \includegraphics[width=\textwidth]{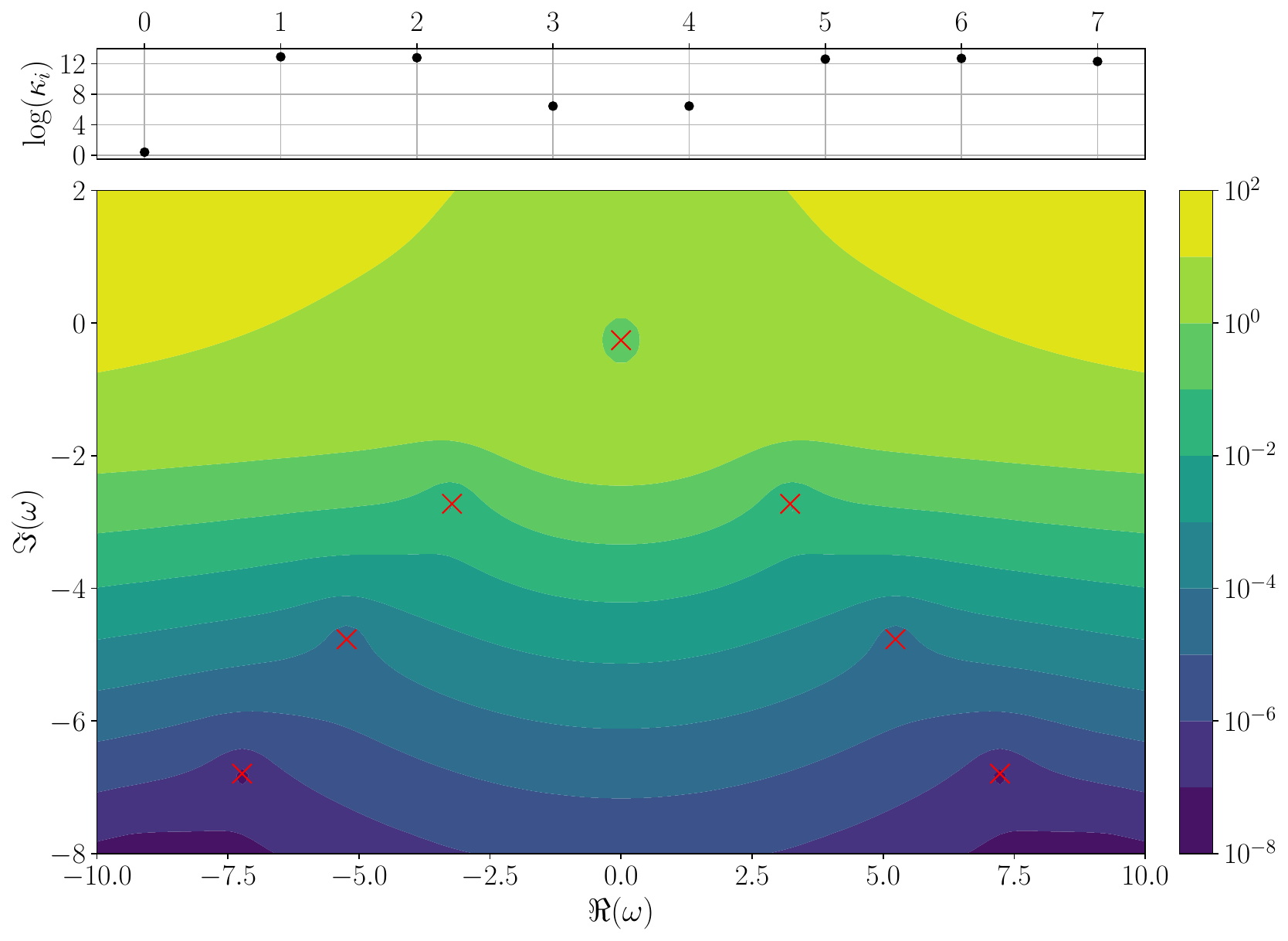}
        \caption{$k=1$, $Q=0.1$.}
        \label{fig:sheark1q0.1}
    \end{subfigure}
    \;
    \begin{subfigure}[t]{0.48\textwidth}
        \centering
        \includegraphics[width=\textwidth]{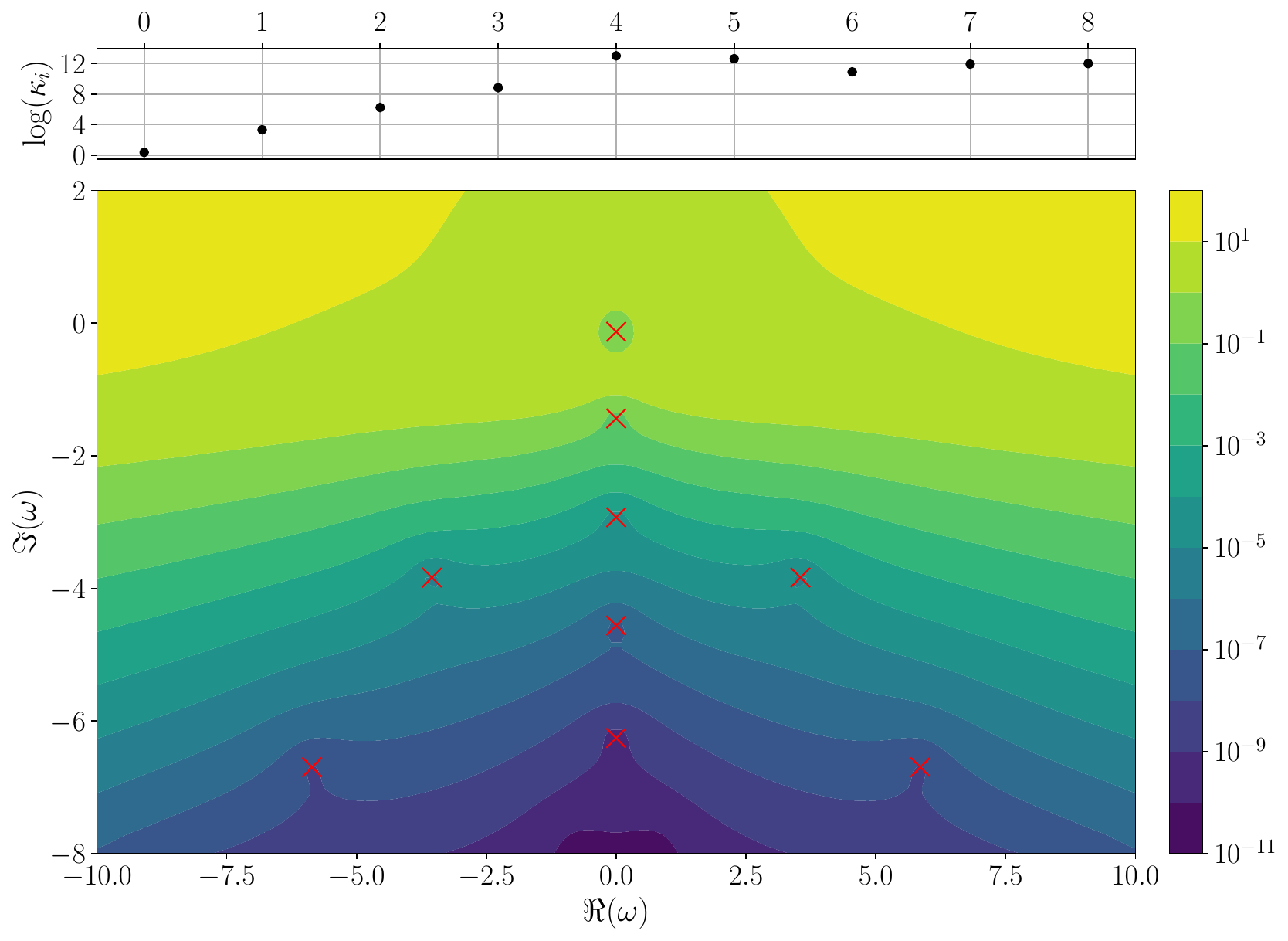}
        \caption{$k=1$, $Q=1=0.7\, Q_\mathrm{max}$.}
        \label{fig:sheark1q1}
    \end{subfigure}
    \caption{The $\epsilon$-pseudospectrum for various values of momentum $k$ and charge $Q$ in the vector channel, shear sector. Recall that the extremal solution is obtained for $Q_\mathrm{max}=\sqrt{2}$. Above each plot, the condition numbers $\kappa$ of the eigenvalues are plotted with increasing overtone number. The red crosses mark the spectrum points.}
    \label{fig:shear}
\end{figure}

\begin{figure}[thb]
    \centering
    \begin{subfigure}[t]{0.48\textwidth}
        \centering
        \includegraphics[width=\textwidth]{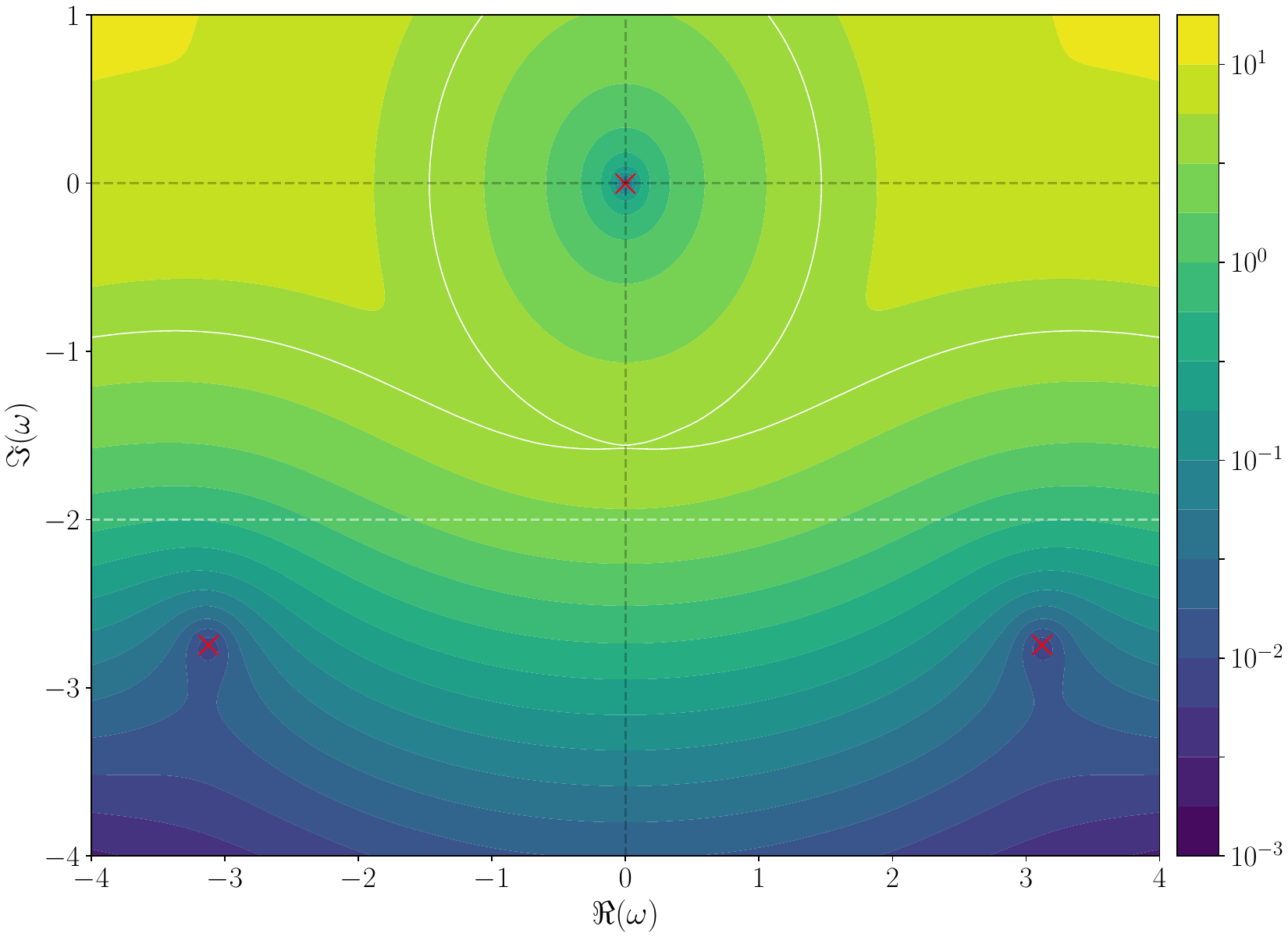}
        \caption{$k=0$, $Q=0$.  The highlighted contour lies at a value of  $\sigma^\epsilon \sim 10^{0.64}$.}
        \label{fig:shearZoomk0q0}
    \end{subfigure}
    \;
    \begin{subfigure}[t]{0.48\textwidth}
        \centering
        \includegraphics[width=\textwidth]{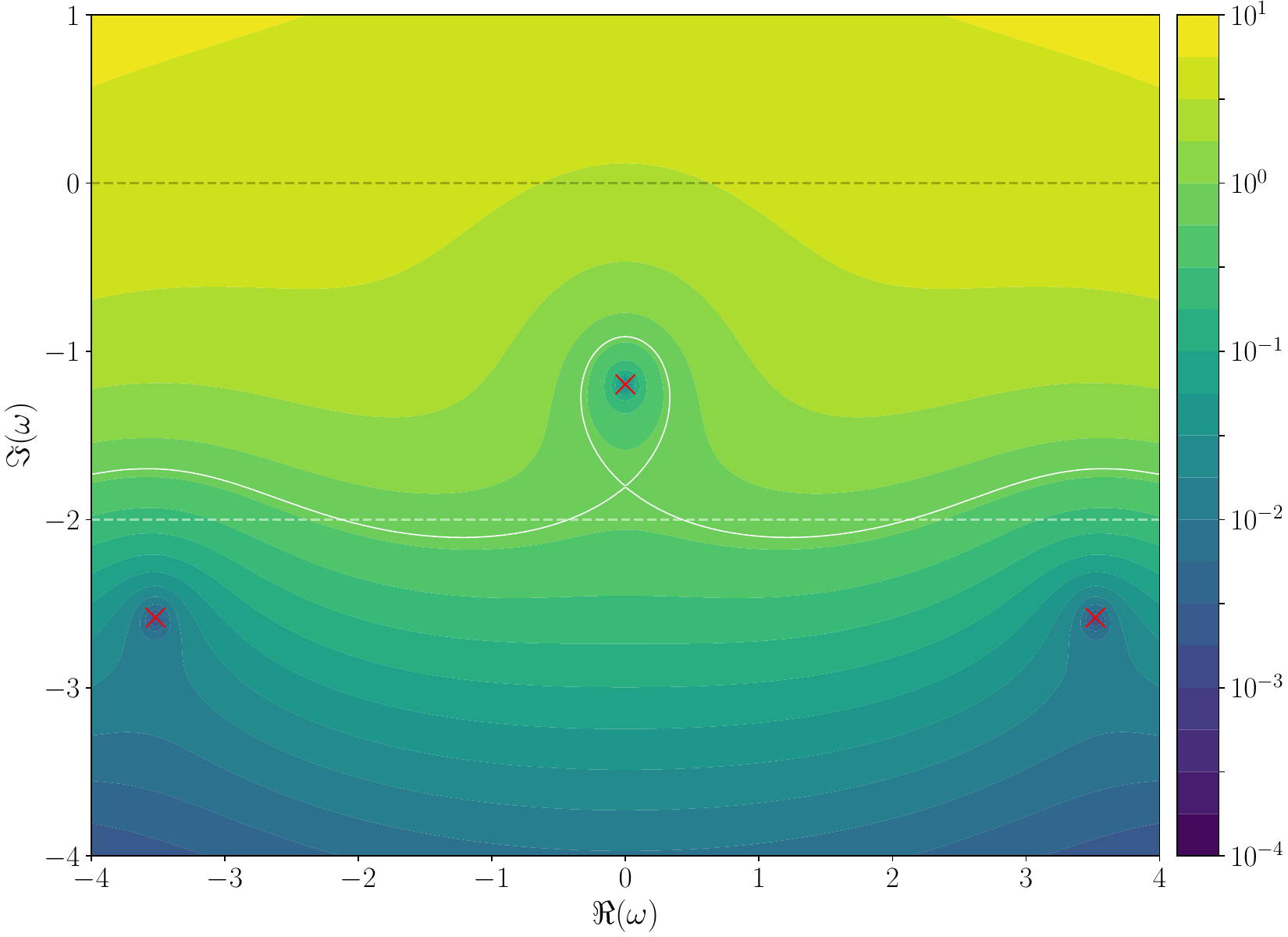}
        \caption{$k=4.0$, $Q=0$. The hydro mode is now less dominant. The highlighted contour lies at a value of  $\sigma^\epsilon \sim 10^{-0.19}$.}
        \label{fig:shearZoomk4q0}
    \end{subfigure}
    \caption{Zoomed-in plot of the $\epsilon$-pseudospectrum in the shear sector of two $Q=0$ black branes possessing hydro modes. The red crosses mark the spectrum points.}
    \label{fig:shearZoom}
\end{figure}

\begin{figure}[thb]
    \centering
    \begin{subfigure}[t]{0.48\textwidth}
        \centering
        \includegraphics[width=\textwidth]{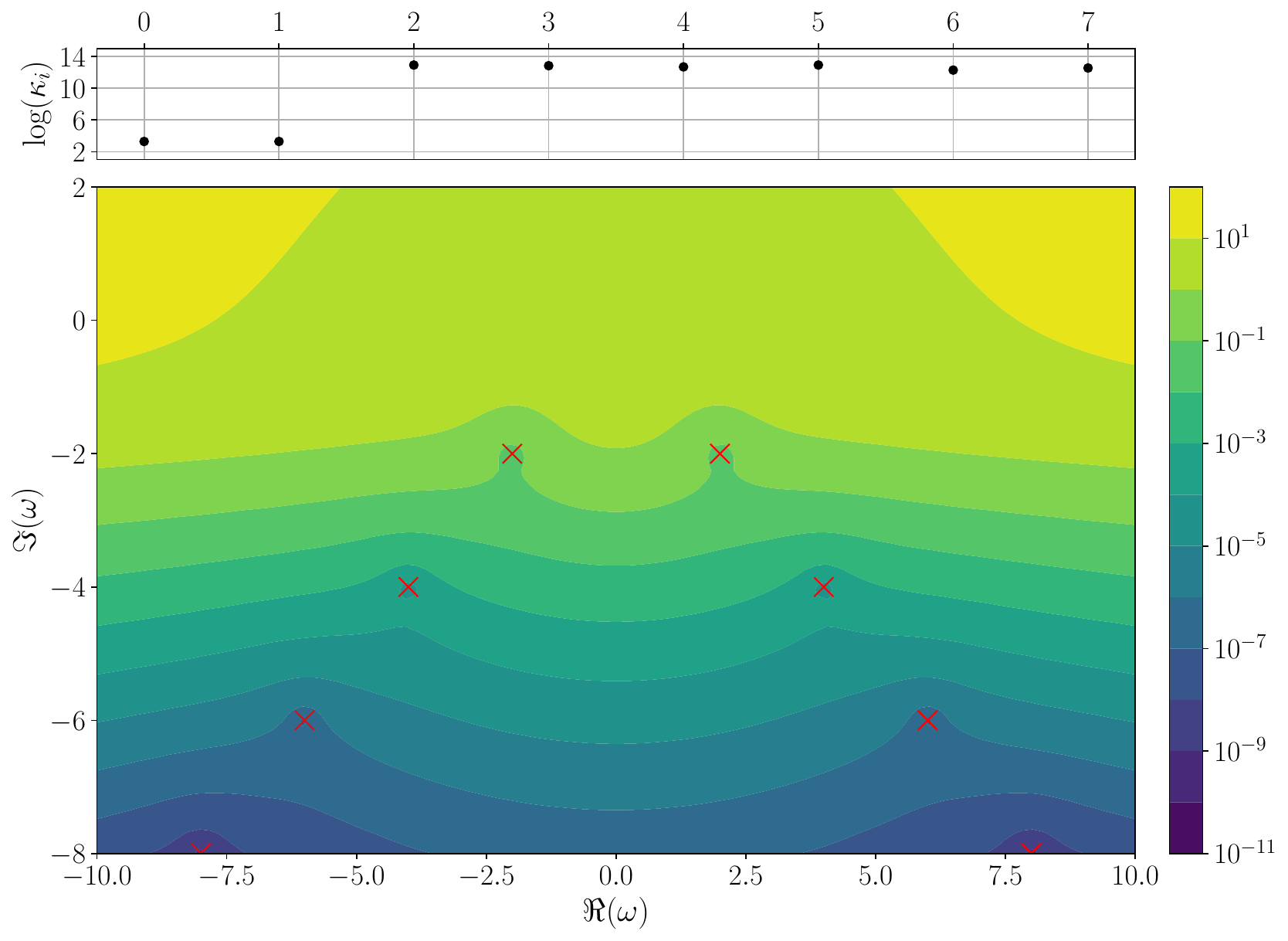}
        \caption{$k=0$, $Q=0$.}
        \label{fig:transk0q0}
    \end{subfigure}
    \;
    \begin{subfigure}[t]{0.48\textwidth}
        \centering
        \includegraphics[width=\textwidth]{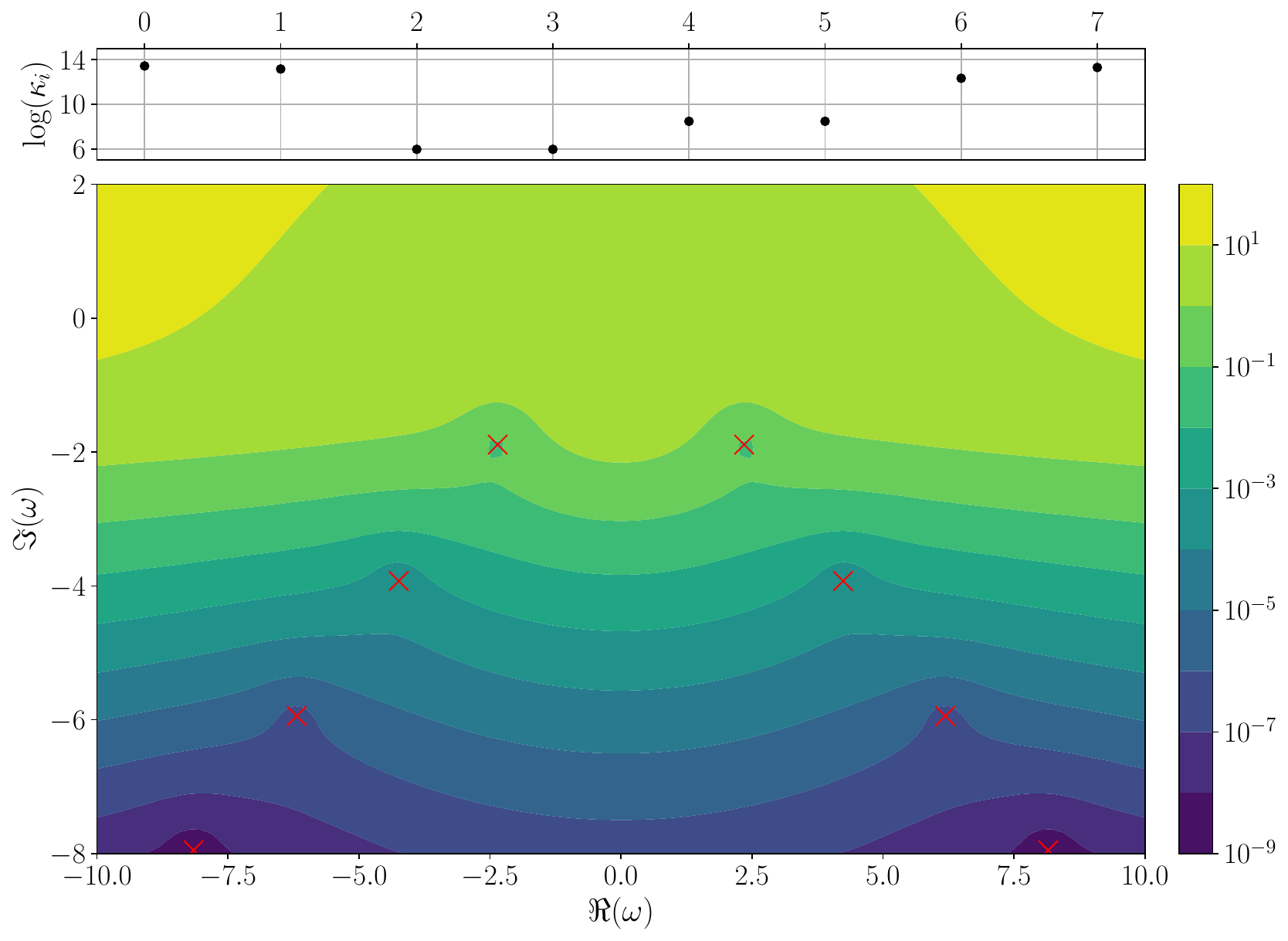}
        \caption{$k=1$, $Q=0$.}
        \label{fig:transk0q1}
    \end{subfigure}
    \\
    \begin{subfigure}[t]{0.48\textwidth}
        \centering
        \includegraphics[width=\textwidth]{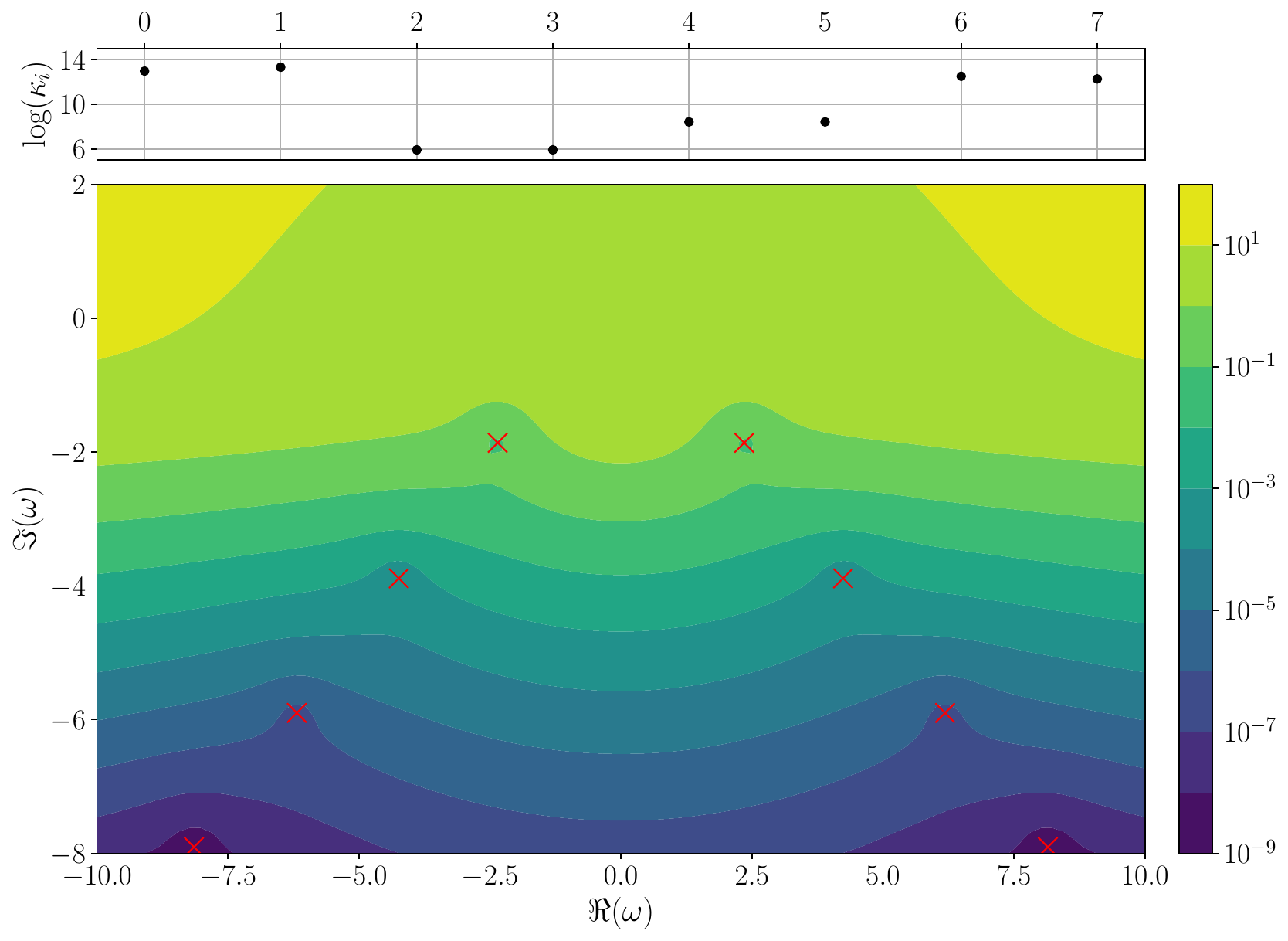}
        \caption{$k=1$, $Q=0.1$.}
        \label{fig:transk1q0.1}
    \end{subfigure}
    \;
    \begin{subfigure}[t]{0.48\textwidth}
        \centering
        \includegraphics[width=\textwidth]{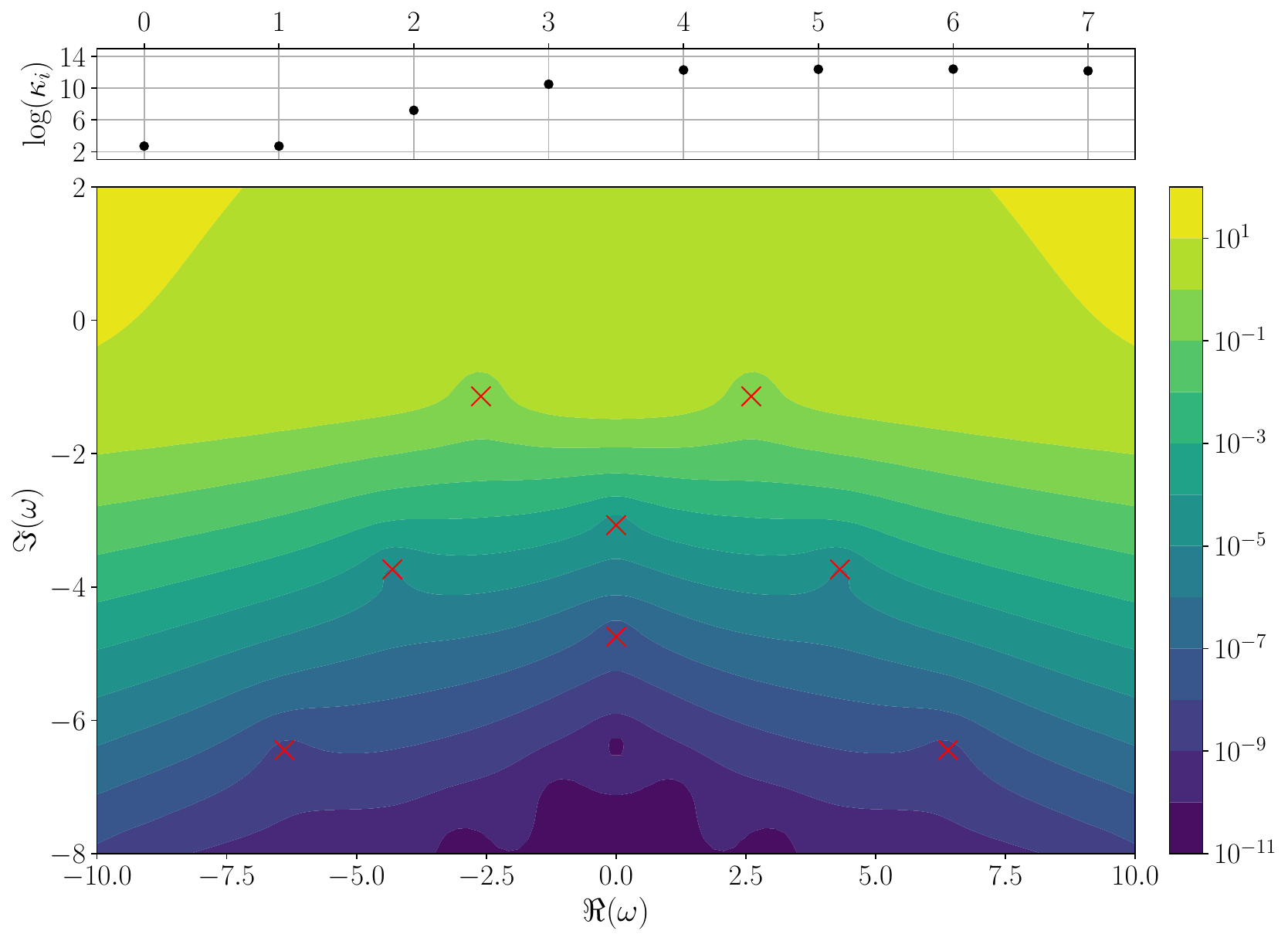}
        \caption{$k=1$, $Q=1=0.7\,Q_\mathrm{max}$.}
        \label{fig:transk1q1}
    \end{subfigure}
    \caption{The $\epsilon$-pseudospectrum for various values of momentum $k$ and charge $Q$ in the vector channel, transverse gauge sector. Recall that the extremal solution is obtained for $Q_\mathrm{max}=\sqrt{2}$. Above each plot, the condition numbers $\kappa$ of the eigenvalues are plotted with increasing overtone number. The red crosses mark the spectrum points.}\label{fig:transgauge}
\end{figure}

\begin{figure}[thb]
    \centering
    \begin{subfigure}[t]{0.48\textwidth}
        \centering
        \includegraphics[width=\textwidth]{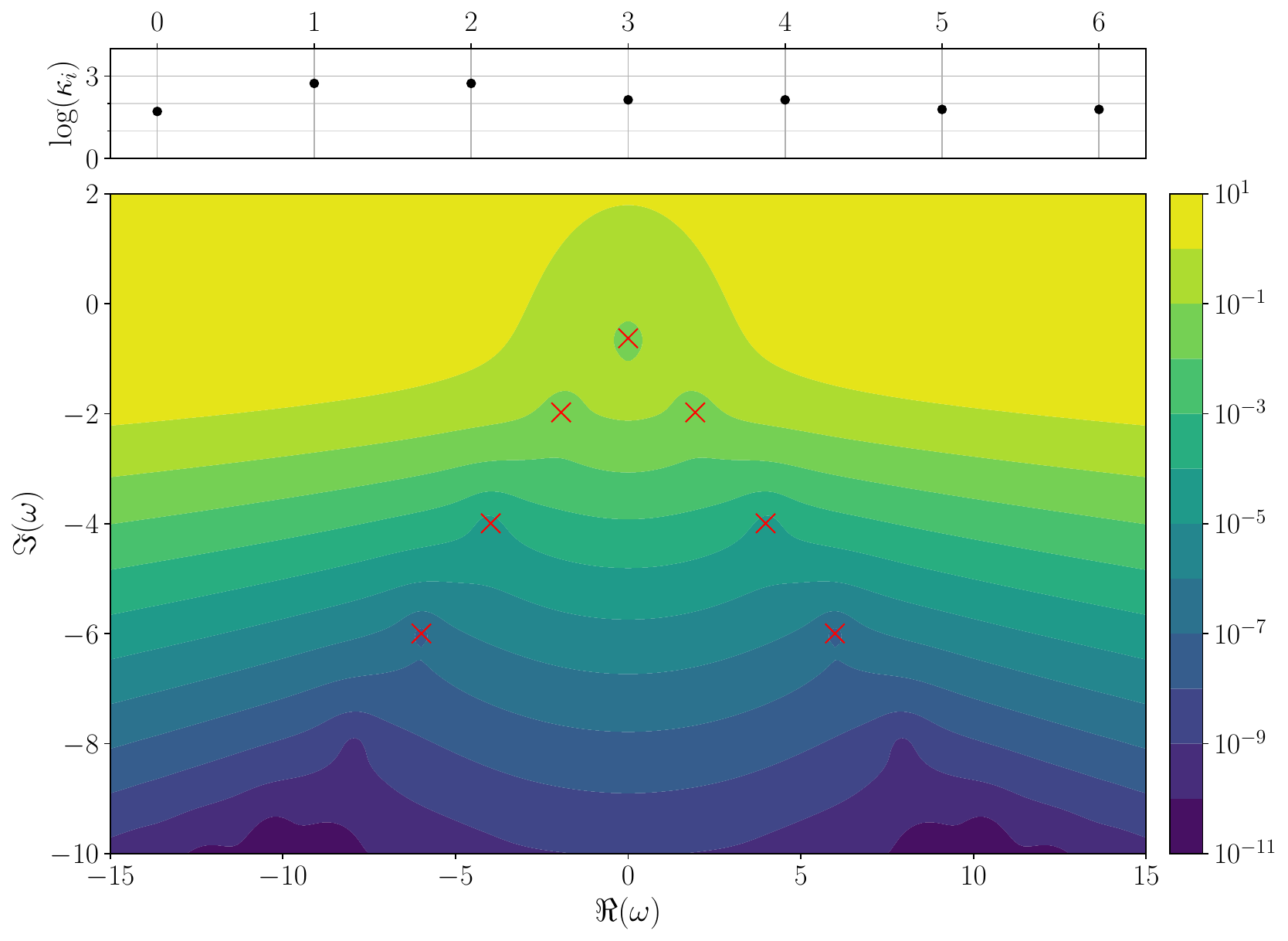}
        \caption{$k=1$, $Q=0$.}
        \label{fig:diffk1q0}
    \end{subfigure}
    \begin{subfigure}[t]{0.48\textwidth}
        \centering
        \includegraphics[width=\textwidth]{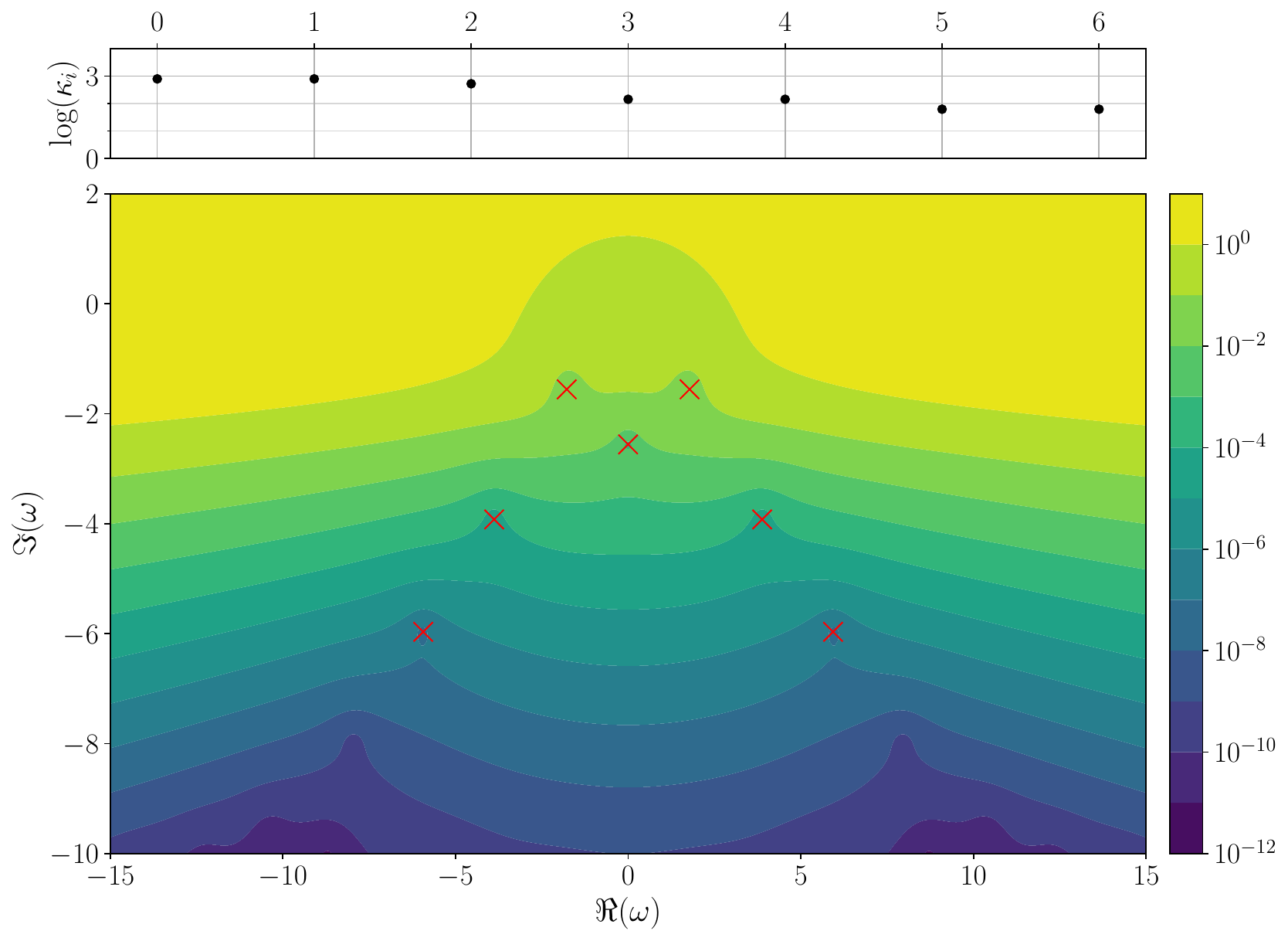}
        \caption{$k=1.5$, $Q=0$.}
        \label{fig:diffk15q0}
    \end{subfigure}
    \;
    \\
    \begin{subfigure}[t]{0.48\textwidth}
        \centering
        \includegraphics[width=\textwidth]{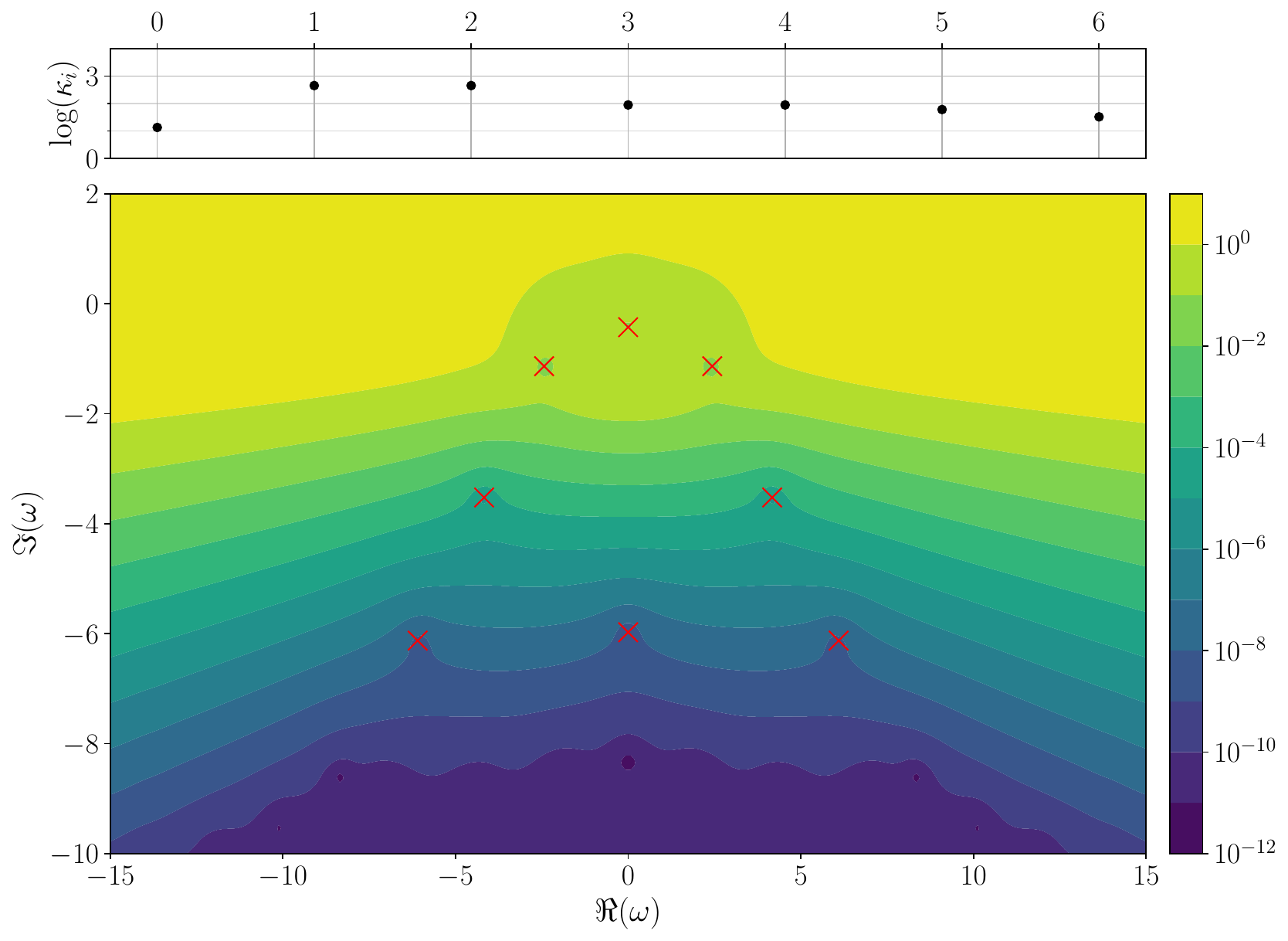}
        \caption{$k=1$, $Q=0.8$.}
        \label{fig:diffk1q0.8}
    \end{subfigure}
    \;
    \begin{subfigure}[t]{0.48\textwidth}
        \centering
        \includegraphics[width=\textwidth]{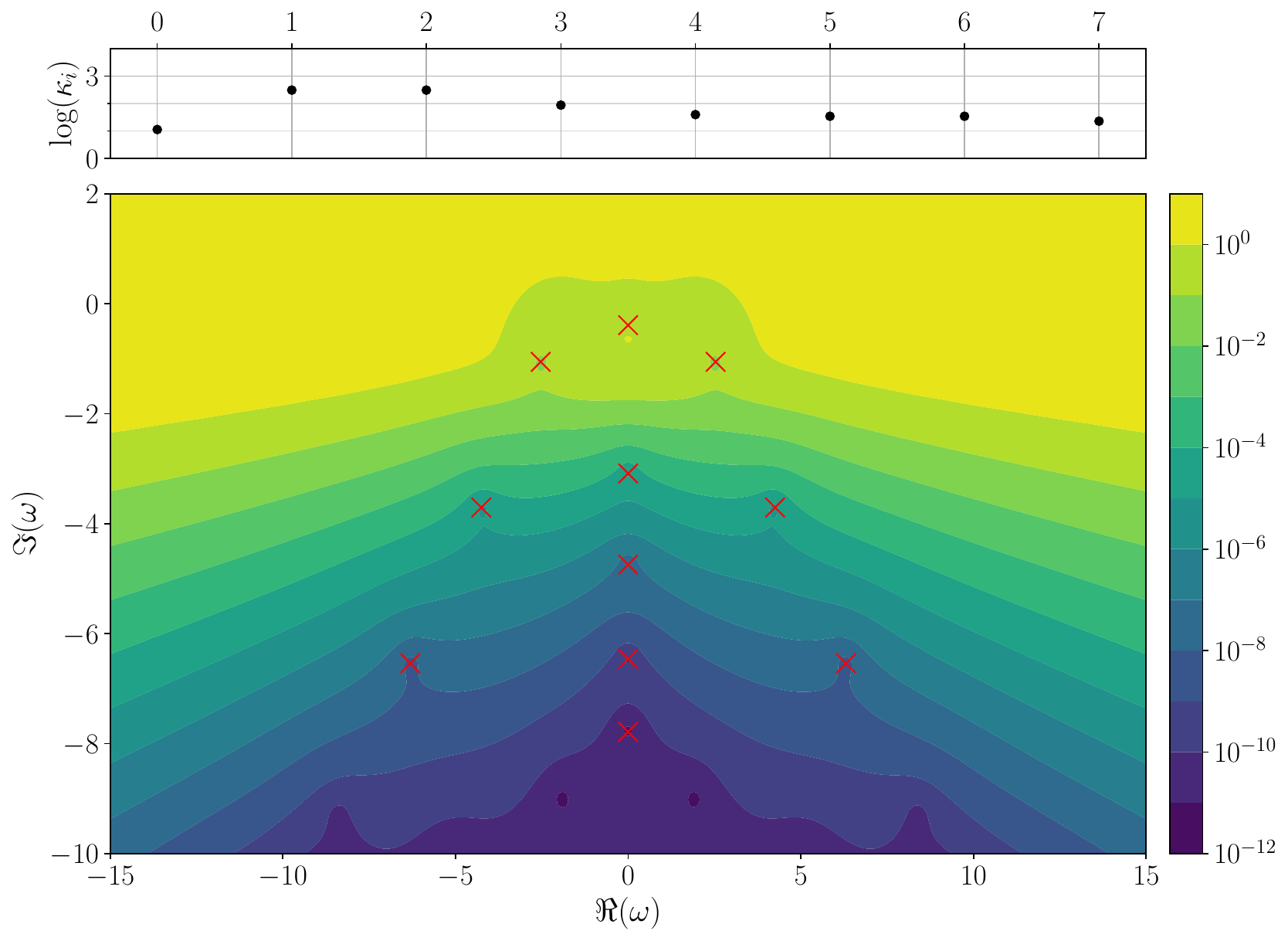}
        \caption{$k=1$, $Q=1=0.7\,Q_\mathrm{max}$.}
        \label{fig:diffk1q1}
    \end{subfigure}
    \caption{The $\epsilon$-pseudospectrum for various values of momentum $k$ and charge $Q$ in the scalar channel, charge diffusion sector. Recall that the extremal solution is obtained for $Q_\mathrm{max}=\sqrt{2}$. Above each plot, the condition numbers $\kappa$ of the eigenvalues are plotted with increasing overtone number. The red crosses mark the spectrum points.}\label{fig:diffusion}
\end{figure}

 \begin{figure}[pthb]
    \centering
    \begin{subfigure}[t]{0.48\textwidth}
        \centering
        \includegraphics[width=\textwidth]{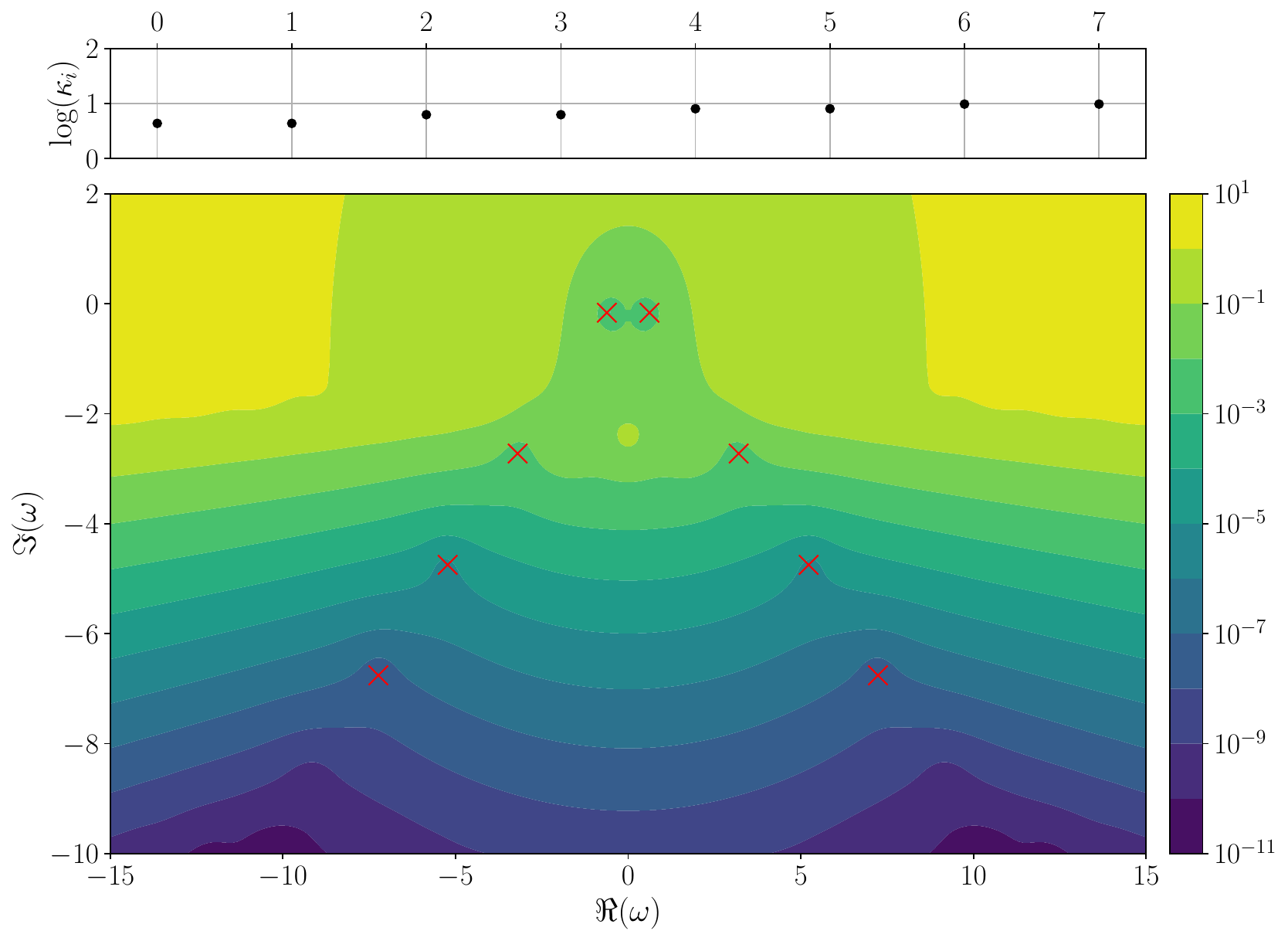}
        \caption{$k=1$, $Q=0$.}
        \label{fig:soundk1q0}
    \end{subfigure}
    \;
    \begin{subfigure}[t]{0.48\textwidth}
        \centering
        \includegraphics[width=\textwidth]{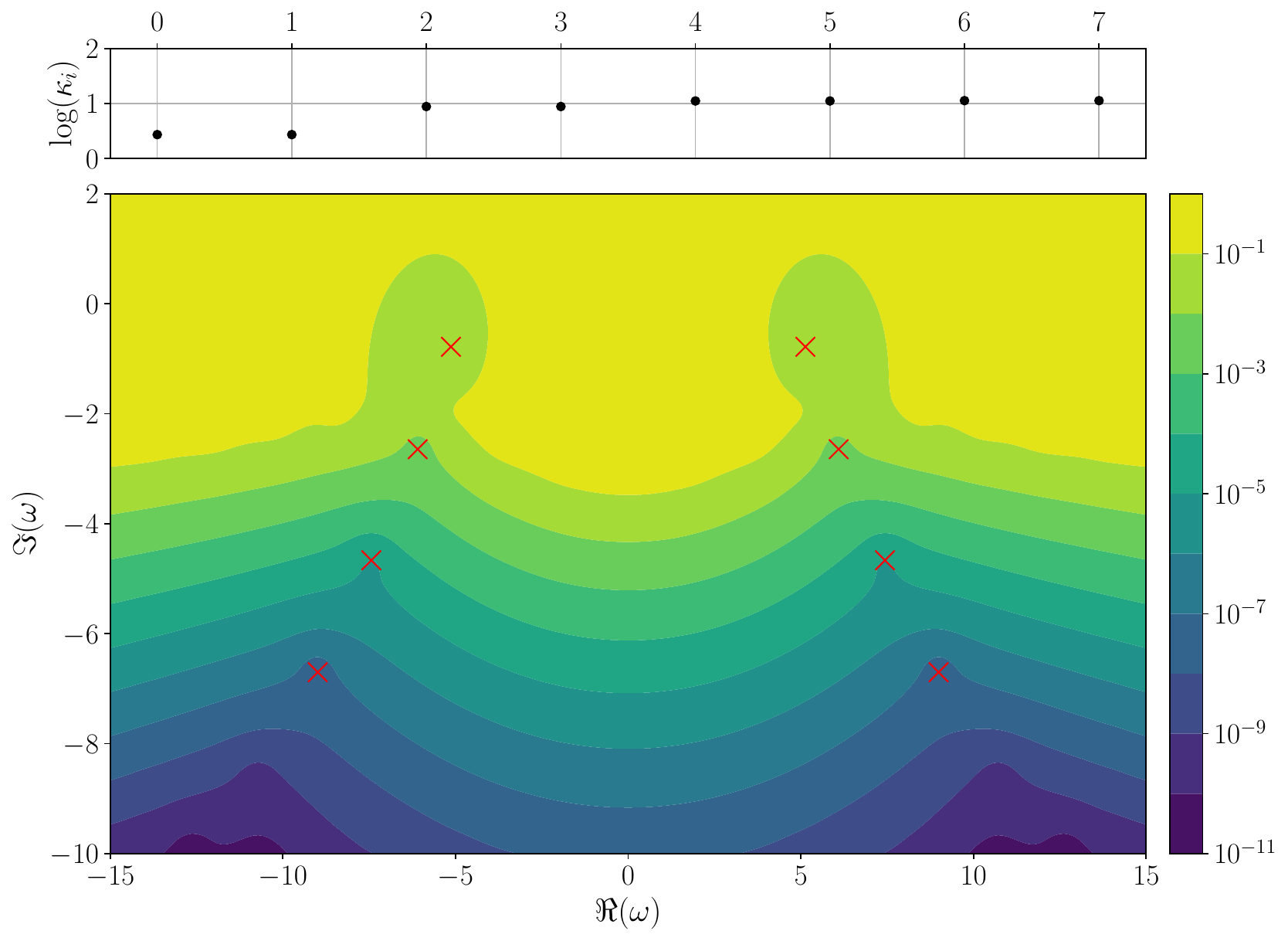}
        \caption{$k=5$, $Q=0$.}
        \label{fig:soundk5q0}
    \end{subfigure}
    \\
    \begin{subfigure}[t]{0.48\textwidth}
        \centering
        \includegraphics[width=\textwidth]{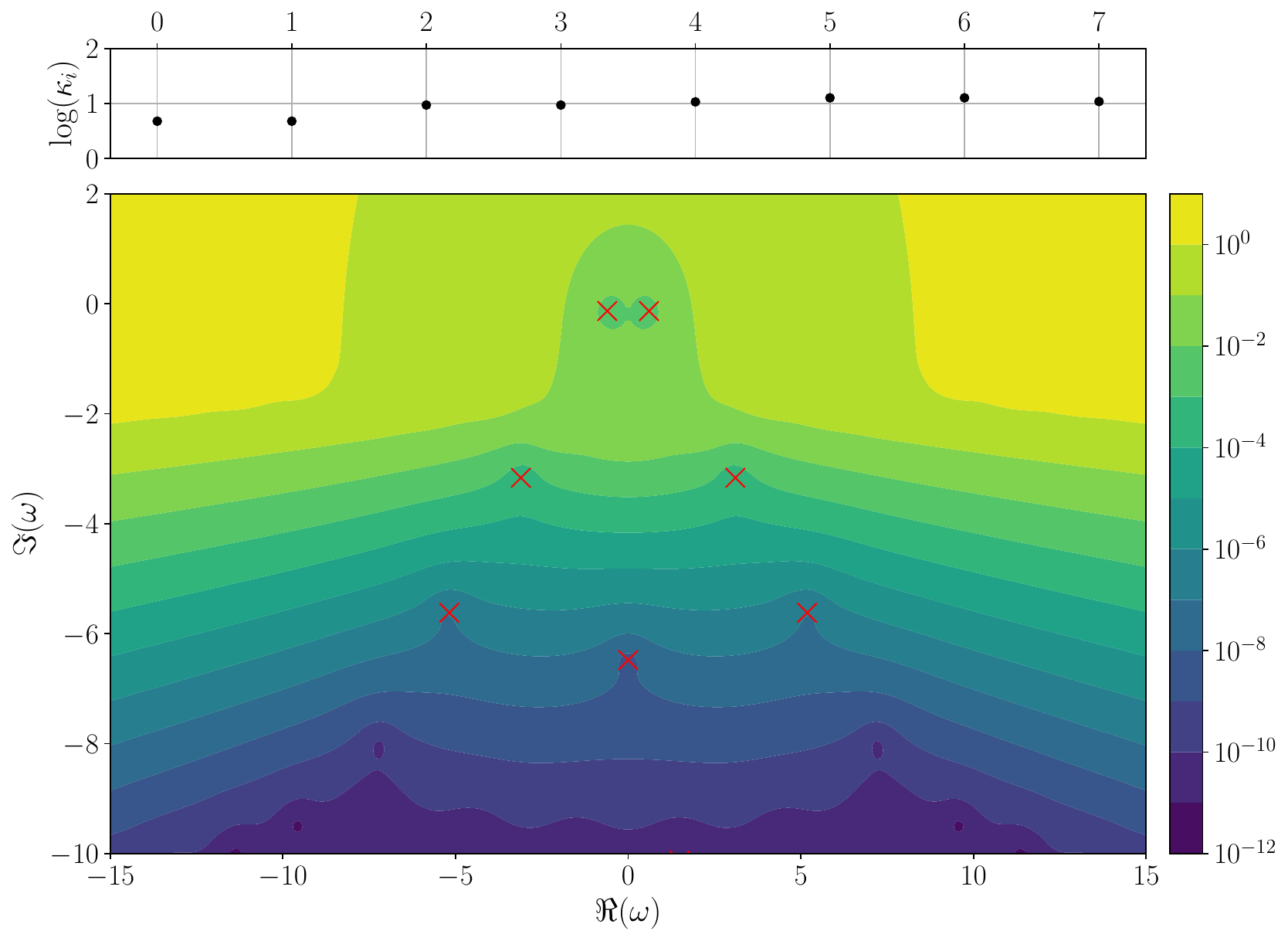}
        \caption{$k=1$, $Q=0.5$.}
        \label{fig:soundk0q0.1}
    \end{subfigure}
    \;
    \begin{subfigure}[t]{0.48\textwidth}
        \centering
        \includegraphics[width=\textwidth]{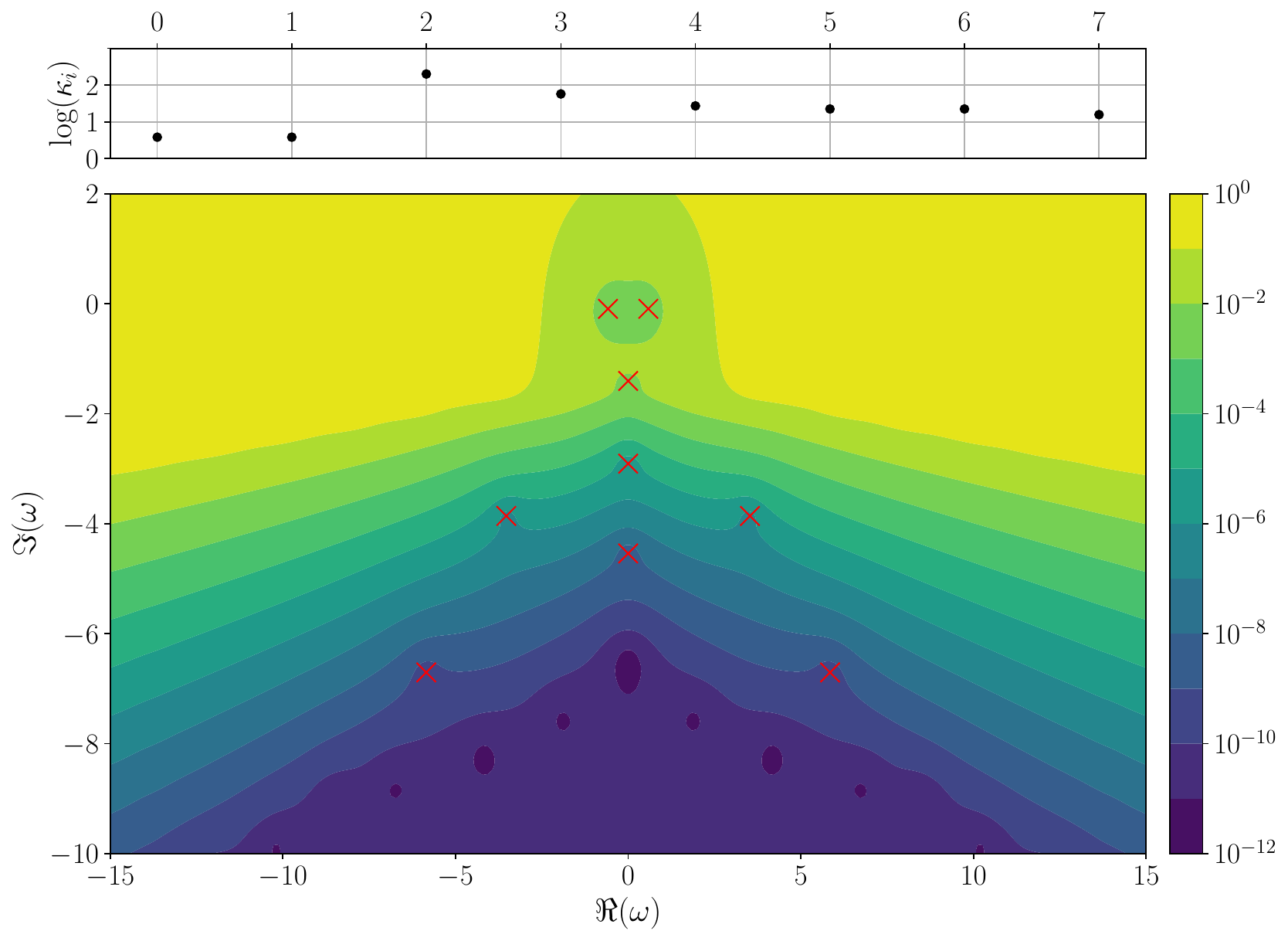}
        \caption{$k=1$, $Q=1=0.7\,Q_\mathrm{max}$.}
        \label{fig:soundk1q1}
    \end{subfigure}
    \caption{The $\epsilon$-pseudospectrum for various values of momentum $k$ and charge $Q$ in the scalar channel, sound sector. Recall that the extremal solution is obtained for $Q_\mathrm{max}=\sqrt{2}$. Above each plot, the condition numbers $\kappa$ of the eigenvalues are plotted with increasing overtone number. The red crosses mark the spectrum points.}\label{fig:sound}
\end{figure}

The pseudospectrum for the shear, transverse gauge, charge diffusion and sound channels are shown respectively in figures~\ref{fig:shear},~\ref{fig:shearZoom},~\ref{fig:transgauge},~\ref{fig:diffusion} and~\ref{fig:sound} for different values of the momentum $k$ and the charge $Q$. The red crosses correspond to the locations of the QNMs for the respective sector. Just like for the scalar field discussed in section~\ref{ssec:scalar}, the pseudospectra form circular
sets if we zoom arbitrarily close to the spectrum, but their large scale global structure present open sets signaling spectral instability in all channels. We also observe that instability increases as the imaginary part of the QNM frequency increases, with implications for the various modes as the momentum $k$ and the charge  $Q$ are tuned.  For a discussion on the holographic interpretation of these results  see section \ref{ssec:dual}.%From the dual field theory side, this can be interpreted as shorter-lived excitations being more unstable. 

With the exception of the transverse gauge, all other sectors contain hydrodynamic modes, namely modes that reside in the origin of the complex frequency plane for zero momentum, i.e.\ $\omega\to0$ as $k\to0$. A zoomed-in version of the pseudospectrum for the shear mode for $Q=0, k=0$ and $Q=0, k=4$ are shown in figure~\ref{fig:shearZoom} --
similar behaviour appears for hydrodynamic sound and charge diffusion modes. In the left panel, the hydrodynamic shear mode is sitting at the origin on the complex plane and we see that even the circular contour line cross in the upper half plane. The change from circular contour lines to open sets occurs at $\epsilon\sim 10^{0.64}$, signaling that one needs a fairly large perturbations in order to destabilise the hydrodynamic mode. In the right panel, the hydrodynamic shear mode is still the dominant mode but it has now moved further down the imaginary axis. In this case, it is only the ``open set'' contour lines that cross to the upper half plane (as we saw in the previous subsection for scalar perturbations), corresponding to $\epsilon\sim 10^{-0.19}$, which is again fairly large.  Note that we have confirmed that the values of $\epsilon$ quoted above do not change with the numerical resolution, i.e.\ the pseudospectrum (associated with the energy norm) is convergent in this region of the complex frequency plane.

Just like for the scalar case studied earlier, the fact that the  pseudospectral contour levels cross to the upper half complex plane signals an unstable perturbed spectrum for large enough perturbations  and potentially connecting with transient instabilities~\cite{treften90hydro,Boyanov:2022ark,Jaramillo:2022kuv}. Further analysis is needed to establish whether these are physical or not. A similar behaviour was observed in 4-dimensional asymptotically flat RN black hole in the extremal limit~\cite{Destounis:2021lum}, where a zero-frequency QNM exists associated to the Aretakis instability~\cite{Aretakis:2011hc}.  Regardless of the fate of transient manifest, we can already argue for the existence of pseudo-resonances at the points where the pseudospectrum crosses the real line~\cite{Boyanov:2022ark,Jaramillo:2022kuv} \textemdash these also give rise to non-linear dynamical instabilities.

Some attention was also given to the extremal limit $T/\mu \to 0$ (or equivalently $Q\to Q_\mathrm{max}$, where $Q_\mathrm{max}=\sqrt{2}$). In terms of the spectrum for the scalar and vector channels, we see an accumulation of poles along the negative imaginary axis that moved upwards from $-i \infty$; the consensus is that these modes will give rise to a branch cut at $T=0$~\cite{Edalati:2010hk,Edalati:2010pn}. We see that the pseudospectrum is deformed in that region to reflect the existence of these new QNMs, while its qualitative behaviour away from the negative imaginary axis remains mostly unchanged; note, in particular, that the asymptotic behaviour of the contour lines at large real part of the frequency stays the same. No other significant features appear.

The pseudospectrum in the extremal limit of asymptotically flat RN black hole was studied in~\cite{Destounis:2021lum}. A significant difference there is that the spectrum contains a branch cut at all values of the temperature and at extremality a zero-frequency mode, associated to
the Aretakis instability~\cite{Aretakis:2011hc}, appears. It was observed that the pseudospectral contour levels cross to the upper half plane around the origin of the complex frequency plane, where the Aretakis mode lives. The extremal limit of Kerr was considered in~\cite{Yang:2022wlm} for energetically infinitesimal perturbations finding instability of the zero-damping modes. Interestingly,~\cite{Yang:2022wlm} found stability of zero-damping modes in near-extremal RN-de Sitter black holes under energetically infinitesimal perturbations.

\subsubsection{Testing the stability of gravitoelectric QNMs}

\begin{figure}[!h]
    \centering
    \begin{subfigure}[t]{0.49\textwidth}
        \centering
        \includegraphics[width=\textwidth]{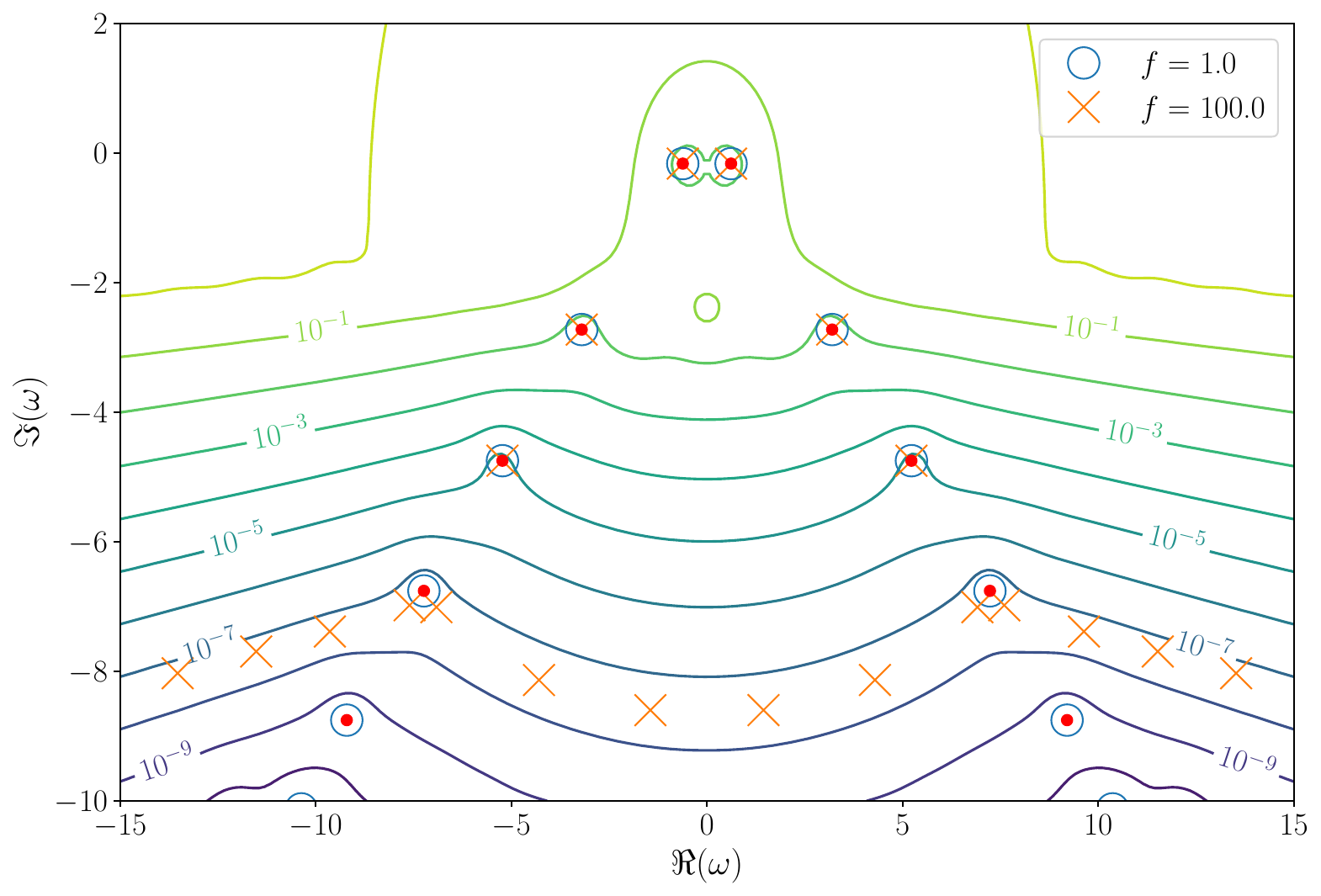}
    \end{subfigure}
    \hspace{-8pt}
    \begin{subfigure}[t]{0.49\textwidth}
        \centering
        \includegraphics[width=\textwidth]{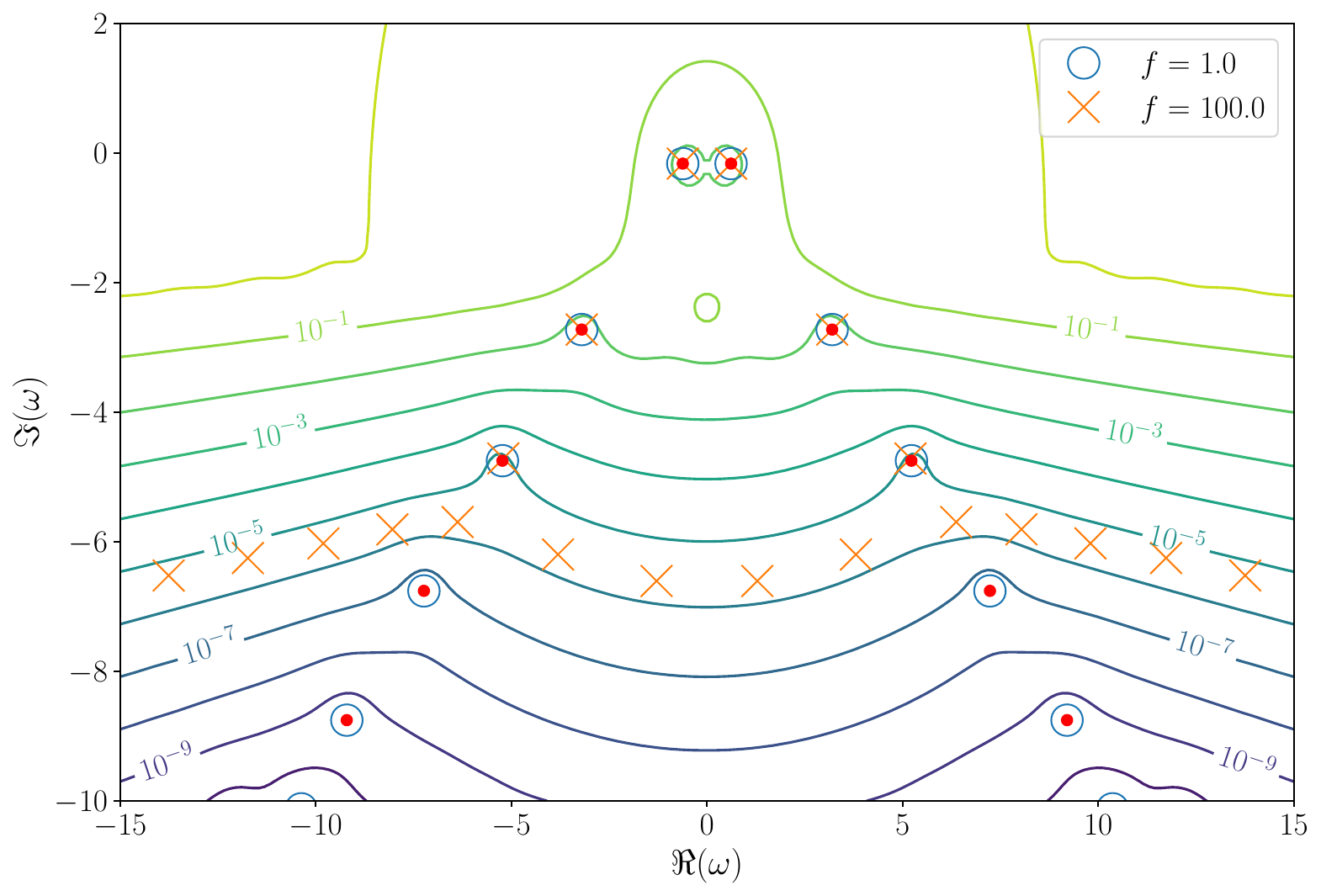}
    \end{subfigure}
    \\ \vspace{-6pt}
    \begin{subfigure}[t]{0.49\textwidth}
        \centering
        \includegraphics[width=\textwidth]{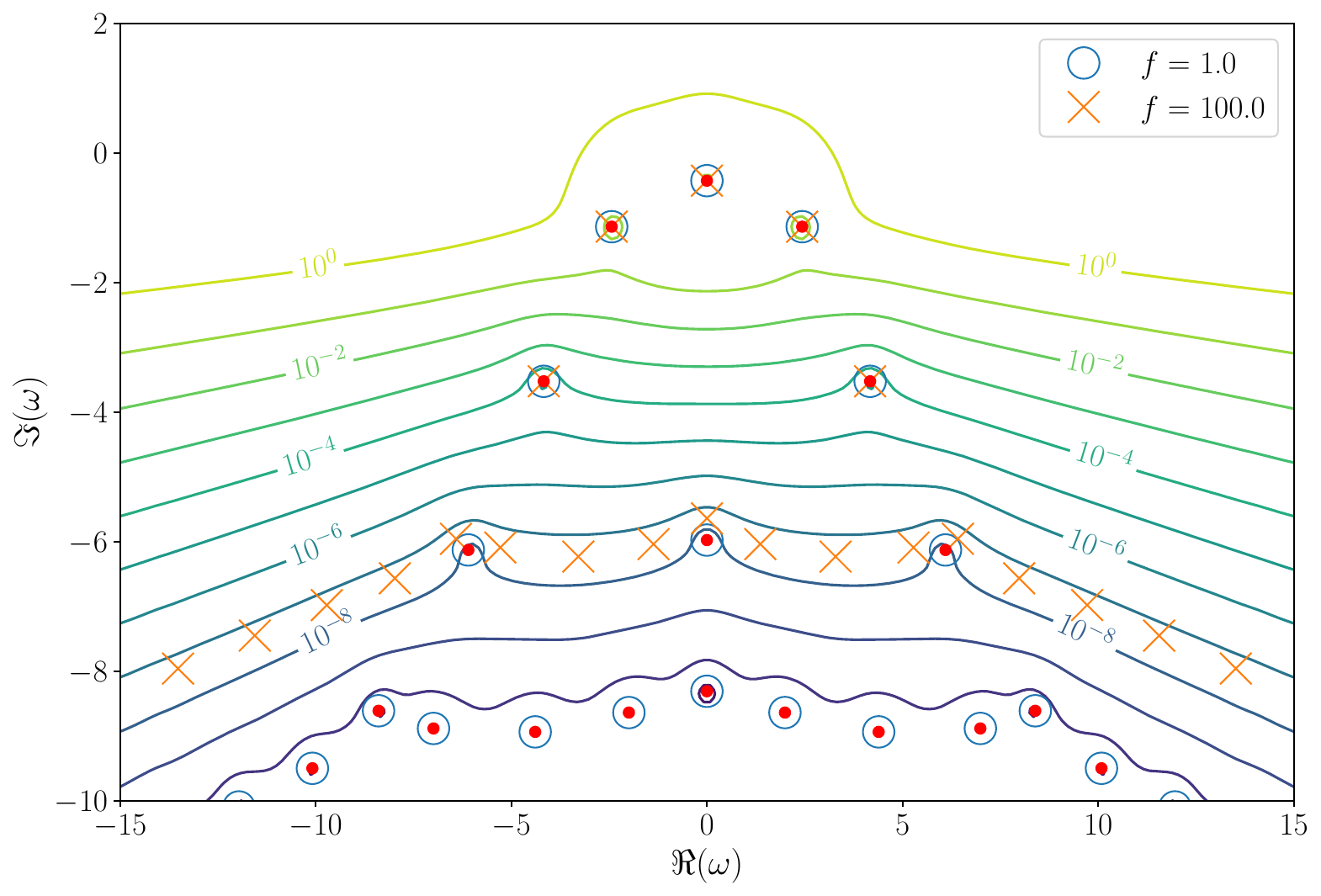}
    \end{subfigure}
    \hspace{-8pt}
    \begin{subfigure}[t]{0.49\textwidth}
        \centering
        \includegraphics[width=\textwidth]{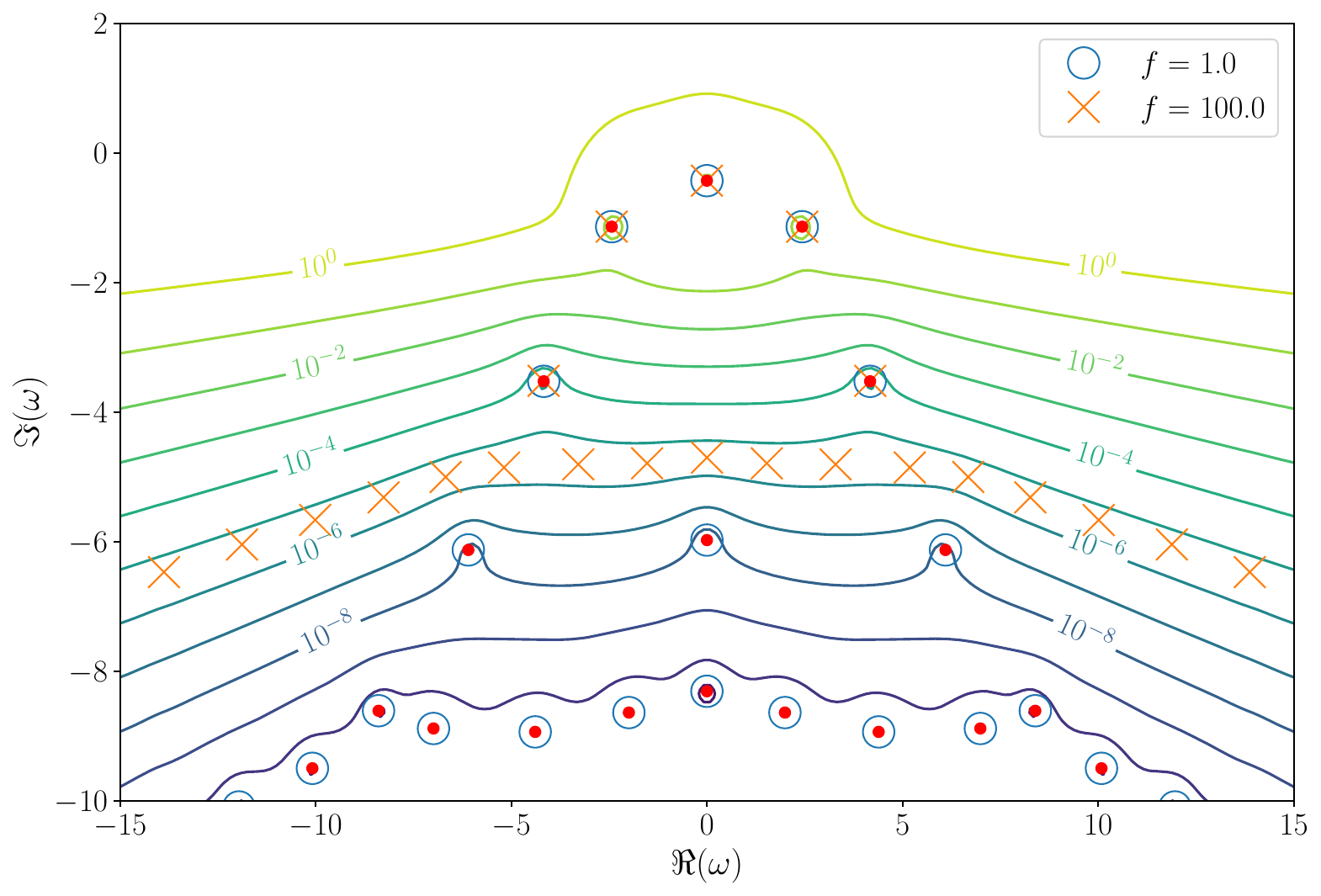}
    \end{subfigure}
    \caption{The perturbed spectra of QNMs in the scalar sector, for both the sound and charge diffusion channels for various AdS-RN configurations. Red dots denote the unperturbed spectra. \emph{Left column}: perturbations of order $\epsilon = 10^{-8}$. \emph{Right column}: perturbations of order $\epsilon = 10^{-6}$. \emph{Rows, from top to bottom}: sound sector, $k=1$ $Q=0$; 
    charge diffusion sector, $k=1, Q=0.8$.}
    \label{fig:pertsScalarSector}
\end{figure}

 Just like the scalar case discussed in the previous subsection, we consider the response of the spectrum under specific perturbations in the potential. In figure~\ref{fig:pertsScalarSector}, we show the effects of adding a sinusoidal perturbation of magnitude $\epsilon =10^{-8}$ (left column), $\epsilon=10^{-6}$ (right column) to the potential in the sound and charge diffusion sectors (scalar channel). We see that the hydrodynamic mode remains stable within this class of perturbations as the perturbation frequency increases up to and including $f=100$ for all values of $k,Q$ that we have considered. Note that in all cases the hydrodynamic modes have smaller condition numbers compared to the non-hydrodynamic modes, hence their increased stability is to be expected. The non-hydrodynamic overtones are progressively more sensitive as the frequency and the overtone number increase. Note that, once again, that the shifted QNMs follow the contours of the $\epsilon$-pseudospectrum.

\subsection{Universality of the pseudospectrum}
\label{ssec:asymptBehaviour}

In this subsection we comment on the behaviour of resonance-free regions: these are regions in the complex frequency plane where no resonance can appear for given potential and boundary conditions and \textit{arbitrary} perturbations.  Following a mathematical analysis  of scattering resonances over general potentials and boundary conditions, it was formally shown that resonance-free regions should belong to one of following four ``universality'' classes depending on the regularity properties of the potential and boundary conditions ~\cite{Zworski}
\begin{equation}
    \Im(\omega)\sim F(\Re(\omega))\,,\qquad \Re(\omega)>>1
\end{equation}
where

\begin{equation*}
F(x)=\begin{cases}
    e^{a x}        & \text{(i)}  \\
    C              & \text{(ii)} \\
    B \, \ln(x)    & \text{(iii)}\\
    \gamma\, x^{b} & \text{(iv)}
    \end{cases}
\end{equation*}
where $a>0$, $b$, $\gamma>0$, $B$ and $C$ are constants and are controlled by the qualitative properties of the underlying system.  Typical behaviours in our setting belong to case (iii) or (iv)~\cite{Zworski}, with logarithmic  asymptotics (case (iii)) appearing when considering potentials and/or boundary conditions allowing low regularity and polynomial asymptotics (case (iv)) in settings with enhanced regularity. 

\begin{figure}[thb]
    \centering
    \includegraphics[width=0.45\textwidth]{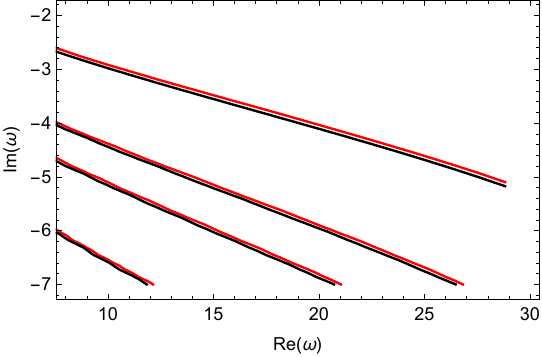}\,\includegraphics[width=0.45\textwidth]{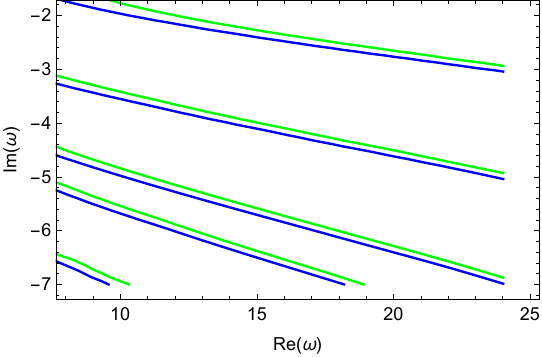}
    \caption{Overlay of asymptotic pseudospectral contour levels for the $Q=0.5, k=1$ AdS-RN black hole. The levels correspond to $\sigma^\epsilon = \{100, 1, 0.005, 0.0005, 0.00001\}$ level curves. \emph{Left:} For the sound (red) and charge diffusion (black) sectors. \emph{Right:} For the shear (blue) and transverse gauge (green) sectors.}
    \label{fig:asymptoticOverlay}
\end{figure}

Figure~\ref{fig:asymptoticOverlay} displays the asymptotic structure of  the pseudospectrum, for the scalar and vector channels for $Q=0.5, k=1$. We see that the pseudospectral levels coincide with each other at large $\Re(\omega)$, demonstrating asymptotic universality. This observation is in agreement with the analysis done in 4-dimensional asymptotically flat spacetime for the Schwarzschild~\cite{Jaramillo:2020tuu,Jaramillo:2021tmt} and RN black holes~\cite{Destounis:2021lum}, where universality was also seen. This can be understood intuitively by re-framing the computation as a ``scattering of a perturbation off a potential in the presence of an obstacle''\footnote{An ``obstacle'' differs from a ``potential'' in the sense that the scattered field cannot penetrate it.}: for sufficiently large frequencies, the potential becomes negligible and thus we expect results that are actually independent from it.

Given the universality observed, it is natural to try to fit the pseudospectrum contour line in an attempt to extract this asymptotic behaviour.
Analysis of the asymptotic structure of pseudospectral contour levels (computed in hyperboloidal slicing) in  4-dimensional asymptotically flat spacetime for both Schwarzschild~\cite{Jaramillo:2020tuu} and RN~\cite{Destounis:2021lum} revealed a logarithmic function. A preliminary analysis of the contour lines in our case  (computed in null slicing) seems to hint towards a polynomial behaviour (case (iv))
\begin{equation}
\Im(\omega)\sim C_1+C_2 \Re(\omega)^\alpha\,
\end{equation}
where $C_1, C_2, \alpha$ are constants, with $\alpha>0$ \textemdash this behaviour seems to persist even in the extremal limit.

In asymptotic flatness the logarithmic behaviour at large $\Re(\omega)$ is supported by a well-motivated conjecture for the existence of logarithmic QNM branches under infinite-frequency perturbations~\cite{Gasperin:2021kfv} (see also~\cite{Zworski}). Given that QNMs can reach logarithmic branches, then necessarily pseudospectrum contour lines must be logarithmic. These results are less robust when considering reflective boundary conditions (such as in AdS spacetimes)  making the deviation from logarithmic asymptotics plausible. More analysis is required to reach a stronger conclusion. Note that, given the different convergence properties in null and hyperboloidal slicing,  comparison with the literature should be done with care.

\subsection{Implication for the dual theory}
\label{ssec:dual}
Other than shedding light on the properties of AdS black branes and on the pseudospectrum itself, this calculation is also relevant for strongly coupled systems via the gauge/gravity correspondence. According to the duality, QNMs correspond to the spectrum of collective excitations of the dual strongly coupled system. Therefore, the spectral instability seen in this work can be interpreted as instability of the excitation spectrum of the dual strongly coupled holographic field theory under small modifications of the theory. The significance of this is twofold. On the one hand, it implies that the frequencies and damping rates computed through an idealised model (such as holography) will potentially differ from those observed in experiments. On the other hand, it indicates that transport properties in these systems (since they are holographically extracted from QNM dispersion relations) could potentially be heavily model dependent, with a small deformation of the original theory leading to potentially very different transport coefficients.  Rephrasing the above in terms of correllation functions, it is well known that QNMs correspond to poles of the retarded Green’s function (in momentum space) of the dual strongly coupled theory \cite{Kovtun:2005ev,Horowitz:1999jd}. The spectral instability observed here signals sensitivity of the location of these poles given external perturbations to the system.

The pseudospectra also indicated possible transient instabilities. This means that under small perturbations of the theory, a quasinormal mode may move into the upper half plane, signaling an instability of the background. From the dual theory perspective,  this means that a small error in the theory may lead in a completely different equilibrium state. Put differently, carrying out our computations in idealised set-ups may lead to a very different equilibrium state than what is seen in practice.

Two more comments are in order.  Firstly, in our results we see that the strength of the spectral instability increases with the overtone; for the dual theory, this implies that short-lived excitation are more unstable, i.e.\ more sensitive to perturbations of the theory. Conversely, we also see that the hydrodynamic mode, residing closest to the real axis, is the most stable, meaning that a stronger perturbation of the model is required to destabilise it. Secondly, and in relation to the hydrodynamic modes once again, we have carried out pseudospectrum calculations at zero momentum (see figure~\ref{fig:shearZoomk0q0} for the shear mode), where we see that the hydrodynamic mode can be pushed in the upper half plane indicating a potential transient instability of flow patterns. Note that instabilities in shear flows are know to exist in hydrodynamics and it has been argued that they explain the early onset of turbulence~\cite{treften90hydro}.

%The spectral instability found herein implies that the spectrum of the dual strongly coupled field theory is also unstable under perturbations and that one should focus more on correlation functions in position-space rather than momentum-space.

\section{Conclusions and outlook}
\label{sec:Conclusions}
In this work we have studied the spectral stability of AdS RN black branes as a function of momenta and charge (or equivalently, $T/\mu$) by means of the pseudospectrum~\cite{trefethen2005spectra}. We considered perturbations corresponding to a neutral, massless scalar field as well as gravitoelectric perturbations in all sectors~\cite{Kovtun:2005ev,Jansen:2019wag}. For all the perturbations considered we find spectral instability for large enough perturbations, including the channels that contain hydrodynamic modes \textemdash the implications of the spectral instability is that the equilibration process of perturbed black branes is sensitive to external perturbations of the system. We observe that the strength of the instability increases as one moves further away for the real axis. Conversely, this means that the fundamental modes/hydrodynamic modes are the most stable. In fact, we find that the perturbations needed to destabilise them can be of order 1. We also see that the pseudospectral contour lines  cross to the upper half plane indicating possible transient instabilities -- further analysis is required to establish if these are physical or not. Such analysis will involve bounding the evolution  operator from above and from below in order to determine whether or not exponential growth takes place. In the case of standard eigenvalue problem these ideas have been explored in \cite{Boyanov:2022ark,Jaramillo:2022kuv, Jaramillo:2021tmt}. However, for generalised eigenvalue problem, defining the evolution  operator is non-trivial, and thus the study of transient is not as developed;  a definition for the evolution operator was put forward in \cite{embree2017pseudospectra}. The existence of potential transients is absent in hyperboloidal slicing~\cite{Arean:2023ejh}. Note that transients are in fact known to arise in hydrodynamics and it has been argued that they explain the early onset of turbulence~\cite{treften90hydro}.  

Regardless of the outcome of the analysis outlined above for transient instabilities, given our results we can already argue about the presence of another mechanism that can give rise to non-linear instabilities: pseudo-resonances~\cite{Boyanov:2022ark,Jaramillo:2022kuv}. The latter can be triggered by external forces when the Fourier decomposition of these forces gives frequencies  close to values where pseudospectrum contour lines cross the real axis into the upper half plane. This is particularly relevant for the non-linear interactions, since the first order perturbation acts as a source at second order. The formal solution at second order is obtained by convolving the source with the Green's function, which is given by the resolvent of the linear operator. Thus, at the crossing points, where the resolvent is infinite, the second order perturbation exhibits growth which breaks down the perturbative expansion and leads to the non-linear dynamical instabilities.

We have also investigated the asymptotic behaviour of pseudospectral contours and we find universality across different perturbation sectors. This extends the universality observed for asymptotically flat black holes \cite{Jaramillo:2020tuu,Destounis:2021lum}. Some preliminary analysis indicated that our curves may be fitted better by a polynomial at large real frequency, rather than a logarithmic function as an asymptotic behaviour; further analysis is required in order to reach a solid conclusion, however, this result is substantially different from what has been seen in the pseudospectrum literature in general relativity so far. The difference might be due to the fact that we are considering a ``scattering problem over a potential and with an obstacle", rather than a ``scattering problem over a potential" as in the case of asymptotically flat black holes.

In this work, we have have formulated the pseudospectrum calculation in ingoing Eddington-Finkelstein coordinates for the first time. The equation of motion for the perturbations become first order in time and the time derivative appears in a mixed derivative term, which naturally leads to a formulation of the pseudospectrum in terms of a generalised eigenvalue problem~\cite{trefethen2005spectra}. This choice of coordinates together with a suitable rescaling of the field not only substantially simplifies the problem and the imposition of boundary conditions, and significantly reduces the condition number of the associated matrices (defined as the ratio of the largest and smallest eigenvalue) making the numerical implementation less demanding. Crucially,  using these coordinates was shown to have superior convergence properties, making the results shown in this paper stronger and more reliable than other results on the pseudospectrum in general relativity.  
Comparing our results with those of~\cite{Arean:2023ejh} is difficult due to the different convergence properties in the two types of spacetime slicing~\cite{Boyanov:2023qqf} (their results do not exhibit convergence). A notable difference is that, unlike~\cite{Arean:2023ejh}, we find that the contour lines cross in the upper half plane indicating the possibility of transients.  This difference highlights the need for a better understanding of the dependence of the pseudospectrum on the particular slicing chosen.

It would be interesting to investigate the implications of spectral instability in the context of the phenomenon of pole-skipping~\cite{Grozdanov:2017ajz}.  Holographically, pole-skipping has been studied for the AdS-Schwarzschild black brane \cite{Grozdanov:2017ajz}, AdS-RN~\cite{Jansen:2020hfd}  as well as for a neutral black hole that explicitly breaks translation invariance~\cite{Blake:2019otz}. This phenomenon is now understood to be generic, manifesting itself in all hydrodynamic and non-hydrodynamic channels \cite{Grozdanov:2019uhi,Blake:2019otz}, and it has been associated with absence of a unique ingoing solution close to the black brane horizon \cite{Blake:2019otz}. Pole-skipping is particularly interesting in the case of the hydrodynamic sound mode, where it was shown that when the sound mode is driven to instability by a choice of a specific value of imaginary momentum $k=ik_*$, then the retarded two-point function exhibits an exponential growth related to chaos; the frequency and momentum are given by the holographic Lyapunov exponent and the butterfly velocity \cite{Grozdanov:2017ajz}.  If spectral instability persists for imaginary $k$, we expect the physical relevance of this phenomenon to be substantially reduced and one would need to understand the implications of this in the context of quantum chaos.  In any case, it would be interesting to study the contour lines of the pseudospectrum slightly before and slightly after the pole-skipping point. A technical complication here is that the momentum $k$ is taken to be imaginary, but one needs to ensure that the energy norm remains real.

In addition, it would also be interesting to investigate the impact of these results in the context of the radius of convergence of the hydrodynamic expansion, within the framework of holography. Strictly speaking, in order to compute the radius of convergence one needs to perform a perturbative expansion of the dispersion relations in small momenta to a very high order~\cite{Withers:2018srf, Jansen:2020hfd}, extract the coefficients and then perform a fit. In~\cite{Jansen:2020hfd} it was found that the radius of convergence for a charged holographic fluid extracted this way agrees quantitatively with the locations of the branch points obtained by considering the analyticity properties of the dispersion relations of the hydrodynamic modes on the complex frequency and momentum plane. Understanding how the two computations are modified given the instability of the spectrum and how they compare will be valuable.

In this work we have also probed the extremal limit of the RN black brane, by considering $T/\mu\to 0$~\cite{Edalati:2010pn,Edalati:2010hk}. We found that the pseudospectrum was sufficiently deformed along  the negative imaginary axis to accommodate for the additional modes that appear there. It would be interesting to repeat the pseudospectrum analysis starting directly at zero temperature.

\appendix

\section{Massless scalar perturbations in global AdS$_4$}
\label{sec:SAdS}

To contrast with the cases presented in the main text, it is illuminating to analyse a system where there is energy conservation. One such example is that of pure AdS with reflecting boundary conditions. Since there is no black hole horizon, and the (timelike) AdS boundary acts like a box, no energy is lost through the boundaries of the domain.

Let us then consider the familiar case of global AdS$_4$, with metric
\begin{align}
    ds^2 = -f(r)dt^2 + \frac{dr^2}{f(r)} + r^2(d\theta^2 + \sin^2\theta \, d\varphi^2),
\end{align}
where $f(r) = r^2 +1$.

As before, the equation to solve is
\begin{equation}
    \square \Phi = 0 .
\end{equation}

In this case, the ingoing Eddington-Finkelstein coordinates used in the main text are not particularly helpful, so we will proceed with the global $(t,r)$ coordinates instead.
With the ansatz 
\[
\Phi =  \frac{1}{r} \phi(t,r) Y^m_\ell(\theta, \varphi)
\]
and coordinate transformation
\begin{equation}
\label{eq:tortoise-def}
    \frac{dr}{dx} = f(r) \equiv r^2 + 1 ,
\end{equation}
the Klein-Gordon equation reduces to
\begin{align}
\label{eq:KG-tx1}
    & \left(-\partial^2_t + \partial^2_x - V(r) \right) \phi(t,r) = 0, \\
    & V(r) \equiv f(r) \left(2 + \frac{\ell(\ell+1)}{r^2} \right). \notag
\end{align}
We can explicitly solve equation~\eqref{eq:tortoise-def} to find
\begin{equation}
    r = \tan(x),
\end{equation}
so that $x=0$ at $r=0$ and $x=\frac{\pi}{2}$ at $r=\infty$.

Due to the lack of energy dissipation, this spacetime features \emph{normal} modes, and in fact the spectrum for scalar perturbations is given by~\cite{Berti:2009kk}
\begin{equation}
\label{eq:spectrum_AdS4}
\omega = 2n + 1 + \ell, \qquad n = 0, 1, 2, \ldots
\end{equation}

To write equation~\eqref{eq:KG-tx1} in a first-order (in time) form, we introduce $\Pi \equiv \partial_t \phi$. 
It is also convenient to have the compact $x$-domain spanning the interval $x\in[0,1]$, so we further do $x\to \frac{\pi}{2} x$. Equation~\eqref{eq:KG-tx1} then becomes
\begin{equation}
    \partial_t 
    \begin{pmatrix}
        \Pi \\
        \phi
    \end{pmatrix}
    = 
    \begin{pmatrix}
        0           & \frac{4}{\pi^2} \partial^2_x - V \mathbb{1} \\
        \mathbb{1}  & 0
    \end{pmatrix}
    \begin{pmatrix}
        \Pi \\
        \phi
    \end{pmatrix}.
\end{equation}
Defining $u = (\Pi, \phi)^T$, we can write this as 
\begin{equation}
\label{eq:Lop-def}
    \partial_t u = i \hat L \, u, \qquad
    i \hat L \equiv     
    \begin{pmatrix}
        0           & \frac{4}{\pi^2} \partial^2_x - V \mathbb{1} \\
        \mathbb{1}  & 0
    \end{pmatrix}
\end{equation}
where
\begin{equation}
    V \equiv (r^2 + 1) \left(2 + \frac{\ell(\ell+1)}{r^2} \right), \quad
    r = \tan\left(\frac{\pi}{2} x\right)
\end{equation}
$V$ becomes singular at $x=0$ and $x=1$, but note that $V \phi$ is regular (and indeed zero) at these points since
\begin{equation}
    \phi \underset{r \sim \infty}{\sim} \frac{1}{r^2}, \qquad
    \phi \underset{r \sim 0}{\sim} r^{\ell + 1}.
\end{equation}

As in section~\ref{ssec:QNMs and energy norm}, we can define an energy norm given by
\begin{equation}
    \label{eq:energy-tr}
    E = \frac{\pi}{4} \int_0^1 \left(
    \partial_t \bar \phi \, \partial_t \phi
    + \frac{4}{\pi^2} \partial_x \bar \phi \, \partial_x \phi
    + V \bar \phi \, \phi
    \right) dx
\end{equation}
which motivates the definition of the scalar product
\begin{equation}
    \label{eq:scalar-prod-AdS-2}
    \langle u_1, u_2 \rangle
    =  \frac{\pi}{4} \int_0^1 \left(
    \bar \Pi_1 \, \Pi_2
    + \frac{4}{\pi^2} \partial_x \bar \phi_1 \,  \partial_x \phi_2
    + V \bar \phi_1 \, \phi_2
    \right) dx\,.
\end{equation}
With this definition and our choice of boundary conditions $u(x=0) = 0 = u(x=1)$, we can check that $\langle u_1, \hat L u_2 \rangle = \langle \hat L u_1, u_2 \rangle $, and so $\hat L^\dagger = \hat L$ as expected.

We now follow the approach outlined in section~\ref{sec:implementation}, where we define the Gram matrix $G^E$ for the discretised scalar product. In particular, we define 

\begin{align}
\label{eq:ScalarProd_AdS}
\langle u_1,u_2\rangle_{_G} & = u_1^*\cdot G^E \cdot u_2 
= (\bar{\Pi}_1 \ \bar{\phi}_1)
\left(
  \begin{array}{c|c}
    G^E_1   & 0 \\
    \hline 
   0 & G^E_2
  \end{array}
  \right)
  \begin{pmatrix}
  \Pi_2 \\
  \phi_2
  \end{pmatrix}
\end{align}
where
\begin{align*}
 & G^E_1 = \frac{\pi}{4}C_1, \qquad
  G^E_2 = \frac{\pi}{4} \left( C_{V} 
  + \mathbb{D}^t \cdot C_2 \cdot \mathbb{D}
  \right) \\
& (C_1)_{ij} = W_i \, \delta_{ij}, \qquad (C_2)_{ij} = \frac{4}{\pi^2} W_i \, \delta_{ij}, \qquad (C_V)_{ij} = V(x_i) \, W_i \, \delta_{ij}
\end{align*}
(no sum in $i$) and 
$W_i$ are the quadrature weights introduced in section~\ref{ss:chebyshev int}.
Finally, note that the end points where the potential $V$ diverges contribute to zero to the integration due to the boundary conditions used. Therefore the corresponding rows and columns can be removed from the $G^E$ matrix.
We can now proceed as outlined in section~\ref{ss: e-norm to L2-norm} to compute the pseudospectrum, using equation~\eqref{eq:pencil_PS3}.

\begin{figure}[thbp]
    \centering
    \includegraphics[width=0.48\linewidth]{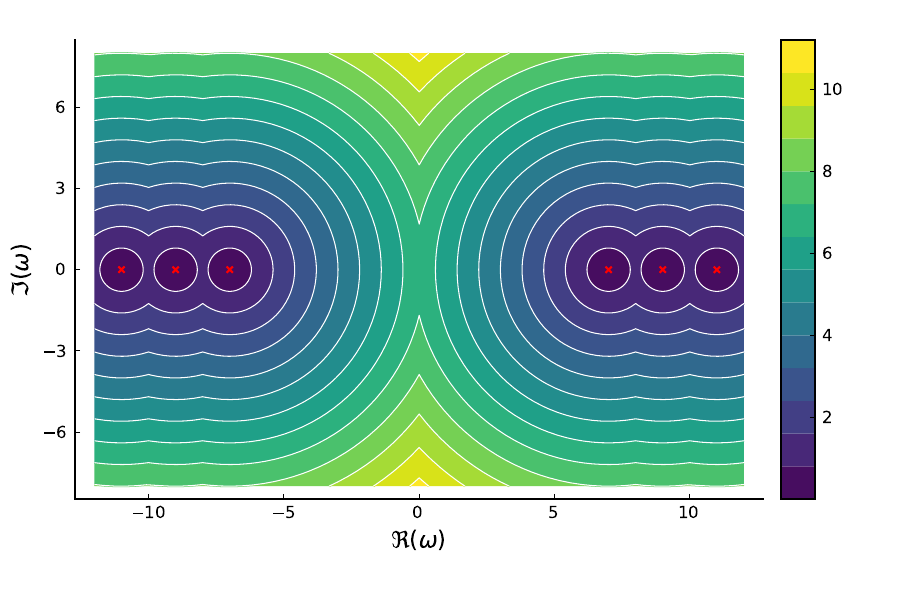}
    \hfill
    \includegraphics[width=0.48\linewidth]{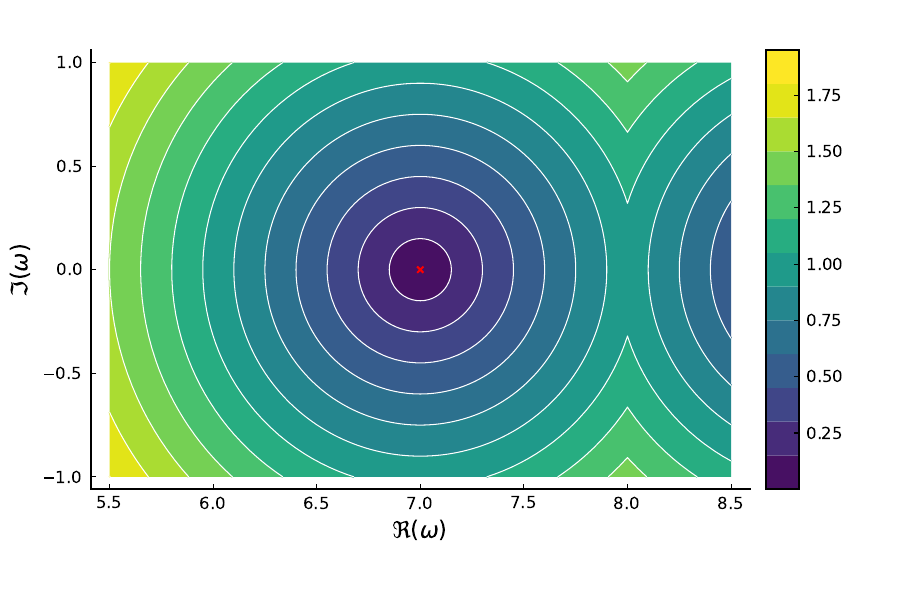}
    \caption{The $\epsilon$-pseudospectrum for the $\ell=2$ case of  global AdS$_4$.
    The contour lines correspond to level sets of $\epsilon$, as defined in equation~\eqref{eq:sigmaG_from_sigma2}, and the red crosses mark the spectrum points from equation~\eqref{eq:spectrum_AdS4}. \emph{Left:} a wide view of the pseudospectrum. \emph{Right:} a zoomed view of the pseudospectrum near the mode $\omega_0 = 3$.
    \label{fig:AdS4_l2_PS}}
\end{figure}

In figure~\ref{fig:AdS4_l2_PS} we observe an example of this computation. In sharp contrast with the case of figure~\ref{fig:AdSplanark0q0_single}, the contour levels are now perfectly spherical, indicating stability of the spectrum.

\section{Convergence}
\label{sec:convergence}

\begin{figure}[t]
    \centering
    \begin{subfigure}[t]{0.45\textwidth}
        \centering
        \includegraphics[width=\textwidth]{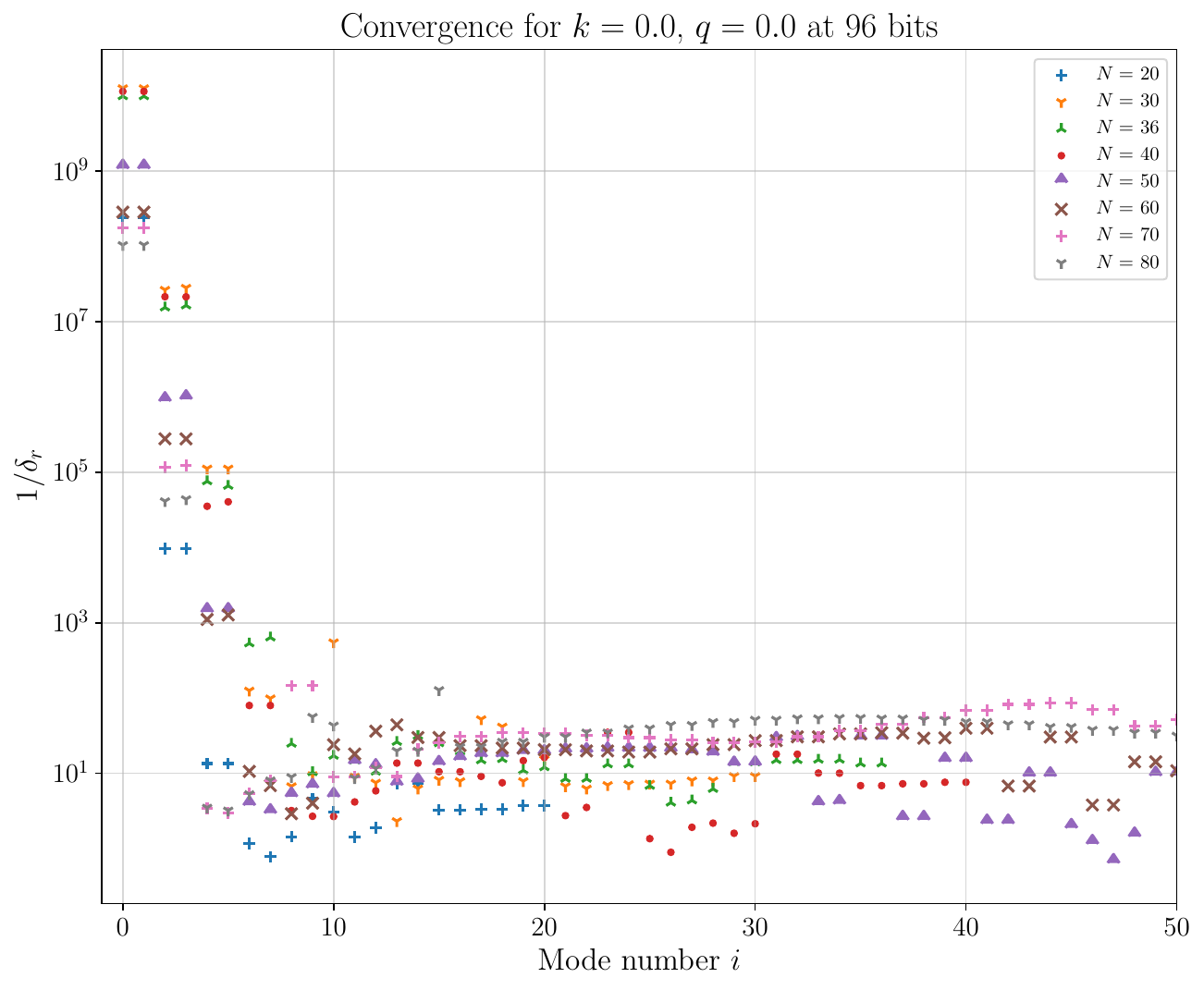}
        \label{fig:convergencek0q0}
        \caption{Convergence of the spectrum for the ${k=0}$, ${Q=0}$ planar black hole in AdS$_5$.}
    \end{subfigure}
    \;
    \begin{subfigure}[t]{0.45\textwidth}
        \centering
        \includegraphics[width=\textwidth]{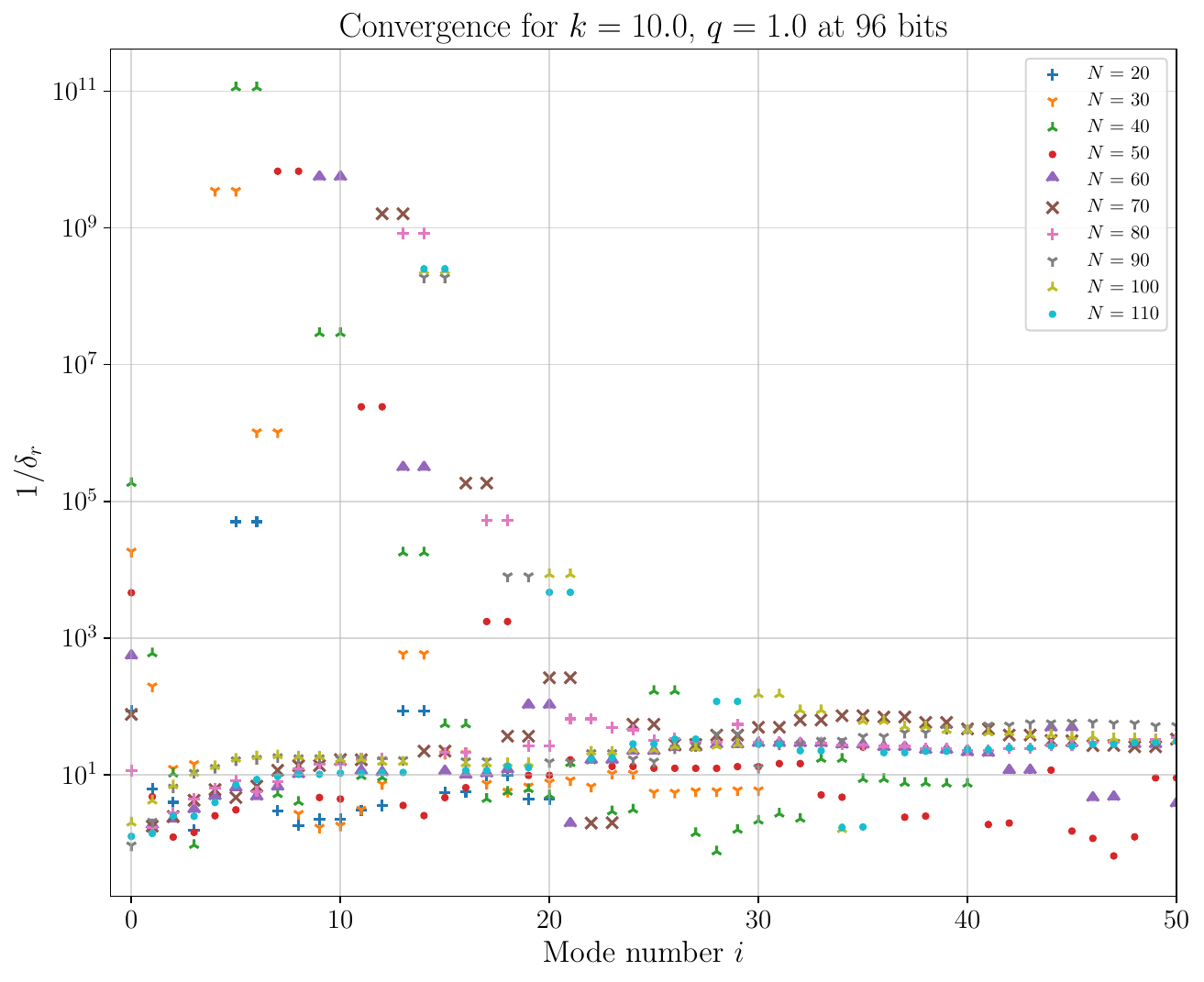}
        \label{fig:convergencek10q099}
        \caption{Convergence of the spectrum for {$k=0$, $Q=1=0.7 Q_\mathrm{max}$} planar black hole in AdS$_5$.}
    \end{subfigure}
    \caption{Convergence of QNM spectra for different planar black holes in AdS$_5$ in terms of the drift ratio, $\delta_r$. In each case, 96 bits of precision were used. Here $Q_\mathrm{max}=\sqrt{2}$.}
    \label{fig:conv}
\end{figure}

\paragraph{Convergence of quasinormal modes.} \hspace{-0.3cm} To measure the convergence of the QNM spectra $\lambda_j$ with increasing resolution $N$, the drift ratio $\delta_r$ was adopted as defined in~\cite{BOYD199611}. This measures the weighted distance between eigenvalues at spectral order $N_1$ and $N_2$ with $N_1 < N_2$ via
\begin{align}
    \delta^{(j)}_r = \frac{\min(|\lambda_j|, \sigma_j)}{|\lambda_j - \bar\lambda_2|}
\end{align}
where $\bar \lambda_2$ is the nearest eigenvalue in the $N_2$ set to the current eigenvalue $\lambda_j$ of the $N_1$ set based on $\min |\lambda_j - \lambda_2|$. The weights $\sigma_j$ are given by
\begin{align}
    \sigma_j = \begin{dcases}
        |\lambda_j - \lambda_1| & \text{if} \; j = 0 \\
        |\lambda_{N-1} - \lambda_j| & \text{if} \; j = N \\
        \frac{1}{2}\left(|\lambda_{j-1} - \lambda_j| + |\lambda_{j+1} - \lambda_j|\right) & \text{if} \; 0 < j < N\, .
    \end{dcases}
\end{align}
In figure~\ref{fig:conv}, we demonstrate this measure of convergence when applied to the scalar field spectrum shown in figures~\ref{fig:AdSplanark0q0}~and~\ref{fig:AdSplanark10q1}. We plot the logarithm of the inverse of the drift ratio against the mode number for several choices of spectral degree, $N$. Each set of points is compared with the spectrum of the next-largest spectral degree, i.e.\ when ${N_1 = 30}$, ${N_2 =40}$, etc. Convergent eigenvalues have large inverse drift ratio compared to the background. In the uncharged case, eigenvalues appear as complex conjugate pairs while in the near-extremal case there is a single non-oscillating mode before the appearance of conjugate pairs. The benefit of the drift ratio measure is that it is robust against spurious modes arising from numerical errors, as these will never remain close to meaningful eigenvalues as the spectral degree changes. 

\begin{figure}[thbp]
    \centering
    \includegraphics[width=0.48\textwidth]{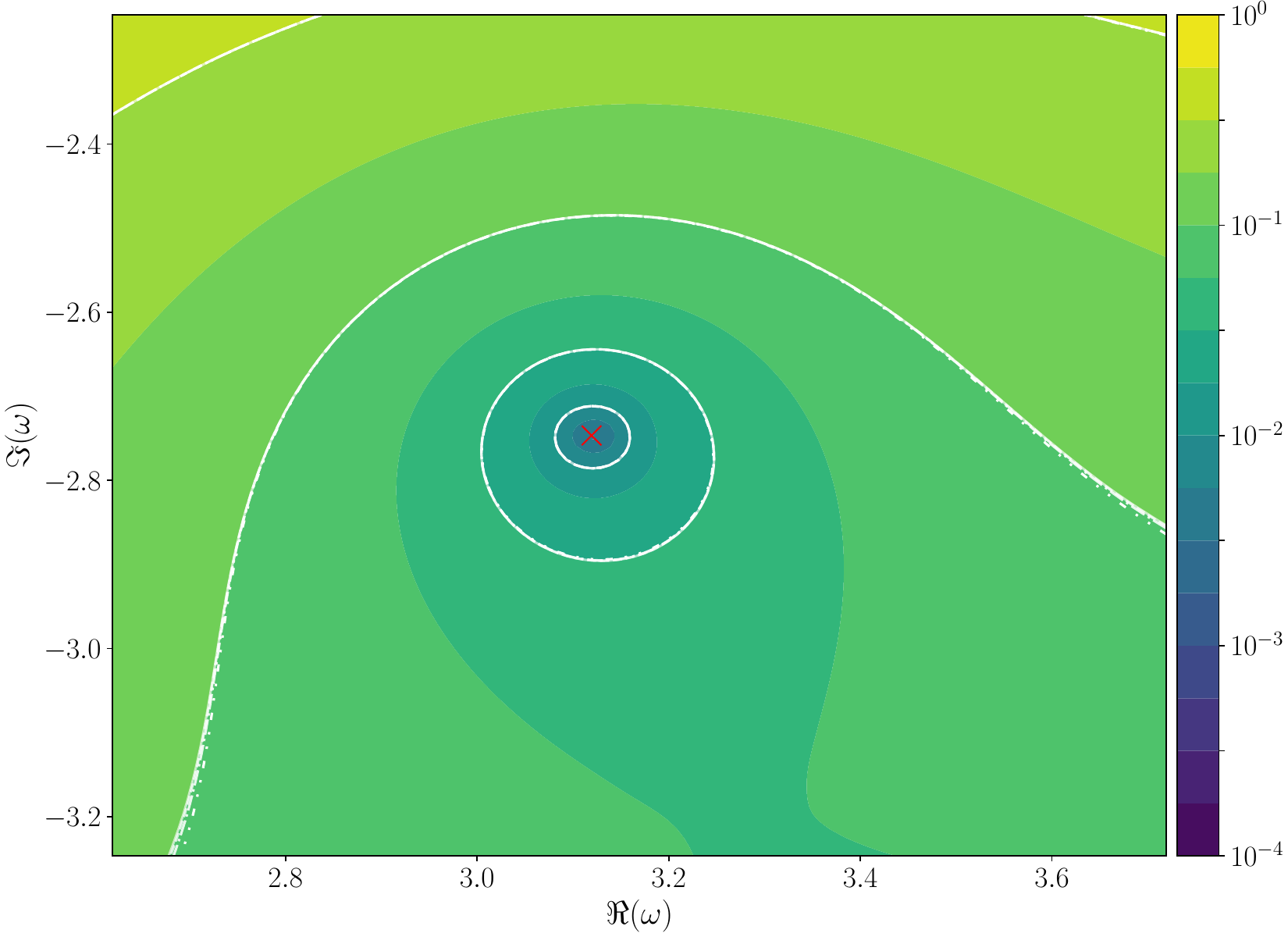}
    \hfill
    \includegraphics[width=0.48\textwidth]{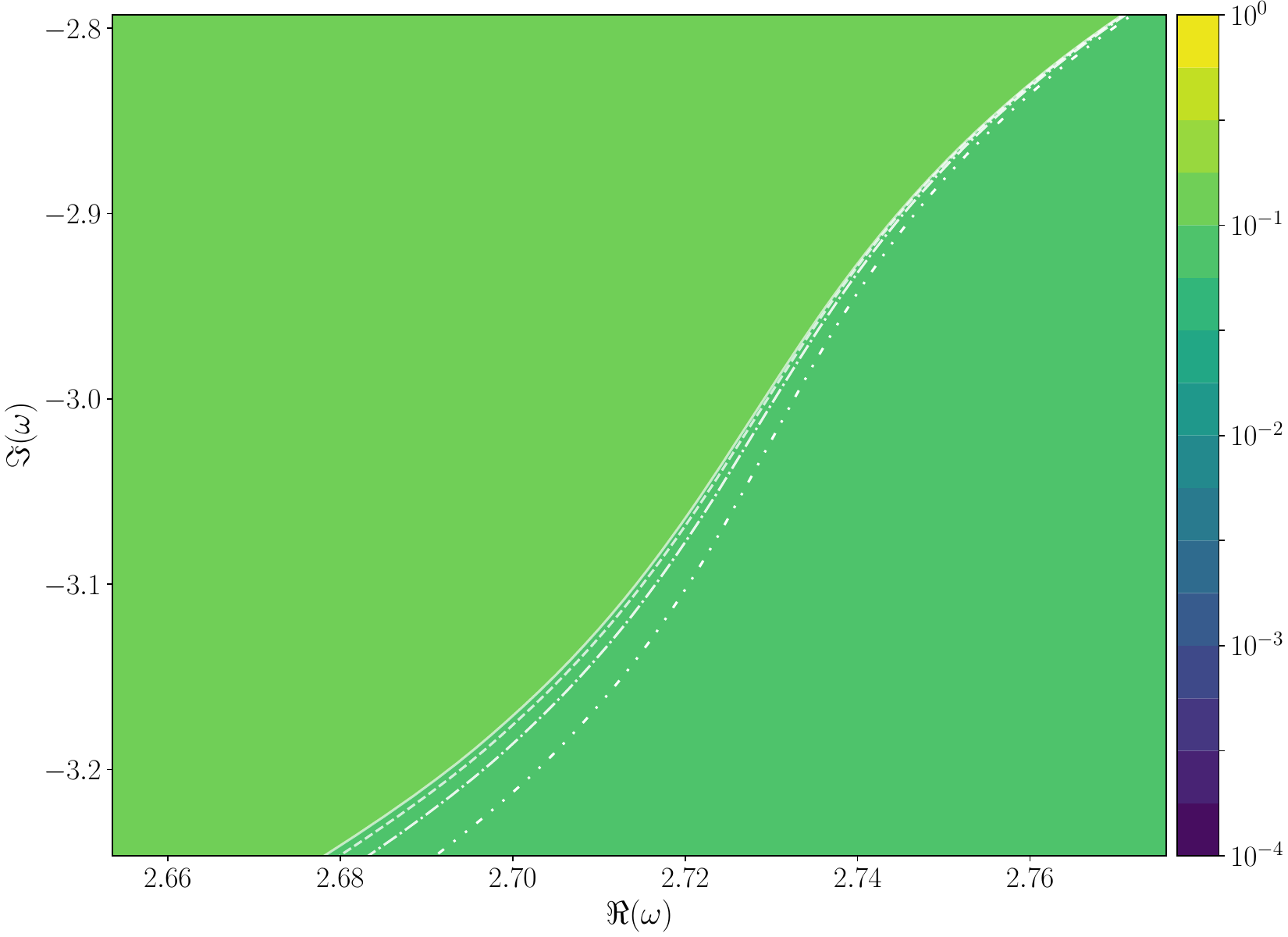}
    \caption{Convergence with increasing spectral resolution $N$ of the $\epsilon$-pseudospectrum $\sigma^\epsilon$ with respect to the $H_2$ norm for the massless, neutral scalar around a $k=0$, $Q=0$ black brane. The contour lines are calculated with $N = 20$ (dash-dot-dot), $N=30$ (dash-dot), $N=40$ (dashed) and $N=50$ (solid) \emph{Left:} the fundamental mode $\omega_0 = 3.119 - 2.747 i$ is represented by the red cross; evenly spaced contours are shown in white. \emph{Right:} a zoomed view of the contours demonstrates clear convergence.}
    \label{fig:pspec_convergence}
\end{figure}

\paragraph{Convergence of the pseudospectrum.} \hspace{-0.3cm} To ensure the convergence of the pseudospectrum with increasing spectral resolution, we calculate a set of contours $\sigma^\epsilon$ for fixed values of $\epsilon$ with increasing resolution and superimpose the results. Figure~\ref{fig:pspec_convergence} demonstrates the convergence of these contours in a region close to the fundamental mode $\omega_0 = 3.119 - 2.747i$ in the massless, neutral scalar spectrum of a $k=0$, $Q=0$ black brane. Note that the fundamental mode lies outside the convergent part of the pseudospectrum when using the energy norm~\cite{Destounis:2023nmb}. For this reason, figure~\ref{fig:pspec_convergence} has been produced using the $H^2$ norm which ensures convergence up to $0\geq\Im(\omega) \geq -4$.

\acknowledgments
It is a pleasure to thank Valentin Boyanov, Jose Luis Jaramillo, Rodrigo Panosso Macedo and Benjamin Withers for discussions.
C.P.\ thanks the Princeton Center for Theoretical Science for hospitality during the programme ``Workshop on Nonlinear Aspects of General Relativity''.
C.P.\ and B.C.\ acknowledge support from a Royal Society -- Science Foundation Ireland University Research Fellowship via grant URF/R1/211027. 
M.Z.\ acknowledges financial support by the Center for Research and Development in Mathematics and Applications (CIDMA) through the Portuguese Foundation for Science and Technology (FCT -- Fundação para a Ciência e a Tecnologia) -- references UIDB/04106/2020 and UIDP/04106/2020 -- as well as FCT projects 2022.00721.CEECIND, CERN/FIS-PAR/0027/2019, PTDC/FIS-AST/3041/2020, CERN/FIS-PAR/0024/2021 and 2022.04560.PTDC.
This work has further been supported by the European Horizon Europe staff exchange (SE) programme HORIZON-MSCA-2021-SE-01 Grant No.\ NewFunFiCO-101086251.

\bibliographystyle{JHEP}
\bibliography{refs}

\end{document}